# A Spline-based Volumetric Data Modeling Framework and Its Applications

A Dissertation Presented

by

**Bo Li**

to

The Graduate School

in Partial Fulfillment of the Requirements

for the Degree of

**Doctor of Philosophy**

in

**Computer Science**

Stony Brook University

August 2012

**Stony Brook University**

The Graduate School

# Bo Li

We, the dissertation committee for the above candidate for the Doctor of Philosophy degree, hereby recommend acceptance of this dissertation.

Hong Qin – Dissertation Advisor
Professor, Computer Science Department

Xiangmin Jiao – Chairperson of Defense
Associate Professor, Applied Mathematics and Statistics Department

Xianfeng Gu
Associate Professor, Computer Science Department

Qiaode Ge
Professor
Mechanical Engineering Department

This dissertation is accepted by the Graduate School.

Lawrence Martin
Dean of the Graduate School

Abstract of the Dissertation

# A Spline-based Volumetric Data Modeling Framework and Its Applications

by

**Bo Li**

**Doctor of Philosophy**

in **Computer Science**

Stony Brook University

2012

The rapid advances in 3D scanning and acquisition techniques have given rise to the explosive increase of volumetric digital models in recent years. This dissertation systematically trailblazes a novel volumetric modeling framework to represent 3D solids. The need to explore more efficient and robust 3D modeling framework has gained the prominence. Although the traditional surface representation (e.g., triangle mesh) has many attractive properties, it is incapable of expressing the interior space and materials. Such a serious drawback overshadows many potential modeling and analysis applications. Consequently volumetric modeling techniques become the well-known solution to this problem. Nevertheless, many unsolved research issues remain when developing an efficient modeling paradigm for existing 3D models: complex geometry (fine details and extreme concaveness), arbitrary topology, heterogenous materials, large-scale data storage and processing, etc.

In this dissertation, we concentrate on the challenging research issue of developing a spline-based modeling framework, which converts the conventional data (e.g., surface meshes) to tensor-product trivariate splines. This methodology can represent both boundary/volumetric geometry and real volumetric physical attributes in a compact and continuous fashion. The regular tensor-product structure enables our new developed methods to be embedded into the industry standard seamlessly. These properties make our techniques highly preferable in many physically-based applications including mechanical analysis, shape deformation and editing, virtual surgery training, etc.

Using tensor-product trivariate splines to reconstruct existing 3D objects is highly challenging, which always involves component-based decomposition, volumetric parameterization and trivariate spline approximation. This dissertation seeks accurate and efficient solutions to these fundamental and important problems, and demonstrates their applications in modeling 3D objects of arbitrary topology.

First, in order to achieve a "surface model to trivariate splines" conversion, we define our new splines upon a novel parametric domain called generalized poly-cubes (GPCs), which comprise a set of regular cube domains topologically glued together.

We then further improve our trivariate splines to support arbitrary topology by allowing the divide-and-conquer scheme: The user can decompose the model into components and represent them by trivariate spline patches. Then the key contribution is our powerful merging strategy that can glue tensor-product spline solids together, while preserving many attractive advantages.

We also develop an effective method to reconstruct discrete volumetric datasets (e.g., volumetric image) into continuous trivariate splines. To capture the fine features in the data, we construct an as-smooth-as-possible frame field based on 3D principal curvatures to align with a sparse set of directional features. The frame field naturally conducts a volumetric parameterization and thus a spline representation.

Next, we focus on promoting broader applications of our powerful modeling techniques. We present a novel methodology based on geometric deformation metrics to simulate magnification lens that can be utilized for the Focus+Context (F+C) visualization. We apply this methodology to both 2D image and 3D volume visualization.

Through our extensive experiments, we demonstrate that our framework is an effective and powerful tool for comprehensive existing models. The great potential of our modeling framework will be highlighted through many valuable applications such as shape modeling, remeshing, finite element analysis, deformation editing, visualization. Furthermore, we also envision further research directions and broader application scopes including many potential theoretical problems for 3D modeling and useful applications.

For my family, advisors, friends and colleagues.

# Contents

# List of Figures

xi

# List of Tables

# Acknowledgements

I would like to express my sincere gratitude to my thesis advisor, Professor Hong Qin for his guidance on research, and his help at various stages of my Ph.D. studies. I cannot imagine myself completing this dissertation without his inspiration, discussion and encouragement. He led me to the field of geometric modeling and processing and showed kindness, patience, and deep understanding about geometry and physics that no graduate students expect more from their advisors.

I would like to thank Professor Xianfeng Gu, Professor Xiangmin Jiao for their kind advice, as well as for serving on various committees.

I would like to thank Professor Qiaode Ge for taking the time to serve as the external member of my dissertation committee.

I would also like to thank my colleagues in our center of visual computing and computer science department for delightful collaborations and discussions we had together, and thank all my friends for their help during the past five years.

My research was supported in part by the following grants awarded to Professor Hong Qin: NSF IIS-0949467, IIS-1047715 and IIS-1049448.

Last but not least, I want thank my family for their endless love and support. Without their support, this thesis would not have been possible.

This dissertation is dedicated to them.

# Publications

**Journal Papers**

1. **Bo Li**, Xin Zhao, and Hong Qin. Four Dimensional Geometry Lens: A Novel Volumetric Magnification Approach. *Computer Graphics Forum*, under revision.

2. **Bo Li**, Xin Li, Kexiang Wang, and Hong Qin. Surface Mesh to Volumetric Spline Conversion with Generalized Polycube. *IEEE Transactions on Visualization and Computer Graphics*, in press.

3. **Bo Li**, and Hong Qin. Component-aware Tensor-product Trivariate Splines of Arbitrary Topology. *Computer & Graphics* , 36(5): 329–340, 2012.

4. Kexiang Wang, Xin Li, **Bo Li**, and Hong Qin. Restricted Trivariate Polycube Splines for Volumetric Data Modeling. *IEEE Transactions on Visualization and Computer Graphics*, 18(5): 703-716, 2012.

5. Xin Zhao, **Bo Li**, Lei Wang, and Arie Kaufman. Texture-guided Volumetric Deformation and Visualization Using 3D Moving Least Squares. *The Visual Computer*, 28(2): 193-204, 2012.

**Conference Papers**

1. **Bo Li**, Xin Zhao, and Hong Qin. A New Geometric Approach to Focus+Context Lens Simulation based on Geometric Deformation Metrics. *Proceedings of Interactive 3D Graphics and Games*, submitted.

2. **Bo Li**, and Hong Qin. Component-aware Tensor-product Trivariate Splines of Arbitrary Topology. *IEEE Proceedings of Shape Modeling and Applications*, 18:1–18:12, 2012.

3. **Bo Li**, and Hong Qin. Generalized PolyCube Trivariate Splines. *IEEE Proceedings of Shape Modeling and Applications*, 261-267, 2012.

4. Xin Zhao, **Bo Li**, Lei Wang, and Arie Kaufman. Focus+Context Volumetric Visualization using 3D Texture-guided Moving Least Squares. *Proceedings of the Computer Graphics International*, 2011.

**Technical Reports**

1. **Bo Li**. A Spline-based Volumetric Data Modeling Framework and Its Applications. *Ph.D. Dissertation Proposal, Stony Brook University (SUNY)*, 2012

2. **Bo Li**. A Survey on Volumetric Splines, Parameterization, and its Applications. *Research Proficiency Exam, Stony Brook University (SUNY)*, 2009.

3. **Bo Li**, and Hong Qin. Physically-based Interactive Volumetric Deformation with Isogeometric Formulations. *Technical Research Paper*, in preparation, 2012.

# Chapter 1

# Introduction

## 1.1 Problem Statements

Since the starting point of computer graphics research and application, surface shape modeling and designing have always been the central issue, mainly because shape design keeps acting as the core applications in industry as while as lacking real 3D data.

In the recent years, we have witnessed a great potential of paradigm shift from surface-only to volume data. Behind this are the rapidly developing 3D data acquisition techniques and the urgent 3D analysis requirement: proliferation of modern 3D scanning devices and shape modeling technologies give rise to the huge number of available high quality 3D dataset. As a result, the need to the ability of making good use of existing models has gained the prominence; Many computer graphics research and industry applications benefit tremendously from this trend: As a direct downstream application, we can now, in the first time, efficiently and robustly adapt real heterogenous material data onto 3D objects, which will significantly improve the physical analysis. Consequently, these newly emerged 3D datasets, as a novel data platform, may lead to a revolutionary transformation and update from existing graphics (e.g., deformation, simulation), visualization (e.g., rendering) and modeling (e.g., multivariate splines) techniques.

Consequently, we now desire to explore more efficient and robust 3D volume data modeling framework to suffice the exciting age of discovery in the above topics. This direction is always accompanied by many challenges. In detail, the difficulties arise from the fact that the quality criteria are diverse and their optimization often requires the consideration of tradeoff on specific applications. The most common quality aspects involve: The representation format must be flexible and powerful to describe complex shape and arbitrary topology; From perspective of analysis brings out the request that it should be simple and analytic; In physical analysis and texturing, we also need to represent both geometry and materials in our data structure; Meanwhile, a volumetric data normally has very large data scales with an explosive growth in the time and memory cost.

In this thesis, we specifically advocate a spline-based framework which can trade-off above requirements well. A key concern of engineering design industry is: These data have to be converted to continuous, compact representations to enable geometric design and downstream product development processes (e.g., finite element analysis, physical simulation, virtual surgery, etc) in CAD environments. As the natural correspondence, spline schemes and relative techniques have been extensively investigated during the recent past to fulfill the aforementioned goal. The material data can also be easily adapted by using multivariate splines. We will achieve a more compact representation of curves, surfaces or volumes at different scales in terms of data size, the number of control points, the user-specified threshold error, and other relevant criteria. We can compute all the differential quantities such as geodesics, curvatures, tensor fields without resorting to any numerical approximations via linear interpolation and/or local algebraic fitting. The rapid and precise evaluations of local and global differential properties will facilitate many applications such as finite element analysis, image registration/segmentation, shape modificaton/integration, surface quality analysis and control, and scientific visualization etc.

We observe that current spline prototypes are frequently based on 2-manifolds geometry and topology (i.e., "surface splines"). Typically, this representation describes the boundary of a solid model. However, a volumetric spline scheme has gathered growing interest from both analysis and CAD research communities, due to its computational advantage over traditional surface-based analysis method and its promise to alleviate the burden of creating effective 3D interior analysis-ready domains in many solid modeling and volume graphics applications. Iso-geometric analysis is an example to illustrate this necessity. NURBS based isogeometric analysis leverages the possible advantages of closer integration of CAD and FEA in isogeometric analysis. However, the critical challenge is how to convert a volumetric NURBS from its original boundary shape NURBS. This is because that any accurate physical analysis approach is based on a volumetric formulation where trivariate NURBS solids are needed for the analysis of three-dimensional (3D) problems while CAD systems use a boundary representation where only a surface representation is available. Although several numerical techniques such as boundary integration are amenable to several specific problems, it is more useful to provide the real trivariate NURBS geometry and material for generalized applications.

Existing volumetric spline techniques generally follow two different trends: (1) Many recent methods divide the volume space into a tetrahedral mesh domain then construct a trivariate spline (like super spline or box spline) on each tetrahedra do-main. These unregular-domain spline theories have just emerged recently, and have not been recognized by the communities outside computer graphics. At present, the regular tensor-product B-splines (NURBS) are still the prevailing industrial standard for freeform surface representation. (2) In contrast, many recent techniques [1], [2], [3] attempt to convert each part into splines defined on a cylinder/tube domain, because they can intuitively use the shape skeleton to produce a tube domain and reveal the global structure and topology. A severe limitation of such approaches is that points on the tube centerline are all singular. Also, the shape of tube is very simple such that it can not support complex shape and preserve any sharp edge and point feature when it serves as the domain.

An ideal volumetric spline modeling framework should have the following properties:

(1) Singularity free. A *singular point* in volumetric domain is a node with valence larger than four on an iso-parametric plane (Fig. 3.1(a-b)). Handling singularity with tensor-product splines is extremely challenging. It is desirable to have a global one-piece spline defined on a globally-connected singularity-free domain.

(2) The proposed domain construction method must be sufficient for surface with boundaries/complex shapes/arbitrary topology/long branches. The only feasible way is to introduce additional cuts and decompose the model into reasonable elements. Each element should abstract a component-aware part in a geometrically meaningful way thus make the following spline fitting process accurate and numerically stable. Also, the separate elements must be glued in a simple and singularity-free fashion.

(3) A practical volumetric parameterization technique must preserve shape feature. Specifically, in areas with well-pronounced consistent curvature directions, patch parametric lines should follow the curvature and patch boundaries should be aligned with sharp features and smooth surface boundaries. Moreover, an improved parameterization method should develop an efficient and systematical framework to better address the heterogenous model with various interior materials.

(4) In our new designed trivariate spline scheme, we desire to inherit the attractive properties of prevailing industrial standard NURBS. For example, NURBS have local support, i.e., moving one control point will only affect its immediate neighborhood. This makes intuitive design with NURBS possible; The basis functions of NURBS are non-negative, have the property of partition-of-unity, thus are qualified as basis functions required by finite element method; Non-uniform knot can confine the basis function inside the domain completely.

(5) We urgently need to design a more efficiently fitting pipeline to handle large scale computation during trivariate spline approximation. For example, a genus-0 solid bounded by 6 simple four-sided B-spline surfaces has originally $6 \times 1024^2$ control points (DOFs). The size of DOFs increases drastically to $1024^3$ or even larger when we naively convert it to a volumetric spline representation. This exponential increase during volumetric spline conversion poses a great challenge in terms of both storage and fitting costs.

In conclusion, our modeling framework involves 3 main challenges and all above requirements can be categorized into them:

(1) Mesh decomposition.

- How to decompose them into component-aware parts?

- How to design a practical or automatic scheme to generate consistent parti- tioning, with a small number of parts and spline-friendly domain shapes and gluing types?

(2) Volumetric parameterization.

- How to reduce the computation complexity of volumetric mapping and make it more robustly?

- How to analyze and restrict the mapping distortion?

- How to integrate the shape feature (like sharp edges, corners), or even various materials (like density value) into our parameterization result?

(3) Trivariate splines.

- How to preserve the critical properties of NURBS surface like partition-of-unity, local refinement and boundary confinement?

- How to decrease the control point number to adapt huge number of degree-of-freedom in trivariate splines?

- How to accelerate fitting efficiency and save fitting cost (time and storage).

- How to handle multivariate splines for many applications like vector volume imaging.

To overcome the above modeling and design difficulties and address the topological issue, we seek novel modeling techniques based on tensor-product spline schemes that would allow designers to directly define continuous spline models over any manifold (serving as parametric domain). Such a global approach would have many modeling benefits, including no need of the transition from local patch definition to global surface construction via gluing and abutting, the elimination of non-intuitive segmentation and patching process, and ensuring the high-order continuity requirements. More importantly, we can expect a true one-piece representation for shapes of complicated topology, with a hope to automate the entire reverse engineering process (by converting points and/or polygonal meshes to spline surfaces with high accuracy) without human intervention.

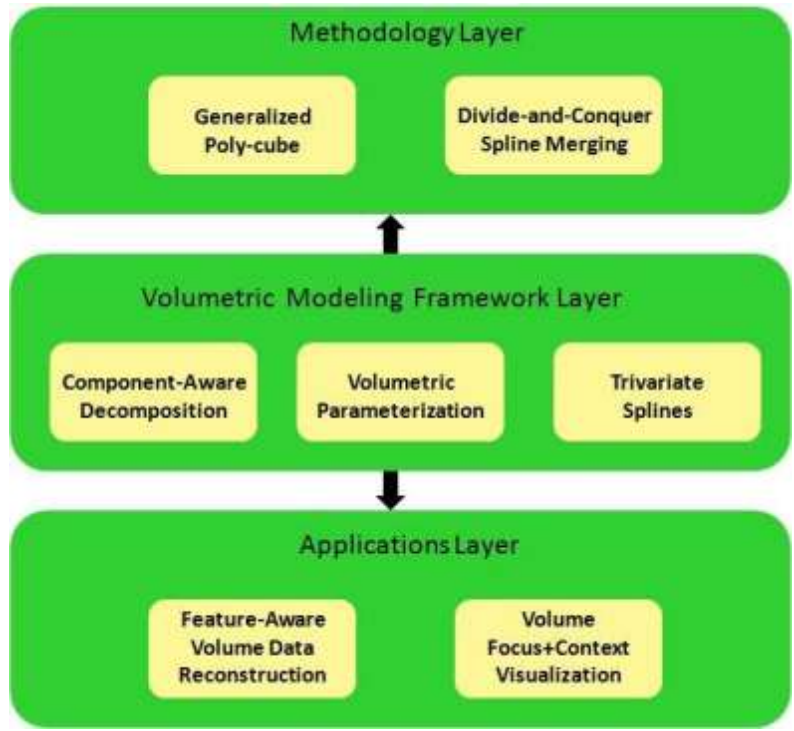

Figure 1.1: Hierarchy of our research contents. Key streamline of our framework (middle row); Main techniques for trivariate spline modeling (upper row); Utilized applications (bottom row).

Towards this goal, we present a novel spline-based solid modeling framework. Figure 1.1 illustrates the conceptual hierarchy of above discussions and the whole proposal. This framework integrates a few projects (first row) and targets on key challenging problems (third row). By solving these key difficulties we have improved the effectiveness and efficiency of shape mapping computation, and are able to utilize this framework into various applications (bottom row).

Through our experiments, we hope to demonstrate that the proposed data modeling framework is very flexible and can potentially serve as a geometric standard for product data representation and model conversion in shape design and geometric processing.

## 1.2   Contributions

In this thesis, we present a spline-based framework to solve 3D objects modeling problems. Particularly, we emphasize our research interest on regular domain ("cuboid") tensor-product splines, because of their favorite advantages. Combining volumetric decomposition, parameterization with trivariate splines, we successfully and effectively solve a variety problems in the areas of geometric shape design and modeling.

Our specific contributions include:

- We develop a novel volumetric parameterization and spline construction

  frame- work, which is an effective modeling tool for converting surface meshes to volumetric splines. Our new splines are defined upon a novel parametric domain called generalized poly-cubes (GPCs). A GPC comprises a set of

  regular cube domains topologically glued together. Compared with conventional poly-cubes (CPCs), the GPC is much more powerful and flexible and has improved numerical accuracy and computational efficiency when serving as a parametric domain. We design an automatic algorithm to construct the GPC domain while also permitting the user to improve shape abstraction via interactive intervention. We then parameterize the input model on the GPC domain. Finally, we devise a new volumetric spline scheme based on this seamless volumetric parameterization. With a hierarchical fitting scheme, the proposed splines can fit data accurately using reduced number of superfluous control points. Our volumetric modeling scheme has great potential in shape modeling, engineering analysis, and reverse engineering applications.

- The next contribution of this thesis aims to bridge the large gap between

  the shape versatility of arbitrary topology and the geometric modeling limitation of conventional tensor-product splines for solid representations. Its contri- bution lies at a novel shape modeling methodology based on tensor-product trivariate splines for solids with arbitrary topology. Our framework advo-

  cates a divide-and-conquer strategy. The model is first decomposed into a set of components as basic building blocks. Each component is naturally modeled as tensor-product trivariate splines with cubic basis functions while supporting local refinement. The key novelty is our powerful merging strategy that can glue tensor-product spline solids together subject to $C^2$ continuity. As a result, this new spline representation has many attractive advantages. At the theoretical level, the integration of the top-down topological decomposition and the bottom-up spline construction enables an elegant modeling approach for arbitrary high-genus solids. Each building block is a regular tensor-product spline, which is CAD-ready and facilitates GPU com-puting. In addition, our new spline merging method enforces the features of semi-standardness (i.e., $\mathbf{\sum_i} w_i B_i(u, v, w) \equiv 1$ everywhere) and boundary restriction (i.e., all blending functions are confined exactly within parametric domains) in favor of downstream CAE applications. At the computational level, our component-aware spline scheme supports meshless fitting which completely avoids tedious volumetric mapping and remeshing. This divide-and-conquer strategy reduces the time and space complexity drastically. We conduct extensive experiments to demonstrate its shape flexibility and versatility towards solid modeling with complicated geometries and non-trivial genus.

- We propose a systematic framework that transforms discrete volumetric raw data from scanning devices directly into continuous spline representation with regular tensor-product structure. To achieve this goal, we propose a novel volumetric parameterization technique that constructs an as-smooth-
  as-possible frame field, satisfying a sparse set of directional constraints, and we compute a globally smooth parameterization with iso-parameter curves following the frame field directions. The proposed method can efficiently reconstruct model with multi-layers and heterogenous materials, which are usually extremely difficult to be handled by the traditional techniques.

- Aiming to promote new applications of our powerful modeling techniques in visual computing, we present a novel methodology based on geometric deformation metrics to simulate magnification lens that can be utilized for Focus+Context (F+C) visualization. Compared with conventional optical lens design (such as fish-eyes, bi-focal lens), our geometric modeling based
  method are much more capable of preserving shape features (such as angles, rigidities) and minimizing distortion.

- We extend this novel methodology and integrates it into a 4-Dimensional space deformation to simulate magnification lens on versatile textured solid models. Compared with other magnification methods (e.g., optical/energy based minimization), 4D differential geometry theory and its practices are much more capable of preserving shape features (angle distortion minimization), and easier to adapt on versatile solid models. The primary advantage of 4D space lies at: we can now easily magnify the volume of regions of interest (ROIs) from the augmented dimension, while keeping the rest region unchanged. To achieve this primary goal, we first embed this volumetric input into 4D space and magnify ROIs in the 4th dimension. Then we flatten the 4D shape back into 3D space to agree with usual applications in the real 3D world. In order to enforce distortion minimization, in both steps we devise the high dimensions geometry techniques from rigorous 4D geometry theory for 3D/4D mapping back and forth to amend the distortion. We demonstrate the effectiveness, robustness and efficacy of our framework with a variety of models ranging from tetrahedral meshes to volume datasets.

## 1.3  Dissertation Organization

The remainder of this thesis is organized in the following fashion. In Chapter 2, we begin with the detailed review prior research work related to component-aware mesh decomposition, volumetric parameterization and trivariate splines with regular structures. In Chapter 3, we present a novel modeling concept "Generalized poly-cube", and develop an automatic modeling framework using GPC to convert a surface mesh into volumetric splines. In Chapter 4, we propose a new bottom-up paradigm that decomposes a surface model into separate spline patches and then integrates them into a global continuous formulation. We design a new spline merging algorithm to guarantee high-order continuities while keeping all other spline properties. In Chapter 5, we propose a trivariate spline-based approach that is able to reconstruct discrete volumetric data directly acquired from scanning devices into regular tensor-product spline representation. We study a new volumetric frame field and parameterization generation method to achieve reconstruction. In Chapter 6, we apply our geometric modeling method into a visualization application: lens design problem. We integrate a flexible geometric metric to simulate the optical lens and our method is much more capable of preserving shape features (angles and rigidities) and minimizing distortion. In Chapter 7, we present a novel methodology that integrates 4-Dimensional space deformation to simulate magnification lens on versatile textured solid models. Finally, we conclude in Chapter 8 with the summery and the discussion on future research directions. We articulate all useful theoretical propositions and proofs about trivariate splines we develop in this thesis.

# Chapter 2

# Background Review

As we have introduced in Chapter 1, the hierarchy of this thesis includes 3 main steps: decomposition, parameterization and spline construction. Spline and parameterization consist of our primary research topics thus we review them first. We notice that many researchers have explored and studied deeply topics in $R^2$, and since our focus is on volumetric modeling, here we only introduce basic techniques and theories about surface study and main review the work on $R^3$.

## 2.1 Splines

Splines normally refer to smooth, piecewise polynomials. They are ideal tools for applications where continuous representations are critical. Their most common quality aspects involve: The fitting can be piece-wised; The data is highly compressed; The analytic computation is very easy; The format is widely accepted by most design softwares.

The first study on splines goes back to 1946 by Schoenberg. Since then, splines become a very active research because of the fast development of industry application and computer science. Between the 1960's and early the 1970's, Birkhoff, Garabedian and deBoor have studied and established a series of theories on Cartesian regular tensor product splines to represent surface. It is well known that now these types of spline functions become the industry standard and play very important roles in many engineering design applications. Although there are huge number of literatures on many extension types of splines to combat the shortcoming of regular splines (like triangular B-splines, Powell-Sabin splines, etc), their applications only exit in theoretical study and the whole industry still insists on regular splines. Therefore, we shall briefly explain the relative concepts of regular tensor-product splines in the following section. Then, we will pay attention on existing trivariate spline techniques.

### 2.1.1 Polynomials and Polar Forms

The most fundamental class of splines is the class of parametric polynomials. In the context of CAGD and computer graphics, splines are best studied with the help of a classical theoretical foundation like "Polar Form" [4],[5]. All spline theories are covered and generated from the polar form theory. Therefore, we here simply brief the basic idea of the polar form.

**Polar Forms.** The parametric polynomials are the fundamental basis for splines. The polar form is a very important tool for polynomials and thus spline study. The definition of polar form are as follows [6]:

**Definition 2.1.1** (**Affine Map**). *A map* $f : R^k \to R^t (k \geq 1)$ *is affine, if and only if it preserves affine combinations, i.e., if and only if f satisfies* $f(\sum_{i=0}^{m} \alpha_i u_i) = \sum_{i=0}^{m} \alpha_i f(u_i)$ *for all scalars* $\alpha_0, \ldots, \alpha_m \in R$ *with* $\sum_{i=0}^{m} \alpha_i = 1$.

**Definition 2.1.2** (**Symmetric, Multi-Affine**). *Let* F *be an n-variable map.* F *is symmetric if and only*

$$F(u_1, u_2, \cdots, u_n) = F(u_{\pi(1)}, u_{\pi(2)}, \cdots, u_{\pi(n)}).$$

*For all permutations* $\pi \in P_n$, *The map* F *is multi-affine if and only if* F *is affine in each argument and the others are held fixed.*

Blossoming principle is a very important express that indicates that any polynomial is equivalent to its polar form [4]:

**Theorem 2.1.3** (**Blossoming Principle**). *Polynomials* $F : R^k \to R^t (k \geq 1)$ *of degree* n, *and a symmetric multi-affine map* $f : (R^k)^n \to R^t$ *are equivalent. Given a map of either type, unique map of the other type exists that satisfies the identity* $F(u) = f(\underbrace{u, \cdots, u}_{n})$. *The map* f *is called the multi-affine polar form or blossom of* F.

The property of blossoming principle is used to define deCasteljau algorithm and de Boor algorithm in the following sections.

### 2.1.2 Regular Tensor Product Splines

**Bézier Splines.** Among all regular splines, a Bézier representation in its most common form is the most widely accepted equation that can be used in any number of useful ways. Bézier curves have obtained dominance in the typesetting industry since 1970's. A Bézier spline can be defined as:

**Theorem 2.1.4** (**Bézier Curve**). *Given a set of* $n + 1$ *control points* $P_0, P_1, \ldots, P_n$, *the corresponding*

**Bézier Curve** *is given by*

$$C(t) = \sum_{i=0}^{n} P_i B_{i,n}(t),$$

*where* $B_{i,n}(t)$ *is a* Bernstein polynomial $B_{i,n}(t) = C_i^n t^i (1-t)^{n-i}$ *and* $t \in [0,1]$.

As we mentioned in the last section, we can also represent Bézier splines of a polynomial $F$ from its polar form like [7]:

**Theorem 2.1.5.** (*Bézier Points and de Casteljau algorithm*) *Let* $\Delta = [r,s]$ *be an arbitrary interval. Every polynomial* $F : \mathbb{R} \to \mathbb{R}^t$ *can be represented as a Bézier polynomial w.r.t.* $\Delta$. *The Bézier points are given as*

$$b_j = f(\underbrace{r, \ldots, r}_{n-j}, \underbrace{s, \ldots, s}_{j}),$$

*where* $f$ *is the polar form of* $F$.

Equation above immediately leads to an evaluation algorithm that recursively computes the values

$$
\begin{aligned}
b_j^l(u) &= f(\underbrace{r, \ldots, r}_{n-l-j}, \underbrace{u, \ldots, u}_{l}, \underbrace{s, \ldots, s}_{j}) \\
&= \tfrac{s-u}{s-r} f(\underbrace{r, \ldots, r}_{n-l-j+1}, \underbrace{u, \ldots, u}_{l-1}, \underbrace{s, \ldots, s}_{j}) + \tfrac{u-r}{s-r} f(\underbrace{r, \ldots, r}_{n-l-j}, \underbrace{u, \ldots, u}_{l-1}, \underbrace{s, \ldots, s}_{j+1}) \; \cdot \\
&= \tfrac{s-u}{s-r} b_j^{l-1}(u) + \tfrac{u-r}{s-r} b_{j+1}^{l-1}(u)
\end{aligned}
$$

from the given control points. For $l = n$ we finally compute $b_0^n = f(u, \ldots, u) = F(u)$, which is the desired point on the curve. This algorithm is called *de Casteljau Algorithm* [7].

Formula above also shows that the de Casteljau Algorithm offers a way to subdivide a Bézier curve: suppose that we wish to subdivide a Bézier curve $F$ over a given interval $\Delta = [s,t]$ at an arbitrary parameter $u \in \Delta$. The new Bézier points of the left and right segments $F_l$ and $F_r$ with respect to the subintervals $\Delta_l = [r,u]$ and $\Delta_r = [u,s]$ are given as

$$b_0^l = f(r, \ldots, r), b_1^l = f(r, \ldots, r, u), \ldots, b_n^l = f(u, \ldots, u),$$

and

$$b_0^r = f(u, \ldots, u), b_1^r = f(u, \ldots, u, s), \ldots, b_n^r = f(s, \ldots, s).$$

**B-Splines.** B-splines (short for Basis Splines) go back to Schoenberg who introduced them in 1946 [8, 9] for the case of uniform knots. B-splines over nonuniform knots go back to a review article by Curry in 1947. De Boor derived the recursive evaluation of B-spline curves [10]. It was this recursion that made B-splines a truly viable tool in CAGD. Before its discovery, B-splines were defined using a tedious divided difference approach which was numerically unstable. Later on, Gordon and Riesenfeld realized that de Boor's recursive B-spline evaluation is the natural generalization of the de Casteljau algorithm and Bézier curves are just subset of B-spline curves. Versprille [11] generalization of B-spline curves to NURBS (non-uniform rational B-spline) which has become the standard curve and surface form in the CAD/CAM industry [12].

**Definition 2.1.6** (**B-Spline**). *Let a vector known as the knot vector defined as*

$$\mathbf{T} = \{t_0, t_1, \ldots, t_m\}$$

*where* $\mathbf{T}$ *is a nondecreasing sequence with* $t_i \in [0, 1]$ *, and define control points* $P_0, \ldots, P_n$ *. Define the degree as*

$$p \equiv m - n - 1$$

*The knots* $t_{p+1}, \ldots, t_{m-p-1}$ *are called internal knots.*
*Define the basis functions as*

$$N_{i,0}(t) = \begin{array}{ll} 1 & \textit{if } t_i \le t < t_{i+1} \textit{ and } t_i < t_{i+1}; \\ 0 & \textit{otherwise.} \end{array}$$

$$N_{i,p}(t) = \frac{t - t_i}{t_{i+p} - t_i} N_{i,p-1}(t) + \frac{t_{i+p+1} - t}{t_{i+p+1} - t_{i+1}} N_{i+1,p-1}.$$

*Then the curve defined by*

$$C(t) = \sum_{i=0}^{n} P_i N_{i,p}(t)$$

*is a **B-Spline**.*

The B-spline basis functions are positive and form a partition of unity. In addition, they have local support given by $N_i^n(u) = 0$ for $u \notin [t_i, t_{i+n+1}]$ .The knot values determine the extent of the control of the control points.

The B-spline can be divided into different types with respect to knot values:

**Uniform B-spline.** When the knots are equidistant the B-spline is called uniform. The uniform B-spline has a succinct definition:

$$b_{j,n} = b_n(t - t_j),$$

with

$$b_n(t) = \frac{n+1}{n} \sum_{i=0}^{n+1} \mu_{i,n}(t - t_i)_+^n,$$

and

$$\mu_{i,n} = \prod_{j=0, j=i}^{n} \frac{1}{t_j - t_i}.$$

where $(t - t_i)_+^n$ is the truncated power function:

$$F_+^n = \begin{cases} F^n & \text{if } F \geq 0 \\ 0 & \text{otherwise} \end{cases}$$

**Open-uniform B-spline.** The difference between uniform spline and open-uniform spline is that there exists k degree at the start and end points of the vector knots. This open-uniform B-spline defines the open-uniform basis function. The motivation of open-uniform B-spline comes from the difference of B-spline and Bézier spline. The B-spline can not preserve one property of Bézier spline that the start and end points of the curve are the same points of the first control point and the last control point. Open-uniform B-spline can solve this problem. For instance, if we set the knot vector as (0,0,0,1,1,1), it can be directly proved that the basis function generated from this vector is equal to the degree-2, with 3 control point Bézier curve's basis function. (0,0,0,0,1,1,1,1) is another example that is the same as cubic, with 4 control point Bézier curve.

**Non-uniform B-spline.** B-spline basis function with arbitrary knot vector that follows the definition requirements. Uniform B-spline is special cases of no-uniform.

**Degree of B-spline.** B-spline allows arbitrary degree of B-spline. In practical use the degree is rarely more than 3. So the basis function computing can be specialized for each degree. Figure 2.1illustrates the basis functions in degree 0,1,2.

- Constant B-spline: The constant B-spline is the simplest B-spline. It is de-

fined on only one knot span.

$$N_{j,0}(t) = 1_{[t_j,t_{j+1})} = \begin{cases} 1 & \text{if } t_j < t < t_{j+1}; \\ 0 & \text{otherwise.} \end{cases}$$

- Linear B-spline: The linear B-spline is defined on two knot spans.

$$N_{j,1}(t) = \begin{cases} \dfrac{t - t_j}{t_{j+1} - t_j} & \text{if } t_j < t < t_{j+1}; \\ \dfrac{t_{j+2} - t}{t_{j+2} - t_{j+1}} & \text{if } t_{j+1} < t < t_{j+2} \\ 0 & \text{otherwise.} \end{cases}$$

- Uniform quadratic B-spline: the un-uniform quadratic B-spline does not have the uniform expression. Here we write out the blending function for uniform type.

$$N_{j,2}(t) == \begin{cases} \dfrac{1}{2}t^2 \\ -t^2 + t + \dfrac{1}{2} \\ \dfrac{1}{2}(1 - t)^2 \end{cases}$$

.

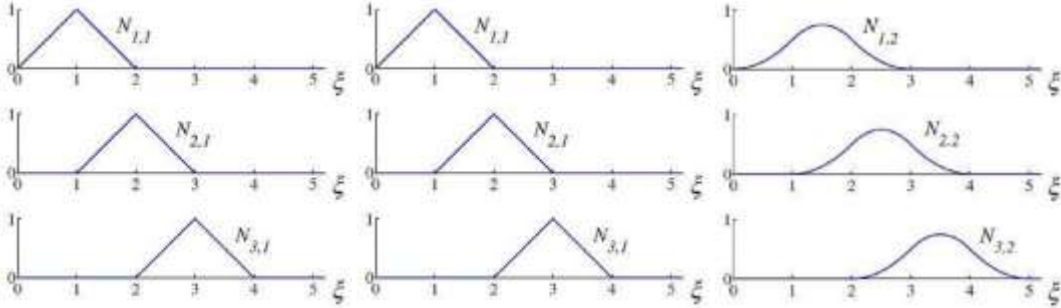

Figure 2.1: Basis functions for B-spline with degree 0,1,2 from left to right.

As we mentioned in the last section, we can also represent B-Splines of a polynomial F from its polar form [13, 14].

**Theorem 2.1.7.** *(De Boor Points and De Boor Algorithm) Every polynomial*

$$F : R \to R^t$$

*can be represented as a B-spline segment over a non-decreasing knot sequence*

$$r_n \leq \ldots \leq r_1 < s_1 \leq \ldots \leq$$
$$s_n.$$

*The de Boor points are given as*

$$d_j = f(r_1, \ldots, r_{n-j}, s_1, \ldots, s_j),$$

*where* $f$ *is the polar form of* $F$.

**Tensor Product B-spline.** We can extend the B-spline from curve to surface. Tensor product surfaces are the most popular surface design method in theory and industry: Given a curve scheme

$$F(u) = \sum_{i=0} B_i(u)b_i, \, b_i \in R^t,$$

the corresponding tensor product scheme is defined as

$$F(u, v) = \sum_{i=0} \sum_{j=0} B_i(u)B_j(v)b_{ij}, \, b_{ij} \in R^t,$$

which can also be written as

$$F(u, v) = \sum_{i=0} B_i(u)b_{i_v},$$

with

$$b_{i_v} = b_i(v) = \sum_{j=0} B_j(v)b_{ij}.$$

The last equation demonstrates that tensor product surfaces may be considered as curves of curves.

**NURBS.** B-spline shows that it is a powerful tool for free form curve and surface shape design. However, it has the drawback that can not express exactly the regular shape. The invention of non-uniform rational B-spline (NURBS) is to solve this problem.

**Definition 2.1.8** (**NURBS**). *Let a vector known as the knot vector be defined*

$$T = \{t_0 \leq t_1 \leq \ldots \leq t_{k+n} \leq$$
$$t_{k+n+1}\},$$

*with the restriction that the interior knots have at most multiplicity* $n$, *that is* $t_i < t_{i+n}$ *for* $i = 1, 2, \ldots, k$, *define control points* $P_0, \ldots, P_k \in E^d$, *and define positive* *weights* $w_0, w_1, \ldots, w_k$, *associated to the control points* $P_i$.

*The analytic representation of the corresponding NURBS curve* R *of degree n in* $E^d$
*is given by*

$$R(u) = \frac{\sum_{i=0}^{k} w_i P_i N_i^n(u)}{\sum_{i=0}^{k} w_i N_i^n(u)}, u \in [t_0, t_{k+n+1}],$$

*where* $N_i^n, i = 0, 1, \ldots, k$ *are the normalized B-spline basis functions of degree n corresponding to the knot vector* T.

Another advantage is that it is invariant under projective transformation (only affine invariance holds for its integral counterpart). Additionally, there are weights which can be used to control shapes in a manner similar to shape parameters. Geometrically, a rational curve can be viewed as the projection of an integral curve from a vector space of one higher dimension. The NURBS curve can be obtained by projecting the B-spline curve $\hat{R}$ in $E^{d+1}$ having the same knot vector and control points $\hat{P}_i = (w_i P_i, w_i)$. As a consequence, the NURBS inherit all the nice properties from B-splines, and can represent conic sections.

**NURBS Surfaces.**   If we extend equation in two parametric directions we obtain a surface with the same properties as the NURBS curve:

$$F(u, v) = \frac{\sum_{i=0}^{n} \sum_{j=0}^{m} w_i P_i B_i(u) B_j(v)}{\sum_{i=0}^{n} \sum_{j=0}^{m} w_i B_i(u) B_j(v)}.$$

The surface does not have to be of equal degree in both directions.  Observe the surface in its rendered form in where we clearly see the local control property.

NURBS generalize the nonrational parametric form. Like nonrational B-splines, the rational basis functions of NURBS sum to unity, they are infinitely smooth in the interior of a knot span, and at a knot they are at least $C^{k-1-r}$ continuous with knot multiplicity $r$, which enables them to satisfy different smoothness requirements. They inherit many of the properties of uniform B-splines, such as the strong convex hull property, variation diminishing property, local support, and invariance under standard geometric transformations. More material of NURBS and further detailed discussion of its properties can be found in [15–18].

## 2.1.3   Hierarchical Schemes

Forsey and Bartels have presented the hierarchial B-spline [19], in which a single control point can be added without covering an entire row or column of control points. In their work two concepts are introduced: local refinement using an efficient representation, and multi-resolution editing. These notions can be generalized to any surface such as subdivision surface. Meanwhile, the localized hierarchical splines have been proposed by Gonzalez-Ochoa and Peters [20], which extend the hierarchial spline paradigm to surfaces of arbitrary topology. Kraft [21] has constructed a hierarchical B-splines with a multilevel spline space which is a linear span of tensor product B-splines on different, hierarchically ordered grid levels. Charms [22] have extended this scheme in a more general setting and adapted it to more applications. Weller et al. [23] have studied spaces of piecewise polynomials with an irregular, locally refinable knot structure (thus it is called "semi-regular bases"). Deng et al. [24] have introduced a new type of splines-polynomial splines over hierarchical T-meshes (called PHT-splines) to model geometric objects. PHT-splines are a generalization of B-splines over hierarchical T-meshes. Song et al. [25] have presented the method to approximate the signed distance function of a surface by using polynomial splines over hierarchical T-meshes. In particular, they compute on closed parametric curves in the plane and implicitly defined surfaces in space.

T-splines, developed by [26], are the most important scheme in our proposal. T-splines are generalizations of NURBS surfaces that are capable of significantly reducing the number of superfluous control points by using the T-junction mechanism. The main difference between a T-spline control mesh and a NURBS control mesh is that T-splines allow a row or column of control points to terminate at anywhere without strictly enforcing the rectangular grid structure throughout the parametric domain. Consequently, T-splines enable much better local refinement capabilities than NURBS. Furthermore, using the techniques presented in [26], we are able to merge adjoining T-spline surfaces into a single T-spline without adding new control points. Sederberg et al. have also developed a simplified algorithm to convert NURBS surfaces into T-spline surfaces, in which a large percentage of superfluous control points are eliminated [27].

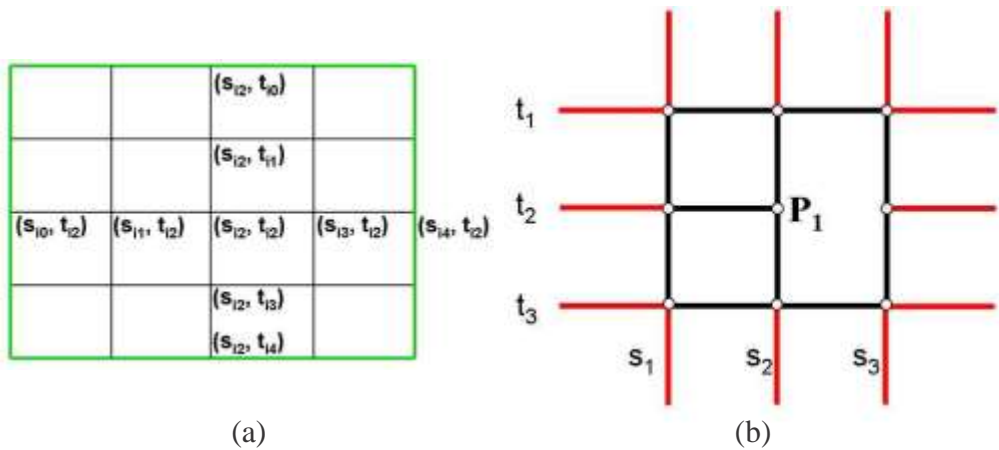

(a)             (b)

Figure 2.2: (a) Local knot lines for basis function $B_i(s, t)$; (b) $P_i$ is a T-junction.

T -spline is a P B-spline for which some order has been imposed on the control points by means of a control grid called a textit T-mesh. A T -mesh is basically a rectangular grid that allows T -junctions. Each edge in T -mesh is a line segment of constant s (which is called s-edge) or constant $\mathbf{t}$ (which is called t-edge). A T -junction is a vertex shared by one s-edge and two t-edges, or by one t-edge and two s-edges. For example, $\mathbf{P}_1$ (see Fig.2.2(b)) is a T -junction. Each edge in a T -mesh is labeled with a knot interval, constrained by the following rules:

1. The sum of knot intervals on opposing edges of any face must be equal.

2. If a T -junction on one edge of a face can be connected to a T -junction on an opposing edge of the face (thereby splitting the face into two faces) without violating Rule 1, the edge must be included in the T -mesh.

In contrast to tensor-product B-spline that uses a rectangular grid of control points, P B-spline is point-based and requires no topological relationship among control points. The equation for a P B-spline is given by:

$$\mathbf{P}(s, t) = \frac{\sum_{i=1}^{n} P_i B_i(s, t)}{\sum_{i=1}^{n} B_i(s, t)} \quad (s, t) \in D,$$

where the $\mathbf{P}_i$ are control points. The $B_i(s, t)$ are basis functions written as

$$B_i(s, t) = N_{i0}^3(s) N_{i0}^3(t),$$

where $N_{i0}^3(s)$ is the cubic B-spline basis function associated with the knot vector $\mathbf{s}_i = [s_{i0}, s_{i1}, s_{i2}, s_{i3}, s_{i4}]$ and $N_{i0}^3$ is associated with the knot vector $\mathbf{t}_i = [t_{i0}, t_{i1}, t_{i2}, t_{i3}, t_{i4}]$ as illustrated in Fig. 2.2(a). Every control point has its influence domain $\mathbf{D}_i = (s_{i0}, s_{i4}) \times (t_{i0}, t_{i4})$. The T -spline equation is very similar to the equation for a tensor-product rational B-spline surface, except that knot vectors $\mathbf{s}_i$ and $\mathbf{t}_i$ are deduced from the T -mesh neighborhood of $\mathbf{P}_i$.

Knot vector $\mathbf{s}_i$ and $\mathbf{t}_i$ for the basis function $B_i(s, t)$ are determined as follows. Let $(s_{i2}, t_{i2})$ are the knot coordinate of $\mathbf{P}_i$. Consider a ray in parameter space $\mathbf{R}(\alpha) = (s_{i2} + \alpha, t_{i2})$. Then $s_{i3}$ and $s_{i4}$ are the s coordinates of the first two s-edges intersected by the ray. The other knots can be found in like manner.

In computer graphics T-splines have been applied to many applications. For example, Song et al. [28] have generalized a T-spline scheme to weighted T-spline and demonstrated its applicability in 3D free-form deformation. Lévy et al. [29] have utilized T-splines for surface reconstruction.

### 2.1.4   Global Splines V.S. Spline Merging

Spline merging techniques always involve the following steps. In order to model an arbitrary manifold in 3D using conventional spline schemes, current approaches will segment the manifold to many smaller open patches, then cover each patch by a single coordinate system, so that each patch can be modeled by a spline surface. Finally, any generic approach must glue all the spline patches together by adjusting the control points and the knots along their common boundaries in order to ensure continuity of certain degree. It requires the merging of splines defined over different local domains. Surface patch merging has been thoroughly discussed first in [26, 30] and later is used in [31], in order to glue the trimmed region to form a single spline. However, it is far more complicated to design semi-standard trivariate splines which demand much more in-depth studies. During spline merging, handling singularity with still high-order continuity is extremely difficult in spline research. For surface modeling, Loop and Scheafer in [32] have given an example of a $G^2$ polynomial construction with general connectivity to accommodate singularities. On the other hand, Peters and Fan [33] have introduced rational linear maps to replace affine linear atlas and handle singularities between charts.

Spline merging also has many shortcomings. The entire segmenting and patching process is primarily performed manually, and it requires users' knowledge and skills, and for non-trivial topology and complicated geometry this task is laborious and error-prone. To overcome the above modeling and design difficulties and address the topological issue, many researchers seek novel modeling techniques that would allow designers to directly define continuous spline models over any manifolds (serving as parametric domains). Such a global approach would have many modeling benefits, including no need of the transition from local patch definition to global surface construction via gluing and abutting, the elimination of a non-intuitive segmentation and patching process, and ensuring the high-order continuity requirements. More importantly, we can expect a true "one-piece" representation for shapes of complicated topology, with a hope to automate the entire reverse engineering process.

Li et al. have presented an automatic technique to convert polygonal meshes to T-splines using periodic global parameterization [29, 34]. Li et al.'s method can be also viewed as manifold splines since the transition functions of the periodic global parameterization are compositions of translations and rotations. Grimm et al. [35] have pioneered a generic method to extend B -splines to surfaces of arbitrary topology, based on the concept of overlapping charts. Cotrina et al. have proposed a $C^k$ construction on a manifold [36, 37]. Ying and Zorin [38] have presented a manifold-based smooth surface construction method which has high-order continuities with explicit nonsingular parameterizations only in the vicinity of regions of interest. Gu et al. [39] have developed a general theoretical framework of manifold splines in which spline surfaces, defined over planar domains, can be systematically generalized to any manifold domain of arbitrary topology (with or without boundaries). He et al. have further developed modeling techniques for applications of manifold splines using triangular B -splines [40].

### 2.1.5   Trivariate Splines

Spline-based volumetric modeling and analysis have gained much attention recently with many applications. For geometric processing, Song et al. [28] have employed trivariate splines with non-uniform weights to model free-form deformation. For physical analysis, Hughes et al. [41] have proposed isogeometric analysis on surface, using bivariate NURBS for modeling smooth geometry and physical attributes together, and conducting physical analysis simultaneously. For virtual surgery, Tan et al. [42] have utilized spherical volumetric simplex splines to model and simulate the human brain. In visualization, Rössl et al. [43] have utilized trivariate super splines to model and render multi-dimensional material attributes for solid objects. A modeling technique introduced in [44] has been developed to model skeletal muscle with anisotropic attributes and conduct FEM analysis directly on NURBS solid. Martin et al. [2] have presented a method to fit a solid model using a cylindrical trivariate NURBS and support continuum force analysis. However, these existing spline schemes tend to handle only simple inputs like genus-0 surfaces. For more complicated shapes, Zhang et al. [3] have proposed the method to convert the long-branch/bifurcations dominant shapes. Martin et al. [1] have studied shapes with a symmetry (called "mid-face") structure. These methods always attempt to transform the model through a top-down scheme, which inspires us to research a new method in a divide-and-conquer fashion.

Compared with surface splines designed to extract features (e.g., [29, 45]), trivariate splines mainly focus on finding part-aware component structures. Besides poly-cube domains, another commonly-used part-aware domain is cylinder (tube) like [2]. Martin et al. in [1] have extended this domain to mimic more complex shapes. However, in terms of spline construction, the cylinder (tube) domain inevitably produces singular points along the tube axis.

## 2.2   Parameterization

Model parameterization is the fundamental basis and powerful geometry processing tool with versatile application, such as detail mapping, such as spline fitting and CAD, meshing processing, FEM analysis, visualization etc. In this thesis research proposal, parameterization is the first and un-avoided step during enabling data-to-spline conversion. In this section, we first briefly outline its mathematical founda-tions and describe recent methods for parameterization. Second, since our research mainly focuses on trivariate spline construction, it is necessary to discuss some recent emerging study interest on volumetric parameterization. Finally, we demonstrate feature-aware parameterization specifically because an efficient feature-aware technique leads to better spline fitting result.

## 2.2.1   Theory and Techniques

In this section, we outline the mesh parameterization including its mathematical foundations, versatile local parameterization techniques on different domains. In [46, 47], authors also have discussed this topic. Our review starts with an introduction to the general idea of parameterization and the state-of-art is reviewed by summarizing the motivation and major idea of several important approaches. Since we mainly consider the representation of volumetric information, we also discuss the emerging tools for regular global parameterization and volumetric parameterization.

**Metric and Distortion Minimization.**   Parameterization can be viewed as a procedure of energy/distortion metric minimization procedure. Energy (distortion metric) gives rise to the solution from the degree of global energy field, that the spring model will converge at a balance state when the global spring energy is minimized. The advantage of these ideas involves that once we set the energy field function , we can solve the parameterization by numerical energy minimization tools directly.

Now we need to specify the energy, or define distortion metrics. The distortion derives from the stretching during the mapping $\mathsf{F}$ between the surface (x, y, z) and the domain (u, v). Suppose $(x, y, z) = \mathsf{F}(u, v)$ is a center point $\mathbf{P}$ of an infinitesimal planar circle. Then, one point on this circle $\mathsf{F}(u + \delta u, v + \delta v)$ is approximated given by first order Taylor expansion:

$$\underline{\mathsf{F}}(u + \delta u, v + \delta v) = \mathsf{F} + \mathsf{F}_u(u, v)\delta u + \mathsf{F}_v(u, v)\delta v,$$

or

$$\underline{\mathsf{F}}(u + \delta u, v + \delta v) = \mathbf{P} + \mathsf{F}_u(u, v)\delta u + \mathsf{F}_v(u, v)\delta v = \mathbf{P} + \mathsf{J}_f(\delta u, \delta v),$$

where $\mathsf{J} = [\mathsf{F}_u, \mathsf{F}_v]$ is a 3 × 2 mapping matrix (normally it is also called Jacobian matrix). Using singular value decomposition, we have:

$$\mathsf{J}_f = \mathsf{U}\Sigma\mathsf{V}^{\mathsf{T}} = \mathsf{U}\begin{bmatrix} \sigma_1 & 0 \\ 0 & \sigma_2 \\ 0 & 0 \end{bmatrix}\mathsf{V}^{\mathsf{T}}.$$

Then we can define the conception of isometric, conformal and equiareal (See details in [48]. The computer language friendly explanation can be found in [22]).

**Theorem 2.2.1.** *For a planar mapping* $\mathbf{f} : \mathsf{R} \to \mathsf{R}$, *the following equivalence gives:*

*1. f is isometric* $\Leftrightarrow \sigma_1 = \sigma_2 = 1$

*2. f is conformal* $\Leftrightarrow \sigma_1/\sigma_2 = 1$

*3. f is equiareal* $\Leftrightarrow \sigma_1\sigma_2 = 1$

So it is $\sigma_1$ and $\sigma_2$ that directly influence the stretch (and the distortion metric energy) of the mapping. So we have

$$\mathbf{E}(\mathbf{f}) = \int_{\sigma} E(\sigma_1(u, v), \sigma_2(u, v)) du dv.$$

This equation should be defined here in different methods. Malliot et al. [49] have proposed the method which minimizes "Green-Lagrange deformation tensor". This tensor is given by:

$$E = (\sigma - 1)^2 + (\sigma - 1)^2.$$

Hormann et al. [50] have presented another method call "Mostly Isometric Parameterization of Surfaces" (MIPS) for parameterization. This method is based on the minimization of the ratio between two direction stretching: $\frac{\sigma_1}{\sigma_2}$. Since minimizing this energy is a difficult numerical problems, they replace it with another simple metric $\frac{\sigma_1^2 + \sigma_2^2}{\sigma_1 \sigma_2}$. Sander et al. [51, 52] have studied a reversed parameterization method that their formalism uses the inverse function to map the parametric space onto the surface. For this reason, their energy can be expressed as $(\frac{1}{\sigma_1})^2 + (\frac{1}{\sigma_2})^2$. Sokine et al. [53] have proposed a method based on the remark that shrinking and stretching should be treated the same. their method uses the following energy to minimize $\mathrm{Max}(\frac{1}{\sigma_1}, \sigma_2)$.

To introduce more flexibility in these methods, some researchers focus on blending these method together in a spectrum. Degener et al. [54] have proposed to use a combined energy, with a term that penalizes area deformations, and another term that penalizes angular deformations. Wang et al. [55] have invented a family of metrics that can flexibly blend the LSCM method [56] and ARSP method [57].

**Barycentric Coordinates.** Barycentric coordinates solve the parameterization procedure from another degree. Retrospect to the simple spring model, barycentric coordinates consider the converge from local region: every vertex and its local neighbors are averaged by the special designed spring force of the connected edge. The motivation of barycentric coordinates derives from the affine combination parameterizing. A succinct idea of this method is based on simple physical model: We constrain the boundary of the mesh onto the boundary of the parameter domain which we target to map to (for simplicity, the domain here is planar rectangular). Suppose two vertices $V_i$ and $V_j$ are connected by Edge $E_{ij}$ and we imagine this edge as a spring. Then, the mesh is transformed to a spring system and the parameterization solving transform to spring energy converge equation: we give each vertex a parameter that where the vertex stop in the domain.

The most important issue here is to specify the spring energy. Barycentric coordinates is one of the spring force representation. Each vertex is represented as the weighted average of the neighbor vertex as:

$$x_i = \sum_{j \in N_i} \lambda_{ij} x_j,$$

and

$$\sum_{j \in N_i} \lambda_{ij} = 1,$$

here the $\lambda_{ij}$ is defined as barycentric coordinates. In some cases the coordinates $w_{ij}$ are determined independently and $\sum_{j \in N_i} w_{ij} = 1$. Then for normalization we set

$$\lambda_{ij} = \frac{w_{ij}}{\sum_{j \in N_i} w_{ij}},$$

where we call $w_{ij}$ homogeneous coordinates. One advantage of inventing $w_{ij}$ includes that we can focus on computing coordinates from geometry information without considering the normalization property.

The earliest generalization of barycentric coordinates goes back to Wachspress [58]. It focuses on finite element analysis and suggests to set the homogeneous coordinates as follows:

$$w_{ij} = \frac{\cot \alpha_{ji} + \cot \beta_{ij}}{r_{ij}^2},$$

where $r_{ij}$ is the edge length. Desbrun et al. [59] have utilized them for parameterization. Meyer et al. [60] for interpolating density values inside convex polygons.

Another set of barycentric coordinates also stems from finite element solving. It actually arises from linear approximation of Laplace equation and is utilized to parameterization, which is given by:

$$w_{ij} = \cot \gamma_{ij} + \cot \gamma_{ji}.$$

Pinkall et al. [61] have also utilized it to compute discrete minimal surfaces. In the area of mesh deformation and interpolation, Sorkine et al. [57] have generalized this coordinates to preserve the surface details.

Another set of coordinates "Mean value coordinates" is proposed by in [62]. The coordinates are given by:

$$w_{ij} = \frac{\tan \frac{\alpha_{ij}}{2} + \tan \frac{\beta_{ji}}{2}}{r_{ij}}.$$

contrary to other coordinates, one advantage of mean value coordinates is that it guarantees that $w_{ij}$ is positive. The negative coordinates may lead to flip-over phenomena and violate injectivity property. Hormann et al. [63] have presented that mean value coordinates have many useful application in computer graphics.

There still exist some other coordinates. [64] have studied and modified the continuity of barycentric coordinates. Lipman et al. [65] have proposed Green Coordinates for closed polyhedral cages. They respect both the vertices position and faces orientation such that it lead to space deformations with shape preserving. Joshi et al. [66] have proposed a character-based barycentric coordinates as practical means to manipulate 3D models by operating to their cages. As indicated in [66], the rigid spatial topological structure of the FFD latices makes the deformation less flexible. Many papers have attempted to analyze the principle of existed coordinates and attempt to give a comprehensive image to all. Ju et al. [67] have analyzed and compared three coordinates (Wachspress, Harmonic, Mean value). They view stokes theory as the root of all three methods. From respect of stock theory, the difference between three coordinates is the chosen of unit element shape: Wachspress use polar dual, mean value use unit circle and Harmonic use original polygon. following the same motivation and pipeline, all 2D polygon barycentric coordinates can extended to arbitrary polyhedron in $R^3$, which is necessary for our volumetric parameterization. [68–70] have extended the mean value coordinates from 2D polygon to 3D polyhedron. [71] have developed the spherical coordinates specifically used for spherical polygons.

## 2.2.2 Volumetric Parameterization Techniques

We have already reviewed many surface parameterization techniques. As a very closely relevant topic to our proposal, here we briefly review the relevant volumetric parameterization techniques. Volumetric parametrization aims to compute a one-to-one continuous map between a 3-manifold and a target domain (or a given surface with interior space) with low distortions. Volumetric parametrization has been gaining greater interest in recent years, a few related techniques have been conducted towards various applications such as shape registration [72, 73], volume deformation [66, 68, 74], and spline construction [2]. Wang et al. [72] have parameterized solid shapes over solid sphere by a variational algorithm that iteratively reduces the discrete harmonic energy defined over tetrahedral meshes, the harmonic energy is rigorously deducted but the optimization is prone to getting stuck on local minima and it only focuses on spherical like solid shapes such as human brain datasets. Ju et al. [68] have generalized the mean value coordinates [62] from surfaces to volumes for a smooth volumetric interpolation. Joshi et al. [66] have presented harmonic coordinates for volumetric interpolation and deformation purposes. Their method guarantees the non-negative weights and therefore leads to a more pleasing interpolation result in concave regions compared with that in [68]. Martin et al. [2] have computed the precise (u, v, w) coordinates for genus-zero tetrahedral meshes, and the target domain is a cylinder. Li et al. [73] have used the fundamental solution method to map solid shape onto general target domains. The current existing methods always attempt to map the model to a standard or simple domain primitives. Thus, how to handle the complex model volumetric mapping is very intriguing. [1] have used a "mid-surface" in combination with harmonic functions to decompose the object into a small number of volumetric tensor-product patches. However, all these methods can not eliminate singularities. Zhang et al. [3] have proposed a method to handle long branches: The algorithm divides possible bifurcations of a vascular system into different cases to solve. Zeng et al. [75] have studied the volumetric parameterization of cylinder wall. In the paper, the differential operator is extended from 2D to 3D. In a similar idea, Xia et al. [76] have utilized Green's function for parameterizing star-shaped volumes. Han et al. [77] have proposed the method to construct the shell space using the distance field and then parameterize the shell space to a poly-cube.

### 2.2.3 Spline-Friendly and Feature-Aware Methods

In this section we briefly review the parameterization techniques that are "Spline-Friendly". "Spline-Friendly" here means "feature-aware". Preserving feature in the parameterization result is very important to spline approximation because it will allow splines to approximate more accurately around the feature region.

Many quadrangulation methods are actually based on parameterization techniques. One important property in quad-mesh generation research is edge-preserving. [45, 78] have constructed an as smooth as possible symmetric cross field that satisfying a sparse set of directional feature edge constraints. Then Daniels et al. [79] have proposed a template-based approach for generating quad-only meshes, which offers a flexible mechanism to allow external input, through the definition of alignment features that are respected during the mesh generation process. [80] have introduced the concept of an exoskeleton as a new abstraction of shapes that succinctly conveys the structure of a 3D model. Here "exoskeleton" actually is the important feature edges on the model surface. Xia et al. [81] have proposed an editable poly-cube parameterization techniques that optional sketched features can be mapped to the corresponding edges on the domain. Huang et al. [82] have presented a extended spectral-based approach. In contrast to the original scheme, it can provide flexible explicit controls of the shape, size, orientation and feature alignment of the quadrangular faces. Zhang et al. [83] have proposed a new method which constructs a special standing wave on the surface to generate the global quadrilateral structure. The wave-equation based method is capable of controlling the quad size in two directions and precisely aligning the quads with feature lines.

## 2.2.4 Global Parameterization and Poly-cube

The motivation of global parameterization comes from the requirement of B-spline. B-Spline fitting demands that the parameter of each local domain keeps regular (tensor-product). It also requires the consistence between different local domains. Another important issue concerns that we expect to construct volumetric spline so that each parameter domain is a $R^3$ space. The surface meshes cover the boundaries of all $R^3$ domains seamlessly and consistently. The way of keeping this property includes choosing a domain (may be composed by a set of sub-domain) that has the same topology but with simplified geometry feature. The most simple way is to map the genus-0 model to a sphere without considering its geometry feature like [84, 85]. However, for more complex topology and geometry feature, more complex domains and parameterization techniques have been developed in the last decades.

The linear discrete harmonic theory is interesting and rich, attractive computationally and enormously useful in applications. The ideas inform contemporary notions of discrete conformality and harmonicity that are based on linear conditions on the vertex coordinates. Examples of applications include [56, 86, 87]. Another set of theories considers the analysis and modification of some key metric (e.g., curvatures). [88–90] have proposed the similar methods based on this theory: First compute a metric for the image mesh and only then a set of vertex positions and then solve the Laplace-Beltrami operator about the metric to flatten a mesh.

Another group of parameterization techniques utilizes curvature directions to drive the parameterization result. For example, in [91, 92], they have proposed an anisotropic polygonal remeshing method, which is the direct application of parameterization,by extracting and smoothing the curvature tensor field and use lines of minimum and maximum curvatures to determine appropriate edges for the remeshed version in anisotropic regions. Meanwhile in some other techniques like [34, 93], they generate two orthogonal piecewise linear vector fields defined over the input mesh (typically the estimated principal curvature directions) and then compute two piecewise linear periodic functions, aligned with the input vector fields, by minimizing an objective function.

Spectral-based parameterization methods study the eigenfunctions of operators (or eigenvectors of matrices in the discrete setting). Dong et al. [94] have used the Laplacian to decompose a mesh into quadrilaterals in a way that facilitates constructing a globally smooth parameterization. Huang et al. [82] have presented an extended spectral-based approach. In contrast to the original scheme, it can provide flexible explicit controls of the shape, size, orientation and feature alignment of the quadrangular faces. Zhang et al. [83] have proposed a new method which constructs a special standing wave on the surface to generate the global quadrilateral structure. The wave-equation based method is capable of controlling the quad size in two directions and precisely aligning the quads with feature lines.

**Poly-cube.**    In [95], they have represented a method to map model with arbitrary shape and geometry to a domain-called poly-cube. Poly-cube is a domain composed by gluing small cubes together. Each segment of input surface mesh maps to one of six surfaces of one cube. The advantage of this mapping method is that the mapping is seamless and each mapping patch is tensor-product regular. The parameter between neighboring patches can transform consistently to each other simply by linear parameter transformation or rotation. So it guarantees consistence between patches by setting the resolution and sampling set of parameter between two patch the same.

Meanwhile, several methods have been developed to improve user control: The user can easily control the mapping by specifying optional features on the model and their desired locations on the poly-cube domain. For instance, Wang et al. [96] have presented a technique where the user can interactively control the desired locations and the number of corners of the poly-cube map; Xia et al. [81] have used user sketches as constraints to control the poly-cube map. Automatic poly-cube construction is always extremely difficult due to the complexity of the input shape. Lin et al. have used Reeb graph to segment the surface and then developed an automatic method to construct poly-cube map [97]. However, their segmentation method may not work for shapes with complicated topology and geometry and does not guarantee a bijection between the poly-cube and the 3D model. He et al. [98] have proposed an automatic algorithm by slicing the model along one horizontal direction and then gluing together. It can only handle the horizontal, planar features from the 3D model. In fact, none of the current techniques constructs the poly-cube simultaneously following all above criteria.

### 2.2.5 Applications on Visualization

In our proposal, one of the important applications about parameterization is on focus+constext (F+C) visualization. Also, a critical part of remaining work involves volume data F+C visualization based on volumetric parameterization. Therefore, it is necessary to introduce and review the related work on this research topic.

Various F+C visualization techniques have already been proposed on many types of informatics inputs, such as trees [99, 100], treemaps [101], graphs [102], tables [103], and city maps [104]. Plaisant et al. [105] have defined the SpaceTree as a novel tree browser to support exploration in the large node link tree. The algorithm applies dynamic re-scaling of branches to best fit the space and includes integrated search and filter functions. For the seamless F+C, Shi et al. [106] have proposed a distortion algorithm that increases the size of a node of interest while shrinking its neighbors. Ying et al. [107] have also presented a seamless multi-focus and context technique, called Balloon Focus, for treemap. Gansner et al. [102] have presented a topological fisheye view for the visualization of large graphs. A method to cope with map and route visualization has been proposed by Ziegler et al. [104]. They depicted navigation and orientation routes as a path between nodes and edges of a topographic network. Recently, Karnick et al. [108] have presented a novel multifocus technique to generate a printable version of a route map that shows the overview and detail views of the route within a single, consistent visual frame. Different from the above methods with specific pre-defined targets, our framework is capable of handling various information or visualization-based applications.

The key component in F+C visualization is to design an efficient lens. Optical effects, such as fisheye [109] for the nonlinear magnification transformation with multi-scale, have been widely used. Fisheye views can enlarge the ROI while showing the remaining portions with successively less detail. Fisheye lens offers an effective navigation and browsing device for various applications [110]. In addition, InterRing proposed by Yang et al. [111] and Sunburst proposed by Stasko et al. [112] have incorporated multi-focus fisheye techniques as an important feature for radial space-filling hierarchy visualization. The major advantage of the fisheye lens is the ability to display the data in a continuous manner, with a smooth transition between the focus and context regions. Although fisheye lens has advantages in preserving the spatial relation, it creates noticeable distortions towards its edges, which fails to formally control the focused region and preserve the shape features in the context region.

Aiming to cope with the shortcomings of the basic fisheye lens, more sophisticated lenses have been proposed. Bier et al. [113] have presented a user interface that enhances the focal interest features and compresses the less interesting regions using a Toolglass and Magic Lenses. Carpendale et al. [114] have proposed several view-dependent distortion patterns to visualize the internal ROI, where more space is assigned for the focal region to highlight the important features. LaMar et al. [115] have presented a fast and intuitive magnification lens with a tessellated border region by estimating linear compression according to the radius of lenses and texture information. Pietriga et al. [116] have provided a novel sigma lens with new dimensions of time and translucence to obtain diverse transitions. Later, they provided in-place magnification without requiring the user to zoom into the representation and consequently lose context [117]. Their representation-independent system can be implemented with minimal effort in different graphics frameworks. Meanwhile, the deformation methods are recently used for the complicated 3D datasets, including volume data [118] and mesh model [119]. Wang et al. [119] have presented a method for magnifying features of interest while deforming the context without perceivable distortion, using an energy optimization model for large surface models. Later, they further extended this framework into 3D volumetric datasets [120]. Inspired by these methods, we utilize geometric deformation that applies to visualization of 2D data sets, targeting to eliminate the local angle distortion and keep the visual continuity.

Many image deformation techniques have been successfully studied and used for various image manipulation applications like image editing and resizing. For example, Schaefer et al. [121] have utilized moving least squares to fit transformations and achieve image editing. Also, many blending polynomial coordinates have been developed for better shape interpolation with boundary deformation constraints (e.g., biharmonic weights [122], green coordinates [65]). Meanwhile, image resizing [96, 123] is introduced in the literature for retargeting images to displays of different resolutions and aspect ratios. Note that, image resizing has a completely different goal from lens design, since the resizing task requires that important image regions are optimized to scale uniformly while regions with other contents are allowed to be distorted. Also, we observe the fact that all of the above techniques confine their operations as energy minimization in the 2D space only. Therefore, it is very attractive to explore a new deformation method that utilizes 3D geometric modeling techniques and broaden the scope of geometric modeling to help the visualization process.

## 2.3 Component-Aware Decomposition

Segmenting 3D surface meshes has been widely studied in graphics and digital geometry processing community. A thorough and detailed discussion on these surface segmentation techniques is beyond the scope of this work, we refer the interested readers to Shamir's great survey [124]. Among these segmentation methods, our volumetric spline conversion task demands to decompose shapes into meaningful volumetric parts, simulating how our vision identifies perceptual parts. "Percep-tual" stresses that part-aware decomposition is inspired by research in perception, in particular by the idea that the human visual system understands shapes in terms of parts [125–127]. Guided by this observation, a lot of part-aware decomposition methods have attempted to encode the appropriate parts-aware metrics to agree with human visual perception and thus get the part-aware parts. For instance, these methods include the slippage [128], shape diameter function [129], interior visual region difference [130], intrinsic symmetry [131–133], modal analysis [134], etc. Meanwhile, particularly relevant to our requirement, skeletons are commonly used global perceptual-part structure representation tools. A lot of skeleton extraction techniques have been presented and thus can be used for part-aware decomposition (e.g., Mesh contraction [135], Reeb graphs [136], Thinning [137], etc). Finally, a part-aware decomposition can be manually edited by simple user interactions on the original surface [138, 139]. However, these methods mainly focus on designing suitable part-aware metrics, none of them has analyzed the segmentation results from the spline modeling view, with respect to criteria such as regularity, controllable corners, patch numbers, etc.

# Chapter 3

# Generalized Poly-cube Splines

As we have introduced in Chapter 2, the engineering design industry frequently pursues data transformation from discrete 3D data to spline formulations because of their compactness and continuous representation. As the newly merged research topic, we want to study the method to construct the volumetric splines in this chapter. The main challenge here is to handle arbitrary topology and complex geometry, which gives rise to our novel idea of "Generalized Poly-cube".

## 3.1 Motivation

Compared with the commonly-used *"surface model to surface spline"* paradigm, volumetric splines can represent both boundary geometry and real volumetric and physical/material attributes. This property makes volumetric representation highly preferable in many physically-based applications including mechanical analysis [41], shape deformation and editing, virtual surgery training, etc. However, converting arbitrary meshes to volumetric splines is extremely challenging because of many conflicting requirements for volumetric parametric domain construction. Attractive volumetric splines should have the following properties.

1. **Structural Regularity.** Tensor-product splines (e.g., NURBS) are defined over regular "cube-like" domains. Compared with the unstructured domain (e.g., polygonal regions covered by tetrahedral meshes), regular domain supports more efficient evaluation and refinement, and GPU acceleration can also be applied directly to spline representation with regular structure. Also, spline-based physical analysis (e.g., isogeometric analysis [41]) has a preference for "cube-shaped" domain.

2. **Singularity-free.** Singularity here means an inability to produce a locally consistent parameterization in the neighborhood. Specially in trivariate splines, a global volumetric model is locally parameterized onto several tensor-product charts. Like Fig. 3.1(a-b), a singular point locates where local charts merge, if its valence number along one iso-parametric plane is larger than four (note that from this definition, singularity in volumetric domain is of difference from surface geometry). Handling singularity with tensor-product splines is very challenging. Therefore, it is desirable to have a global one-piece spline defined on a globally-connected singularity-free domain.

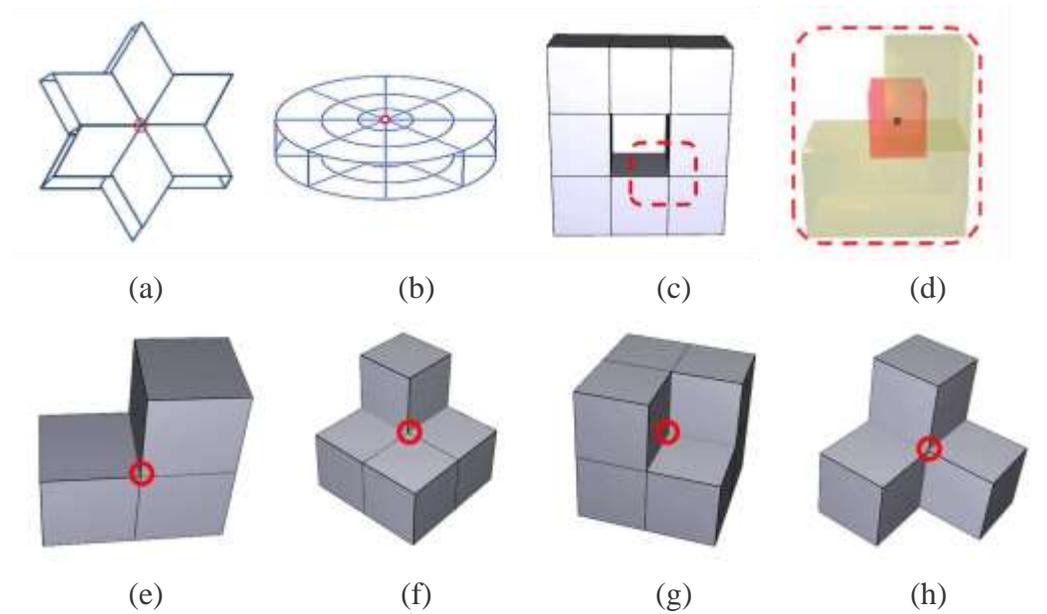

Figure 3.1: Singularity and ill-point distribution in the volumetric domain is very critical to spline construction. (a-b) show two cases of singular points. (c) highlights one ill-point. (d) shows that the basis function around the ill-point has influence outside the cube boundary. (e-h) show different types of ill-points: "Type-1" to "Type-4", which are the concave points in the domain.

3. **Controllable Ill-points.** In a volumetric parameterization over the poly-cube domain, we call the corner point in a concave corner of the poly-cube an ill-point. On such a point, the basis function spans across nearby cubes through outside space (see Fig. 3.1(c-d)). Fig. 3.1(e-h) illustrate all possible types of ill-points in red (note that they are not singularities in volumetric parameterization but singularities in surface parameterization). Being harmless to usual parameterization-related applications, ill-points, however, have an undesirable side-effect on spline construction and subsequent tasks like physical analysis, boundary confinement and partition-of-unity control (see [140-141], [175-178] for more details). Therefore, it is desirable to control the number and types of ill-points. In practice, we hope to restrict ill-points to "Type-1" only, as shown in Fig. 3.1(e), since it is the easiest type and we can simply modify and restrict its "boundary" basis function [141].

4. **Shape Awareness.** Each spline patch should abstract the shape in a geometrically meaningful way, reveal the shape's key perceptual parts and topological structures (e.g., skeleton-like representation). Most importantly, spline construction on large volume data heavily depends on spline gluing in practice. Therefore, one desirable parameterization scheme should try to reduce patch number to cut off spline gluing processing.

Existing volumetric spline techniques generally follow two different trends: (1) Many recent methods [1], [2], [3] convert each part into splines defined on a cylinder/tube domain (e.g., Fig. 3.1(b)), because they can intuitively use the shape skeleton to produce a tube domain and reveal the global structure and topology. A severe limitation of such approaches is that points on the tube centerline are all singular. (2) In contrast, poly-cube splines [142], [98] are defined on domains assembled by multiple cubes, which avoid the central line singularity problem. Such splines are flexible to resemble the shape of the given mesh and are capable of capturing the large scale features with low-distortion mapping. However, gluing of many cubes may produce many uncontrollable ill-points. Limitations from both categories of splines have inspired us to develop a new method that is superior to both types of splines.

The main contributions of this work are as follows. (1) We propose a novel concept of *Generalized poly-cube (GPC)* to serve as the parametric domain for spline construction. Particularly, GPC combines advantages of existing primitives to support splines: (a) GPC is powerful and flexible for representing complex models; (b) GPC provides a simple and regular domain with no singularity and controllable ill-point numbers/types, yet very spline-friendly domain structure. (2) We develop an effective GPC construction and parameterization framework to achieve all the above goals, while still respecting both the global structure and the geometric features. (3) We present a global "one-piece" volumetric spline scheme without stitching/trimming for general volumetric models. Unlike conventional spline schemes, our conversion does not require global coordinates everywhere, and piecewise local coordinates suffice. GPC therefore becomes an ideal parametric domain. We also design an efficient volumetric hierarchical spline fitting algorithm to support recursive refinement with improved accuracy and reduced number of control points.

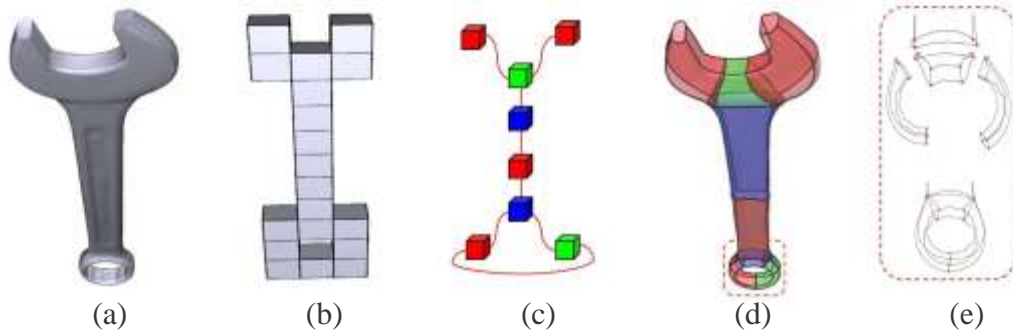

Figure 3.2: Generalized poly-cubes: (a) The wrench model; (b) The conventional poly-cube (CPC); (c) The generalized poly-cube (GPC) as a topological graph; (d-e) The cuboid edges are overlaid onto the model to visualize the GPC global structure.

## 3.2 Generalized Poly-cubes (GPCs)

**Conventional poly-cube (CPC)** is a shape composed of axis-aligned unit cubes that abut with each other. Cubes are glued and realized in a global 3D world coordinate system. CPC usually uses unit cubes as the building block. All cubes are glued together and embedded in the 3D space; any point in a cube is associated with a unique global coordinate. Fig. 3.2(b) shows an example of CPC constructed for a wrench model in (a). Constructing effective (good approximation, coherent topology) CPC for volumetric models with relatively complicated geometry and topology usually requires extensive user involvement. Such a parametric domain is inadequate. A less tedious domain construction with reduced number of ill-points is highly desirable.

**Generalized poly-cube (GPC)** is composed of a set of cuboids glued together topologically. We allow any pair of two distinct cuboid faces to be glued together if these faces have the same size. Fig. 3.2(c-e) show a GPC constructed for the wrench model (Fig. 3.2(a)).

From above definitions, GPC is less restrictive from CPC to be a better spline-friendly domain. First, GPC cuboid is not just a unit box. It can be a general cuboid with rectangular faces. Each cuboid has its local coordinate system; a cuboid is not axis-aligned but can deform (bend or twist) in order to glue with each other to form a global topological structure. Second, cuboids in GPC can be glued together through arbitrary two faces, and it is even possible that they are from the same cuboid. The topology of GPC can be represented using a topological graph, which we denote as a *GPC-graph* (each node represents a cuboid). Fig. 3.2(c) illustrates a GPC graph of Fig. 3.2(d). To represent each cuboid, we project the 12 cuboid edges onto the model to visualize different faces (see Fig. 3.2(d-e)).

A less restrictive GPC has several advantages over CPC, which are very critical to trivariate spline construction: Controlled ill-points, easier domain to simplify spline merging and more general shape modeling.

**Ill-point Controllability.** First, the topological gluing can significantly reduce the number of ill-points (due to the usage of fewer cuboids and simple gluing rules). In a simple shape like Fig. 3.4(a-b), a torus' CPC generates 4 ill-points (in red circles) while a torus' GPC (see the kitten model, Fig. 3.14) has none. Second, our GPC construction algorithm will only generate *Type-1* (Fig. 3.1(e)) ill-points. We can handle them much easier than other types of ill-points [141].

**Easier and Better Domain Construction.** Because of its topological simplic- ity and elegance, the construction of GPC is usually easier than that of CPC. Automatic GPC construction can be developed naturally following the part-aware decomposition of the model. From a spline practitioner's view, CPC requires many redundant cubes (to assemble topological handles in an axis-aligned way, like Fig. 3.4(d)). Cuboids in GPC are similar to the *"generalized cylinder"* so encodes the shape with less cuboids, which can significantly save the cost of spline merging.

When we consider parameterization distortion, less cuboids in GPC may lead to *less distortion* than CPC, because GPC is less restrictive (not axis-aligned) and better mimics shape. For example, a CPC (Fig. 3.4(d)) can merely mimic the genus-3 model (with a narrow top and wide bottom region) in an axis-aligned domain. Consequently, two red-colored parts are parameterized onto the equally-sized domain, introducing large distortion. A GPC (Fig. 3.5(c)) can fit the shape better and significantly improve the parameterization quality, benefitting the final spline construction.

**Highly-twisted and High-genus Shape.** GPC can serve as the parametric domain for a more general category of solid shapes like the twisted or highly curved model, such as the twirl (Fig. 3.3(a)) and möbius band (Fig. 3.3(d)). Unlike axis-aligned CPC, GPC can twist them and glue adjacent cuboids in a topological way so that twisted global shape features can still be modeled as the cuboid edges (b,e), with a very small number of cuboids (c,f). For example, we can hardly construct a useful CPC domain for möbius band; But with GPC, only one cuboid is enough (f). Another category of models includes models with complex topology especially when handle loops/voids are relatively small, such as in the solid bucky model (g). For CPC, not only the above restrictive axis-aligned problem, small handles/voids also make the resulting CPC "over-complex". A less restrictive GPC allows us to model the domain through a correct topological decomposition to small cuboids (h). The pattern of the bucky's GPC-graph around one handle can be decomposed as shown in (i).

The following three sections discuss the algorithmic pipeline to construct GPC and splines (also illustrated in Fig. 3.5). The input model is first decomposed into a few T-shapes. The final output is a global one-piece spline representation.

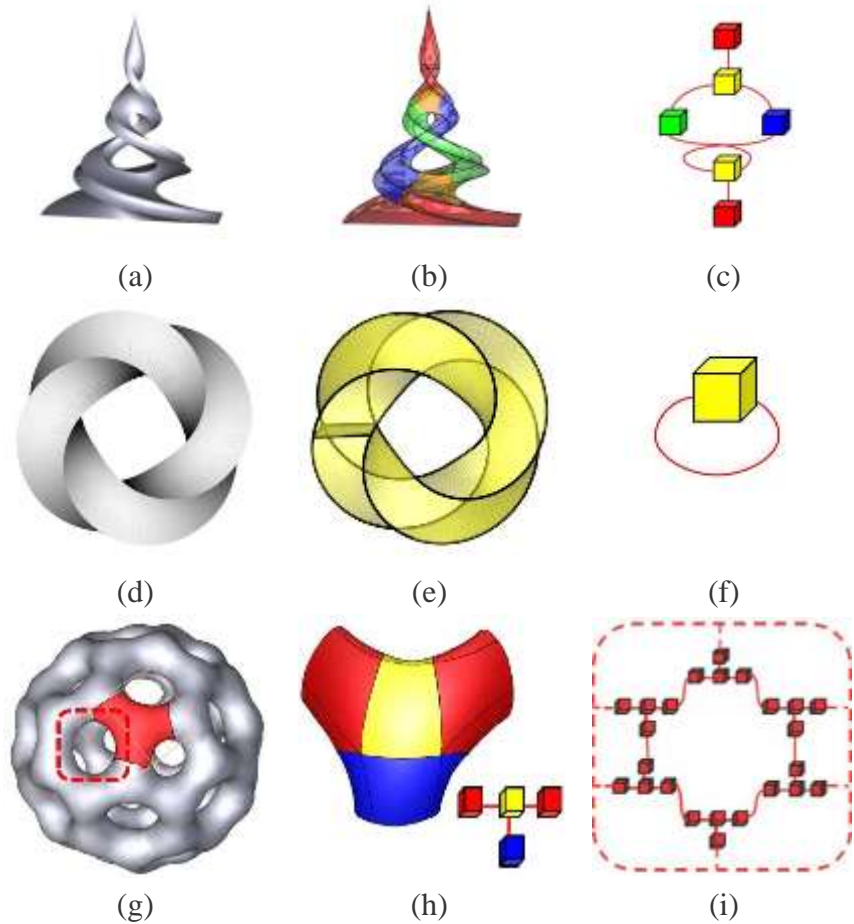

Figure 3.3: GPC can handle more generalized models. Row 1: (a) The highly-twisted swirl model, (b) Its GPC, and (c) Its topological graph. Row 2: (d) The non-axis-aligned möbius model, and (e,f) Its GPC and topological graph. Row 3: (g) The bucky model with complex topology, (h) It is decomposed into small "T-shapes" with 4 cuboids. (i) A subset of the GPC graph around the hole.

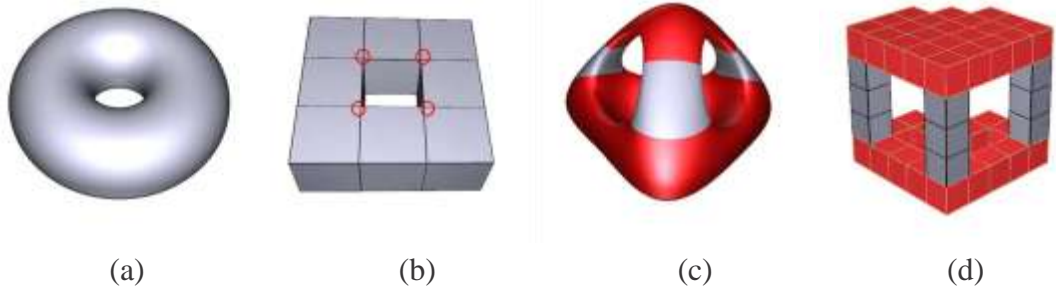

|     |     |     |     |
|:---:|:---:|:---:|:---:|
| (a) | (b) | (c) | (d) |

Figure 3.4: (a) The torus model. (b) Its CPC uses at least 8 cubes and generates 4 ill-points. (c) The genus-3 model with narrow top and wide bottom regions. (d) Its CPC maps two regions onto the equal-sized parameterization domain, leading to large distortion.

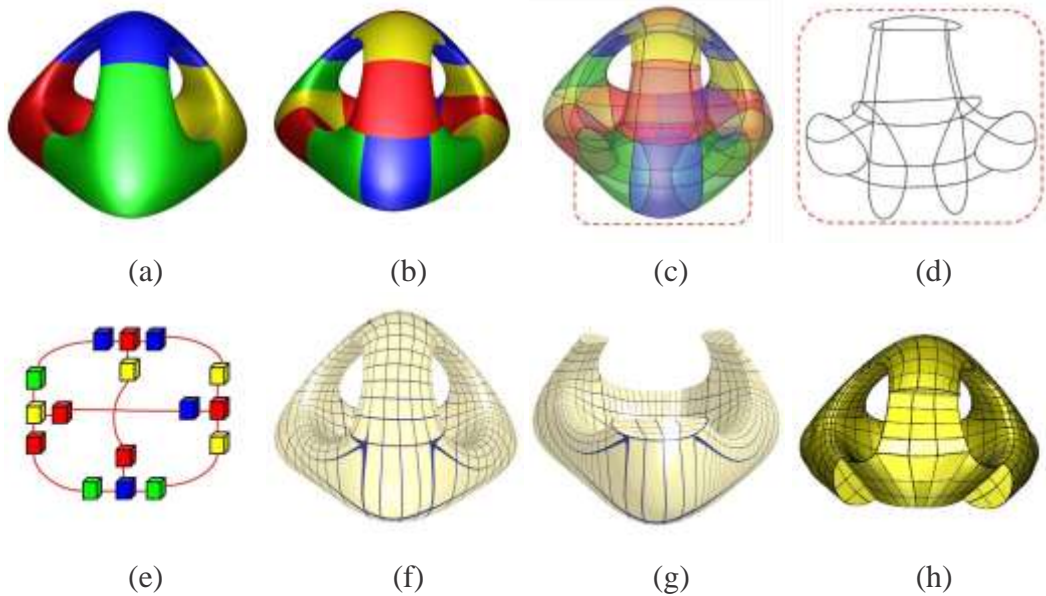

|     |     |     |     |
|:---:|:---:|:---:|:---:|
| (a) | (b) | (c) | (d) |
| (e) | (f) | (g) | (h) |

Figure 3.5: GPC and spline construction pipeline. (a) The input genus-3 model is first decomposed into some "T-shape" patches. (b) Each "T-shape" is further decomposed into 4 cuboids. (c-d) Overlay all cuboid edges onto the model to visualize the global structure. (e) All cuboids comprise a topological GPC. (f-g) Construct the parametric mapping between the input model and its GPC. (h) Transform the model into a volumetric spline representation.

## 3.3   Model Partitioning

Suppose a solid region is bounded by a triangle-meshed surface $\partial M$ (note that $\partial M$ can be of high-genus, but as the boundary of a solid object $M$, $\partial M$ is a closed surface), this section focuses the computation of a group of curves $\{c\}$ on $\partial M$. These curves segment $\partial M$ into sub-patches $\partial M_i$, bounding sub-solid regions $M_i^s$ to be parameterized upon GPC cuboids. We denote these traced curves on $\partial M$ as *poly-edges*, as they will be mapped to edges of GPC cuboids. Our segmentation includes two main steps:

- Partitioning into T-shapes: we decompose the entire model into a group of T-shaped patches.

- T-to-cube decomposition: we generate poly-edges on each T-shape and de- compose it into 4 connected cube-like sub-patches.

T-shapes are used as the basic primitive in our framework to decompose more complicated solid models. A T-shape, which represents the very simple 3-branched volume shape, has trivial topology and only contains Type-1 ill-points.

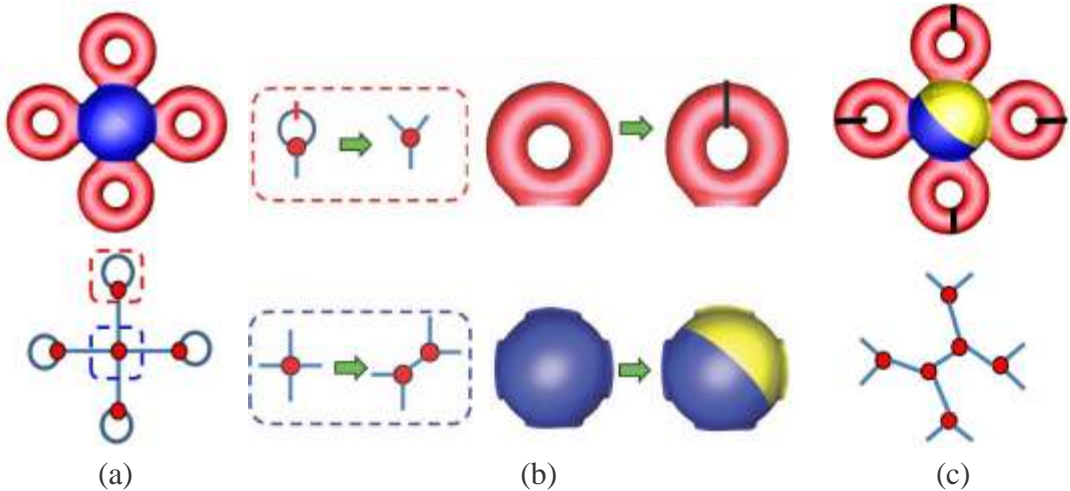

Figure 3.6: Model segmentation into "T-shape" patches. (a) The part-aware segmentation and its abstraction graph. The nodes in the graph have different cases for edge connection (red and blue regions). (b) For each case, we have corresponding operations on the graph and input model. (c) Our operation guarantees that the resulting nodes in the graph are all degree $d = 3$, and the model is segmented into T-shapes.

### 3.3.1 T-shape Segmentation

We use $\partial T_i$ to represent a T-shape surface and $T_i$ for its bounded volume. Our idea is to partition a given model M into several T-shaped sub-regions $\{T_i\}$. We achieve this segmentation through tracing curves on the boundary surface $\partial M$ and partition it to sub-patches $\partial T_i$ or many simpler patches. This pipeline is illustrated in Fig. 3.6. The algorithm has following steps. Note that the challenge is how to ensure the segmented patch is geometrically similar to a "T" in 3D space, not just topologically.

**Step 1.** We first partition the input $\partial M$ into several initial part-aware patches with non-intersecting cutting curves. Any closed surface (the boundary of a solid model) can be partitioned in this way [143]. Different geometric criteria can be integrated in this unified partitioning framework. We choose volume-aware shape descriptors such as the shape diameter functions [129] to guide our cutting curve tracing.

**Step 2.** Upon a complete decomposition, we construct an abstraction graph: a node represents a patch, an edge connecting two nodes indicates their patch adjacency, and an edge connecting a node to itself indicates a handle loop. Fig 3.6(a) shows a 4-torus with colored part-aware segmentation and the resulting abstraction graph.

**Step 3.** We modify each partitioned patch to a standard shape. It means that we split the abstraction graph's nodes with high valance until all nodes have $\leq 3$ incident edges (a graph node with $d = 3$ represents a 3-branch patch, i.e., T-shape, and $d = 1$ or $d = 2$ indicates the patch that bounds a tube). We partition every patch through analyzing all connected edges:

(3.1) *Handle loop* (see Fig 3.6(b), Row 1). We generate the shortest handle loop by [144] and then cut along it. In the abstraction graph, this partitioning cuts the loop into two edges.

(3.2) *High Valence (*$d > 3$*) branch* (see Fig 3.6(b), Row 2). We partition it to two connected nodes $n_1$ (valence-d $-$ 1) and $n_2$ (valence-3). Then we repeat the split until all newly-generated nodes are valence-3. To achieve this idea, we first choose two boundaries (a pair with the closest distance). Then we utilize the technique in [143] to generate a cutting curve that covers two boundaries and avoids any intersection. This curve segments the patch into two patches, one with 3 boundaries (i.e., a T-shape) and another one with $d - 1$ boundaries. We again execute the same partitioning method on the second patch until only 3 boundary patches exist. Row 2 shows an example of the cutting loop.

After repeating the above operations on every node, we can get a decomposition result where every node has its valence equivalent to 3 or less, as shown in Fig. 3.6(c). Compared with existing partition techniques, our segmentation method is uniquely spline-friendly: No prior segmentation result considers the critical is-sues in splines like singularities and ill-points. Without addressing these issues, a segmentation is less suitable for spline conversion. Our T-shape based segmentation, however, is completely singularity-free and ill-point controllable.

Cutting curve loops should be prevented from intersecting each other in our system. This can be ensured by not allowing a newly traced curve hitting (vertices of) existing loops. When two loops are very near and the triangle mesh is very sparse, triangles around this region will be subdivided to ensure the topologically correct tracing without intersection (for mesh refinement to ensure reliable curve tracing, please see [143] for details).

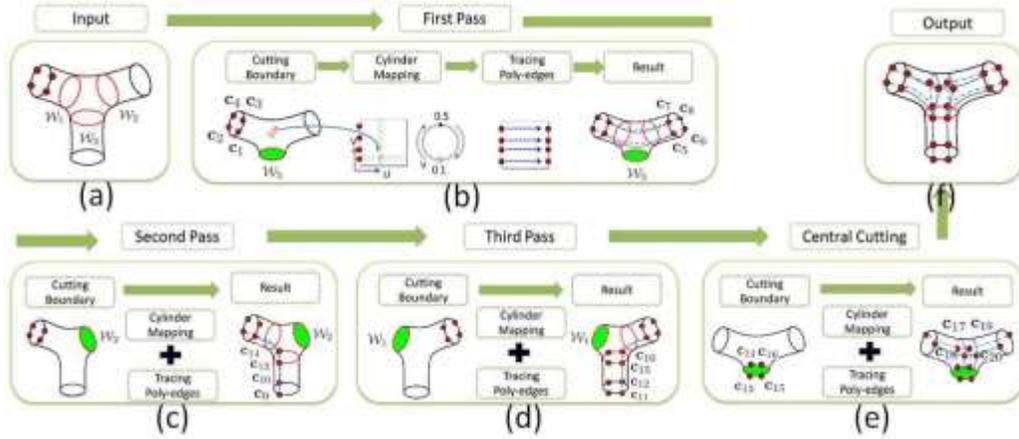

Figure 3.7: Illustration of T-to-cube segmentation.

### 3.3.2 T-to-cube Segmentation

We process a set of T-shapes $\partial T_i$ or tube-shaped (cylinder) patches, one-by-one in an arbitrary order. $\partial T_i$ is first partitioned into 4 sub-patches $\partial M_{ij}$, then we generate corners and poly-edges on each $\partial M_{ij}$ (recall that poly-edges are the traced curves that will be mapped to the edges of cuboid domains), as shown in Fig. 3.7. Meanwhile, for any simple tube-shaped patch, we can generate its corners and poly-edges directly by the Step 2, the first pass, i.e., Fig. 3.7(b). Finally, each resulting patch has 8 corners and 12 poly-edges like a cuboid. To guarantee corner alignment, when we determine one T-shape's result, we transfer its corners on the boundaries to the adjacent T-shapes if they are not processed yet.

**Step 1.** We generate three cutting lines $W_1$, $W_2$, and $W_3$ (See Fig. 3.7(a)). We first find 4 corners on one boundary. We denote this boundary as "left" while arbitrarily denoting other two as "right" and "bottom". Positions of 4 corners are determined by its previously-processed adjacent T-patch (except for the first processed T-shape, on which we manually set these 4 corners). To generate 3 cutting lines, we detect 3 branches of $\partial T_i$ by extracting associated skeleton [145], with 3 resulting cutting lines.

**Step 2.** We generate all poly-edges and corners on a T-shape $\partial T_i$, separately in 3 passes (Fig. 3.7(b-d)). Each time we trace poly-edges between 2 boundaries with 3 sub-steps.

(2.1) We first remove the third long branches by cutting along its cutting lines (e.g., $W_3$ in Fig. 3.7(b)). After filling the cutting hole [146], the resulting surface is a 2-boundary tube-shaped patch $\partial P$.

(2.2) We map the tube shape to a cylinder domain $[u, v]$ following the approach of [2]. We shall briefly describe this algorithm: First, set $u = 0$ for vertices on one boundary and $u = 1$ for the other boundary, solve $\Delta u = 0$ by mean value coordinates [62]. Second, trace an iso-v curve along $\nabla u$ from an arbitrary seed vertex on the boundary $u = 0$ to the other boundary $u = 1$ and slice along this iso-curve and get two duplicated boundary paths, then set $v = 0$ and $v = 1$ on them respectively and solve $\Delta v = 0$. The $\partial P$ is therefore parameterized onto a cylinder domain.

(2.3) We generate poly-edges between possible node pairs based on the cylinder-parameterized patch. For the first pass, we trace 4 edges from all corners on the left boundary to the right. For the second pass, we find 2 corners on the left boundary with shortest Dijkstra distance to the bottom ($c_1$, $c_2$) as shown in Fig. 3.7(c) and trace 2 edges from them to the bottom. For the third pass, we choose pairing corners of $c_1$ and $c_2$ on the right boundary ($c_5$, $c_6$) and trace 2 edges to the bottom (the possible node pairing/poly-edge tracing algorithm is described below).

**Step 3.** We generate poly-edges and corners for the central cuboid cutting. With 4 intersection corners (between the bottom cutting line and the traced paths) generated in the second and the third pass, now we trace poly-edges between two intersection corners in each pass ($c_{13}$, $c_{14}$ and $c_{15}$, $c_{16}$).

**Tracing Poly-edges.** The above algorithm involves tracing edge $[c_1, c_2]$ on a cylinder parameterized patch $[u, v]$. According to the processing queue, $c_2$'s location is either already determined by other precedent patches or is not yet known. For an unknown $c_2$, we trace the poly-edge from the starting corner ($c_1$) along the gradient direction $\nabla u$ to another boundary at a new point $c_2$. For a determined $c_2$, we map both $c_1$ and $c_2$ to the cylinder domain $[u, v]$ and trace the straight line on the domain between them, then project this parametric straight line back to the patch and get the resulting poly-edge. Note that none of poly-edge is restricted to mesh edges. We allow them to cross and split the mesh triangles. This strategy enables more smooth path lines.

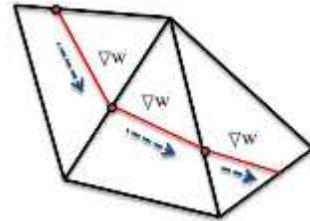

**Node Pairing.** When we trace poly-edges, it is very possible that all corners' locations on two boundaries are predetermined by other precedent patches. In such scenario, we desire to pair two boundaries' corners before tracing edges between unpaired corners. Intuitively, the traced path should be least deviated from the gradient of the harmonic field. Suppose we are tracing paths between boundary $b_1$ and $b_2$. If corners on both $b_1$ and $b_2$ are predetermined, we trace the gradient line from $b_1$'s corners and get ending nodes on $b_2$. Then we compute and find the pairing between ending nodes and corners on $b_2$, satisfying that the sum of total distance between each pair is minimized. In this way, we can get the pairing between corners on $b_1$ and $b_2$; If corners on $b_2$ are not predetermined yet, we directly use the ending nodes as the new determined corners and thus get the pairing. In practice, we can merge edge tracing in the second/third pass (Fig. 3.7(c,d)) together: we determine the 4 node pairing together to avoid possible intersected poly-edges generated between two passes.

**Feature-preserving Segmentation.** Although the above automatic algorithm can handle most of models very well, sometimes users still expect to use several sharp features as the poly-edges. For example, this choice is specially natural and meaningful on the strong symmetric man-made models with sharp features (e.g., CAD models in Fig. 3.3(b-e)). Specifically, a scaling factor is applied to edges on feature curves, so they are considered shorter in the Dijkstra path tracing. Therefore, features will be on the traced curves and poly-edges if we compute shortest path between corners. Fig. 3.3(a, d, e) and Fig. 3.2(d) show the results with feature-preserving poly-edges. In practice, this method can only pick a few major feature lines (like in the twirl model, the poly-edges are sharp features we pick). It is still difficult to handle more complex features. Instead, we can preserve the extra sharp features through the following spline fitting step.

## 3.4 Parameterization

After the input model M is decomposed into sub-patches $\{\partial M_{ij}\}$, bounding topological solid cuboids $\{M_{ij}\}$, we now perform cuboid parameterization of $\{M_{ij}\}$. We first map the patch boundary to the cuboid domain surface. Then we use this mapping as boundary condition and compute the interior volumetric parameterization.

### 3.4.1 Surface Parameterization

The subpatch $\partial M_{ij}$ computed previously has 8 corners and 12 poly-edges (see Fig. 3.8(a)), we partition $\partial M_{ij}$ into 6 topological rectangles, then solve 3 harmonic mappings $\Delta u = 0$, $\Delta v = 0$, $\Delta w = 0$ on all rectangles. Each time we pick 2 opposite rectangles as 2 iso-plane domains on one direction (e.g., $u = 0$ and $u = 1$).

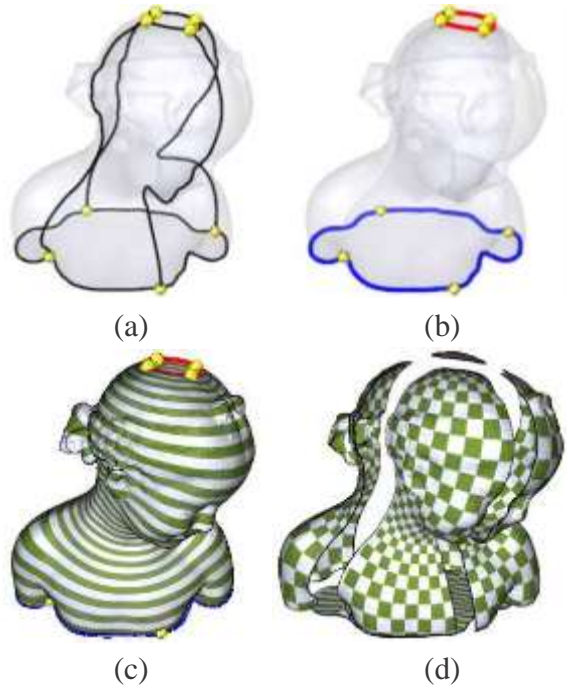

(a)          (b)

(c)          (d)

Figure 3.8: Illustration of surface parameterization.

Then we compute the parameters of this direction (u) on all other 4 rectangles. For example, to solve $\Delta u = 0$, we select 8 poly-edges on two opposite rectangles (see Fig. 3.8(b)). 4 red poly-edges bound an iso-u rectangle (u = 0) and the 4 blue poly-edges bound another iso-u plane (u = 1). Then we compute the approximated discrete harmonic map $\Delta \mathbf{u} = 0$ [62] on other regions. Fig. 3.8(c) illustrates the computed u. Similarly, we can compute the harmonic scalar fields of v and w with $\Delta v = 0$ and $\Delta w = 0$, respectively. After solving 3 harmonic mappings, each vertex on the surface patch is mapped to a coordinate $(u_0, v_0, w_0)$ on the cube surface. The surface parameterization is illustrated in Fig 3.8(d).

### 3.4.2 Volumetric Parameterization

We compute the volumetric parameterization of $M_{ij}$ on a set of $n_0 \times n_1 \times n_2$ grid points. These grid points correspond to the uniformly-sampled coordinates in the parametric space (u, v, w). This volumetric parameterization can be considered as finding the locations of these nodes within $M_{ij}$. Similarly, as we discussed in surface parameterization, we need to find the point locations that minimize the equations $\Delta u = 0$, $\Delta v = 0$ and $\Delta w = 0$ in 3D space.

The $n_0 \times n_1 \times n_2$ grid points include two categories: the surface grid points and interior points. We determine their positions as follows.

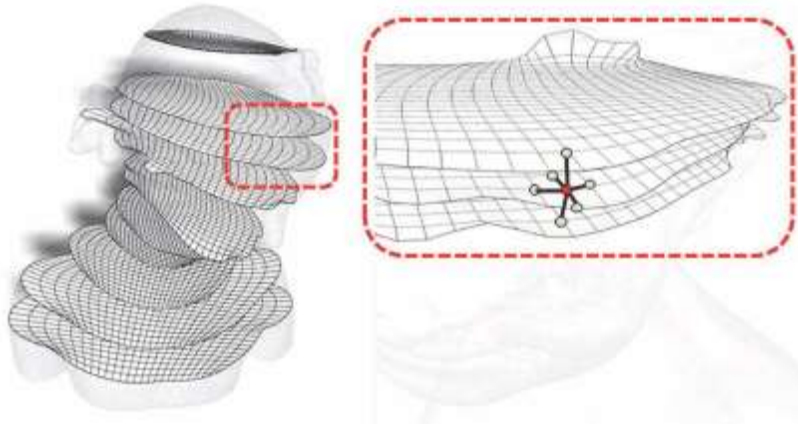

Figure 3.9: Volumetric mapping. We extract sample points as a hexahedral model. Each node has 6 neighbors for solving 3-D Laplacian in Eq. (3.1).

(1) If the parameter of a grid point n falls on the domain surface, we can always find its location on $\partial M$ by the parameter of n, where n's parameter always falls into a triangle $[v_1, v_2, v_3]$ of $\partial M$ on the parametric domain with corresponding barycentric coordinates $\lambda_1, \lambda_2, \lambda_3$, then its spatial location is interpolated as $\sum_{i=1}^{i=3} \lambda_i P(v_i)$, where $P(v)$ denotes the 3D position of vertex v.

(2) Keeping the surface points fixed, we compute the interior point position by minimizing 3D Laplacian Eq. (3.1), where $n_{ijk}$ and $N_{ijk}^{\lambda}$ represent the node and its neighbor's spatial positions in (x, y, z) and $w_{\lambda}$ is the point weight. In practice, each node is moved to the weighted mean center of their six neighbors. Here, the choice of weight $w_{\lambda}$ has been studied in [72], [68]. In our implementation we simply use the uniform weight $w_{\lambda} = 1/6$ as suggested in [146] and [147].

$$E(n_{ijk}) = \sum_{\lambda} w_{\lambda} \times ||(n_{ijk} - N_{ijk}^{\lambda})||, \lambda \in N_b(n_{ijk}). \qquad (3.1)$$

We move grid points iteratively. The update converges when changes of all node positions are smaller than a threshold during one iteration. Fig. 3.10(a-c) show the computation results of the femur model after 20, 60, and 80 iterations.

**Refinement across Cutting Boundary.** Before merging, the parameterization of two adjacent sub-patches are already computed separately. Along the cutting interface, only $C^0$ continuity is guaranteed and the cutting boundary is not smooth. We perform a refinement to improve this smoothness. To reduce computation time, we only extract a small region from each patch. For example, we pick a region from one patch within the parameter $(1 - \alpha, 1) \times (0, 1) \times (0, 1)$, and $(0, \alpha) \times (0, 1) \times (0, 1)$ from another patch if two patches are connected along

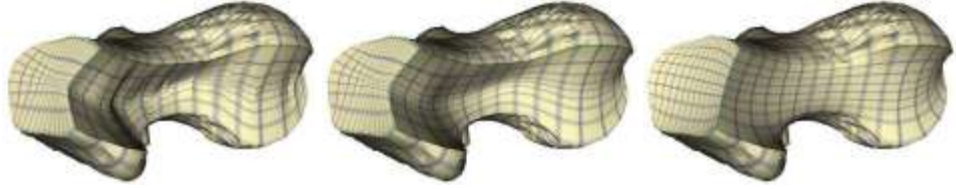

Figure 3.10: Results of the cut-out view of the interior femur model by solving Eq.(3.1) after 20, 60, and 80 iterations.

$\nabla u$ direction ($\alpha$ is a small scalar value). Gluing two extracted region together, the new patch also has 8 new corners and 12 new poly-edges (recall that poly-edges between two adjacent patches are aligned along the boundary), thus we can recompute the surface mapping and volumetric mapping on the new patch. Meanwhile, this recomputing is subject to an extra constraint, on regions that connect to an extra third cuboid. We keep these region's parameter unchanged during recomputing, to avoid our modification destroying global parameter consistency.

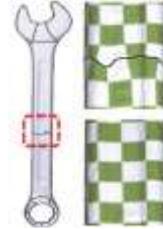

## 3.5   GPC-Splines

Two challenging issues must be addressed when designing the mesh-to-spline transformation over GPC. First, allowing adaptive refinement without significantly increasing control points is highly desirable since volumetric spline fitting usually requires a large number of control points when we seek high approximation accuracy. Second, unlike conventional B-splines that each control point and its knots are associated with global coordinates, GPC provides only locally-defined parameters in each cuboid domain. This is because a global realization of GPC parametric domain in 3D Euclidean space is oftentimes impossible on highly twisted/high genus models. Thus we design a unique GPC-spline algorithm using a point-based scheme.

In principle, a volumetric cubic spline can be viewed as a point-based spline: Each control point $C_i$ (located in parametric cube $D^j$ with local coordinate $q^j$) is associated with three knot vectors along three principal axes: $r = [r_1, r_2, r_3, r_4, r_5]$, $s = [s_1, s_2, s_3, s_4, s_5]$, $t = [t_1, t_2, t_3, t_4, t_5]$, where $c_{ij} = (r_3, s_3, t_3)$. All knots can be determined using a *ray-tracing* strategy [26]. For any sample point with $(u, v, w)$ as its local parameter, the blending function is

$$B_i(u, v, w) = N_r(u) \times N_s(v) \times N_t(w), \tag{3.2}$$

where $N_r$, $N_s$, and $N_t$ are cubic B-spline basis functions associated with the knot vector r, s, and t respectively. The formulation for point-based splines (PB-splines) is

$$P(u, v, w) = \frac{\sum_0^n C_i B_i(u, v, w)}{\sum_0^n B_i(u, v, w)}. \tag{3.3}$$

We modify the above equation to construct GPC splines. The GPC domain comprises a collection of coordinate charts locally defined in individual cuboid. Adjacent local parametric coordinates are transformed coherently by transition functions, which can be encoded in a GPC-graph structure. Consequently, the global PB-splines are piecewise rational polynomials defined on GPC, whose transition functions between adjacent cuboids are compositions of simple cuboid translations and rotations of $n\pi/2$, where n is an integer.

In a cuboid $D^j$, given an arbitrary parameter h, also denoted as $h^j$, the spline approximation can be carried out as follows:

(1) Find all the neighboring cubes $\{D^i\}$ that support h (i.e., it contains control points $C_k$ that may support h);

(2) The spline function is:

$$P(h) = \frac{\sum_{k=0}^n C_k^i B_k(\varphi^{ij}(h^j))}{\sum_{k=0}^n B_k(\varphi^{ij}(h^j))}, \tag{3.4}$$

where $h^j$ is the local parametric coordinate of point h in the cube domain $D^j$, $\varphi^{ij}$ is the transition function from cube domain $D^j$ to $D^i$, and $C_k^i$ denotes the control point k in the cube domain $D^i$.

In theory, a transition function $\varphi^{ij}$ from cube domains $D^j$ to $D^i$ is a composition of translations and rotations following the shortest path from cube $D^j$ to cube $D^i$ in the GPC-graph. Suppose $\overline{D^{ij}} := D_1(=D^i) \rightarrow D_2 \ldots \rightarrow D_n(=D^j)$, and the transition function $\Phi_{(i,i+1)}$ (derived by way of cube-gluing) from $D_{i+1}$ to $D_i$ is already known, then $\varphi^{ij}$ is formulated by

$$h^i = \varphi^{ij}(h^j) = \Phi_{1,2}(\Phi_{2,3}(\ldots\Phi_{n-1,n}(h^j))).$$

In practice, because most control points only influence a very small local region and do not cut across non-adjacent cubes, we observed that only using a neighboring cube transition function is usually enough.

Along any merging region, two connected cubes share the same domain size along the merging face (i.e., we forbidden partial gluing between a large and small cubes). Therefore, when we merge two cuboids' control grid (with the same resolution), all the control points and intervals along the merging faces will merge coherently, without any T-junction before hierarchical fitting.

### 3.5.1 Hierarchical Fitting

Following above GPC-spline definitions, we develop a hierarchical fitting scheme to approximate volumetric models. For a sample point $f(h_i)$ in the model whose parametric coordinate is $h_i$ (defined by the volumetric parameterization computed in previous sections), $P(h_i)$ is our GPC-spline representation. We minimize the following equation:

$$E_{dist} = \sum_{i=0} ||P(h_i) - v_i||^2, \qquad (3.5)$$

which can be rewritten in matrix format

$$\frac{1}{2} C^T B^T BC^T - V^T BC, \qquad (3.6)$$

where $C$ is the vector of control points, $V = v_i$ is the vector of sample points, and $B = B_i(h_i)$ is the matrix of basis functions. This least square problem is not difficult to solve numerically. Given a sample parametric point $h$ in GPC, in order to decide if we need to refine the approximation, we measure the root-mean-square error (RMS) $\sigma(h)$ between its spatial position $f(h)$ and its spline approximation $P(h)$. Algorithm 1 documents the main steps. The input includes all sample points and an initial control grid with control points. The initial control grid mimics the structure of GPC: Each cube corresponds to a local regular control grid. All local grids are topologically glued coherently following the GPC-graph, generating a one-piece global control grid. The function `KnotVectors` collects 3 direction knots for each control point. We use the same "ray-tracing" strategy in [26]. `InfluencedSamples` returns all sample points in the influenced region of a control point. `Transition` transports a local parameter from one cube to another cube. `AssembleMatrix` assembles the matrix for Eq. (3.6) and `SolvingEquation` solves it and determines the control point positions. `FittingError` returns the worst fitting result in a small grid. `Subdivision` divides a grid uniformly into 8 smaller sub-grids. Fig. 3.11 illustrates our hierarchical fitting results.

---

**Algorithm 1** Hierarchical spline fitting.

---

Input: Initial control grid $L_g$,

        List of sample points $L_s$,

        List of control points $L_c$,

        Fitting error threshold Output:

all control points positions. **loop**

  //Update control point knot vectors

  **for all** $L_c$ **do**

    c = $L_c$.next()

    c.knots = KnotVectors(c, $L_g$)

    $L_s^{\emptyset}$ = InfluencedSamples(c, $L_s$)

    **for all** $L_s^{\emptyset}$ **do**

      s = $L_s^{\emptyset}$.next()

      s.ctrlist.push_back(c)

    **end for**

  **end for**

  //Compute basis functions for samples

  **for all** $L_s$ **do**

    s = $L_s$.next() $B_{total} = 0$

    $L_c^{\emptyset}$ = s.ctrlist $L_B = \{\}$

    **for all** $L_c^{\emptyset}$ **do**

      c = $L_c^{\emptyset}$.next()

      param=Transition(s.cube#,c.cube#,c.param)

      B= BasisFunction (param,c.knots)

      $L_B$.push_back(B) $B_{total} = B_{total} + B$

    **end for**

    AssembleMatrix ($L_B, B_{total}, s$)

  **end for**

  //Fitting and evaluation

  SolvingEquation()

  **for all** $L_g$ **do**

    g = $L_g$.next()

    **if** $FittingError(g) >$ **then**

      $L_g^{\emptyset}$ =Subdivision(g)

      $L_g$.delete(g) $L_g$.insert($L_g^{\emptyset}$)

    **end if**

  **end for**

  Stop if no updated grid

**end loop**

---

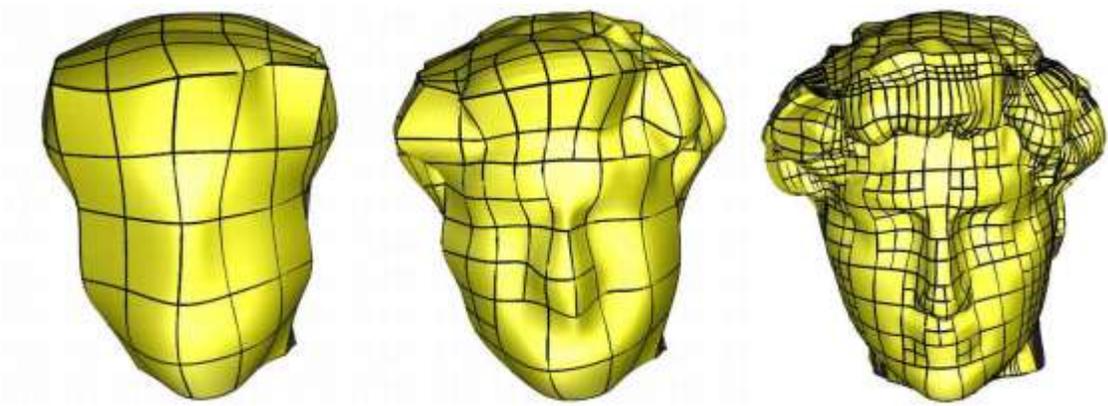

Figure 3.11: Hierarchical spline fitting results at levels 0, 1, and 2, respectively.

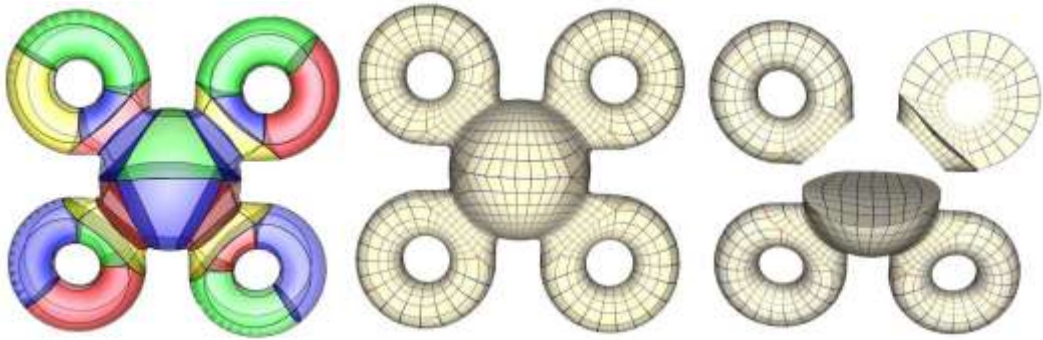

Figure 3.12: The 4-sphere model visualized with cuboid organization, poly-edge structure, surface parameterization, and volumetric parameterization.

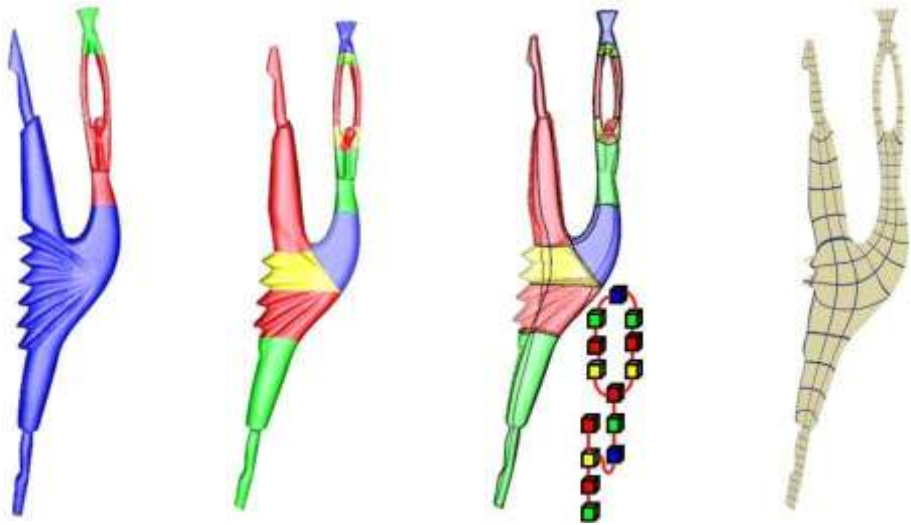

Figure 3.13: The dancer model visualized with T-shape decomposition, cuboid organization, poly-edge structure, GPC-graph, and volumetric parameterization.

## 3.6   Implementation and Discussion

Our experimental results are implemented on a 3GHz Pentium-IV PC with 4Giga RAM. To demonstrate the versatility of our approach (therefore, the flexibility of our computational framework), we construct GPC splines for many models. Our experiments include models with twisted shape: twirl (Fig. 3.3(Row 1)), möbius solids (Fig. 3.3(Row 2)); and with complex topology: bucky (genus 31, Fig. 3.3(Row 3)), genus-3 (Fig. 3.5), 4-sphere (genus 4, Fig. 3.12); and with complex conceptual parts: wrench (Fig. 3.2), dancer (Fig. 3.13), and greek and david (Fig. 3.15). Table 3.1 summarizes the statistics of the GPC construction, including every model's properties (genus, twisted/not twisted), the number of T-shapes, cuboids and ill-points.

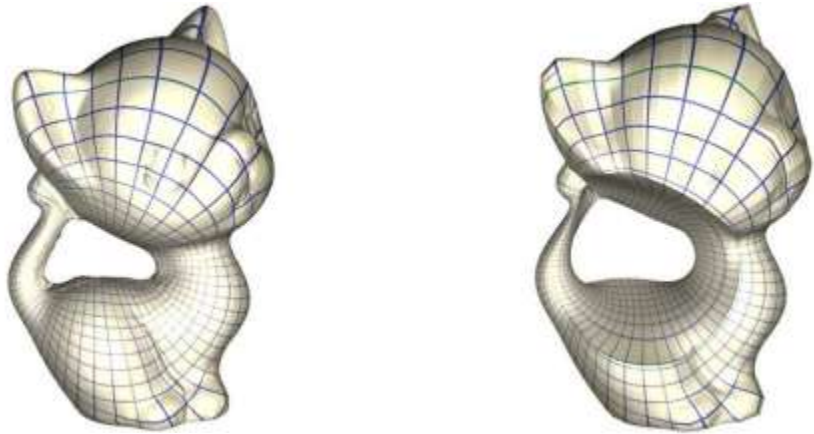

Figure 3.14: The kitten model visualized with surface and volumetric parameterization.

It may be noted that our parameterization algorithm may not guarantee a globally-minimized angle and volume distortion. However, since our algorithm decomposes the input into part-aware patches, each of which is parameterized on a geometrically similar cuboid, the distortion is satisfactory for our spline construction. The models of dancer, 4-sphere, kitten, greek and david (Fig. 3.12, Fig. 3.13, Fig. 3.14, and Fig. 3.15) demonstrate several surface and volumetric GPC parameterization results. Fig. 3.16 shows several volumetric spline approximation results. We overlay the control grid line (black lines) onto the fitting results, and the T-junctions on the control grid reduce the control point greatly while still preserving the shape details. The statistical results are given in Table 3.2. The table shows that the vertices' number increases dramatically when we convert a surface model into a volume data. Our spline scheme can significantly reduce control points for shape representation. In most of our experiments, approximation with good quality can be achieved within 3 levels of hierarchical refinement. The fitting qualities are measured by RMS errors normalized to the overall sizes of solid models.

Table 3.1: Statistics of various test examples

| Model | Genus | Twisted | # T-shape | # Cuboid | # Ill-points |
|---|---|---|---|---|---|
| genus-3 | 3 | no | 4 | 16 | 8 |
| bucky | 31 | no | 60 | 240 | 120 |
| mobius | 1 | yes | 1 | 1 | 0 |
| twirl | 1 | yes | 2 | 6 | 4 |
| 4-sphere | 4 | no | 6 | 24 | 12 |
| bimba | 0 | no | 1 | 1 | 0 |
| femur | 0 | no | 1 | 1 | 0 |
| wrench | 1 | no | 2 | 8 | 4 |
| dancer | 1 | no | 3 | 14 | 6 |
| david | 3 | no | 4 | 12 | 24 |
| greek | 4 | no | 6 | 19 | 12 |

Table 3.2: Statistics of various spline examples.

| Model | #. Surface vertices | #. Volume vertices | #. Control points | RMS error | Running time |
|---|---|---|---|---|---|
| kitten | 12403 | 40000 | 3020 | 0.35% | 202s |
| wrench | 7550 | 12000 | 2966 | 0.2% | 105s |
| 4-sphere | 2042 | 22800 | 1088 | 0.2% | 47s |
| genus-3 | 6632 | 51200 | 1280 | 0.17% | 162s |
| david body | 15572 | 81600 | 5956 | 0.37% | 890s |
| greek body | 20109 | 91900 | 7265 | 0.4% | 1096s |

**Comparisons.** We compare our method with other volumetric parametric domain construction and mapping approaches: [95], [98], [1], [3], [97], and [72]. As shown in Table 3.3 and Fig. 3.3, our method has advantages in the following aspects. First, our method works well for volumes with complex topology and structure. Second, our domain does not have any singularity and can control the type and number of ill-points (which is highly desirable for spline construction). Our domain construction does not require tedious design, even for very complex shape input. Meanwhile, we can also flexibly edit the cube domain to better approximate the shape interactively.

We also test our system on the rocker-arm model, which also appears in other papers (e.g, CubeCover in [148]). As the comparison shown in Fig. 3.17, our parameterization has the same quality as in [148]. However, 26 cube domains and 4 singular points are used in [148], while we only have eight cuboids and no singularity. In Fig. 3.18, we compare our domain (Middle Left) with the methods in [98] (Middle Right) and [149] (Right) using the fertility model. Our domain significantly decreases the number of cuboids (19) as while as ill-points (only on cuboids with more than two edges in the GPC-graph).

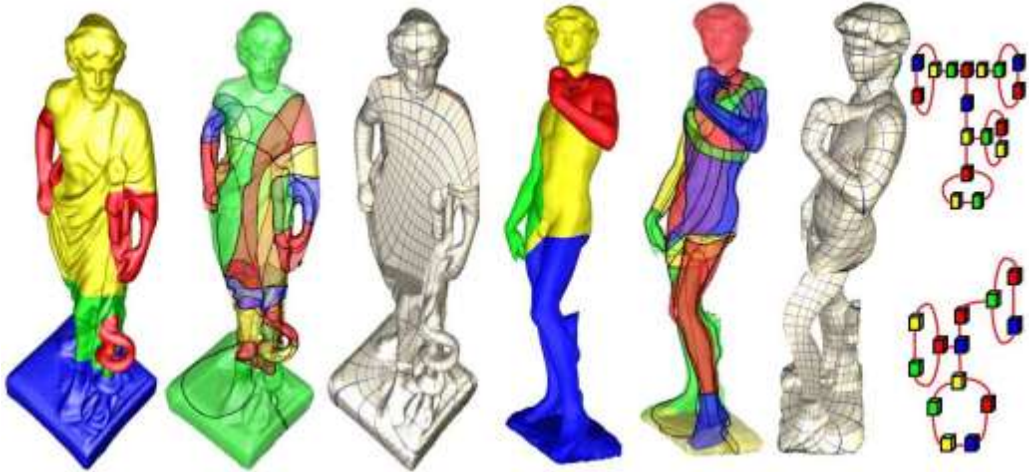

Figure 3.15: The greek and david model visualized with T-shape decomposition, cuboid organization, poly-edge structure, volumetric parameterization and their GPC graphs, respectively.

**Discussions.** Since singularity-free and ill-point simplification is the first priority in our spline-oriented system, this enforcement may lower mapping quality in certain region. According to users' requirement, we can always change it on the fly based on a hybrid system. Inside the current partitioning framework, we may further allow extra local segmentation to improve its geometry awareness. Upon initial partitioning we detect long branches, and construct additional cuboids to parameterize these branches. For example, we map the axial shaft and handle of the screwdriver (Fig. 3.19) to separate cuboids. Compared with using only one cuboid, the distortion (e.g., the extrusion effect) around the handle top is significantly reduced. However, as mentioned above, this modified GPC decomposition will bring extra singularities, ill-points, and merging cases. In this example we add four extra "type-4" ill-points.

Our system decomposes the input model mainly according to global shape and topology. This implies that it fails to handle the model with complex features if they are everywhere. Enforcing poly-edges covering features (Section 3.3.2) can only recover major features which are globally dominant. For spline construction, this is not a critical issue since we can always improve the fitting quality hierarchically around any sharp feature. However, many feature based applications may require features to be retained. We will investigate how to preserve the feature as much as possible.

Table 3.3: Comparison with the existing approaches.

| Method | Tarini [95] | He [98] | Martin [1] | Zhang [3] |
|---|---|---|---|---|
| Primitives | cube | cube | cylinder | cylinder |
| Topology | axis-aligned | axis-aligned | symmetric | long branch |
| Twisted model | no | no | yes | yes |
| Singularity | no | no | center | center |
| Ill-points | no control | large number | no | no |
| Domain construction | Artiest design | Axis scan | Simple | Simple |
| Editable domain | yes | no | no | no |
| Method | Lin [97] | Wang [72] | Hex-mesh [150] | Ours |
| Primitives | cube | sphere | numerous small no parameter | cube |
| Topology | reeb-graph | genus-0 | arbitrary | arbitrary |
| Twisted model | no | no | yes | yes |
| Singularity | no | center | large number | no |
| Ill-points | no control | no | large number | controllable |
| Domain construction | Simple | Sphere only | No domain existed | Simple |
| Editable domain | no | no | no | yes |

Our poly-edge tracing algorithm can not prevent them from intersecting with each other. Fortunately, our tracing algorithm can avoid intersection on a well partitioned part-aware patch. However, intersection may happen on a very poorly-shaped T-shaped patch. We will develop an automatic method to detect degeneration and correct it.

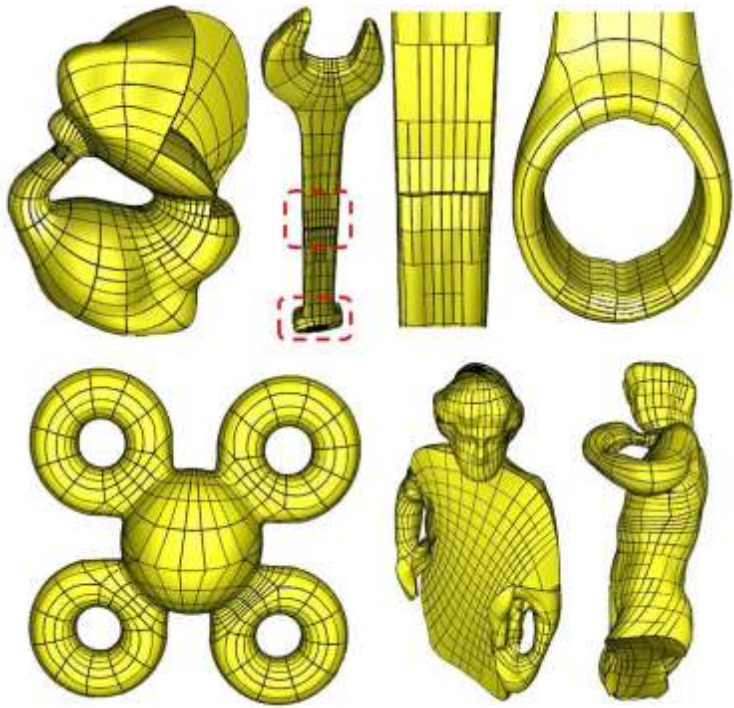

Figure 3.16: The volumetric spline approximation results.

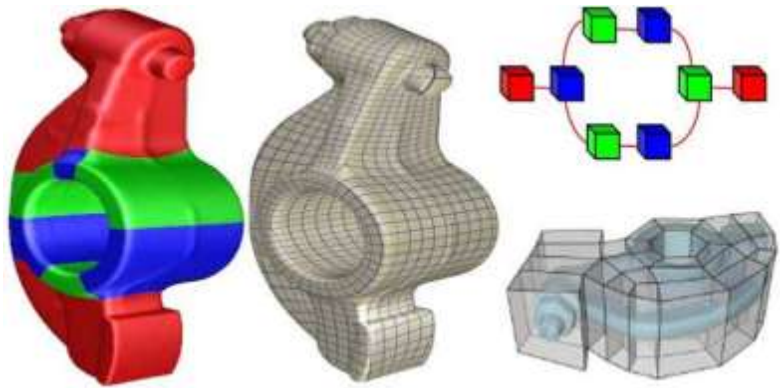

Figure 3.17: Our segmentation/mapping result of the rocker-arm model (left/middle). Our GPC (right up) has only 8 cuboids/no singularity, compared with 26 cubes/4 singular points (right bottom, courtesy of [148]).

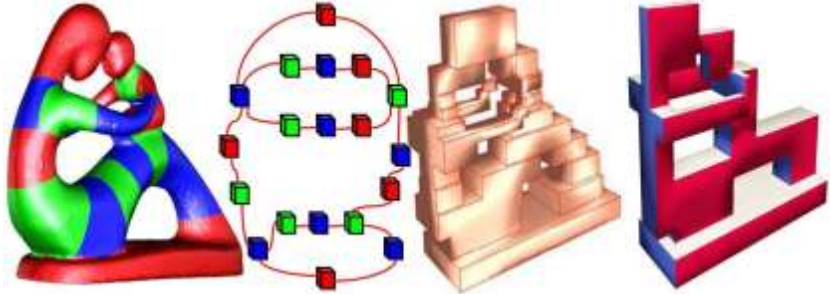

Figure 3.18: Comparisons of different methods on the fertility model (courtesy of [98] and [149]). Our domain has significant improvement on cuboid and ill-point number.

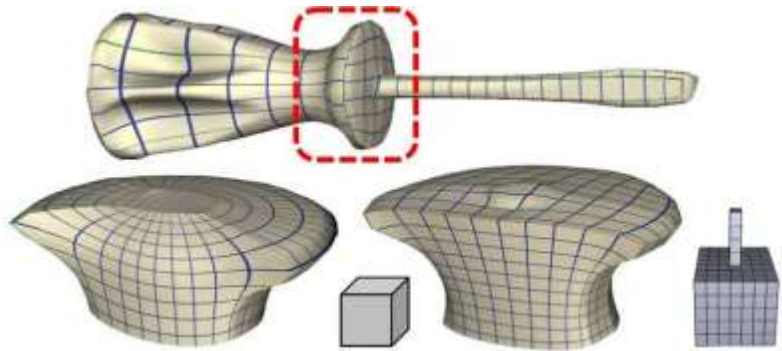

Figure 3.19: Modified result of the screwdriver model (up). Mapping it to two separate domains (bottom right) instead of one cuboid domain (bottom left) can moderate distortion like extrusion round the handle top region.

## 3.7　Chapter Summary

In this chapter, we have presented a GPC spline framework for data transformation from surface meshes to continuous volumetric splines. The novelty in this chapter lies at the systematic handling of generalized poly-cube (GPC) parametric domain without any strong assumption. Compared with conventional poly-cube (CPC), GPC provides more generalized shape domain and better numerical stability to represent complicated models of arbitrary structure. We design a volumetric parameterization procedure based on GPC, which better handles solid objects with general topology and structure than existing volumetric parameterization techniques. We then devise a global "one-piece" volumetric spline based on GPC parameterization. The GPC construction enables a novel and desirable mechanism that facilitates the "one-piece" spline representation. Using local point-based strategy, global volumetric T-splines can be constructed on piece-wise GPC because transition functions can be effectively computed from the GPC's topological structure. The entire spline framework affords hierarchical refinement and level-of-detail control. Our GPC volumetric splines have great potential in various shape design and physically-based analysis applications. Our GPC is of great value to a wide range of geometry processing tasks, including volumetric isogeometric analysis [41], volume deformation, anisotropic material/texture synthesis.

# Chapter 4

# Component-aware Trivariate Splines

In the last chapter, we have proposed the technique to map the model into a generalized poly-cube domain. That means the integral model is decomposed into several cube components. Subsequently, it is natural to construct splines on each component and then glue them together. However in the previous chapter, we still use the global parameterization to approximate the global splines integrally for numerical reasons. Very naturally, this phenomena intrigues us to answer the question: "How to apply divide-and-conquer schemes onto decomposition-already inputs?" In this chapter, our primary goal is to develop efficient methods for arbitrary solids undergoing spline transformation, with local spline construction and global spline merging.

## 4.1   Motivation

To achieve this goal, we must address the following key challenges.

(1) **High genus.** An attractive spline representation must accommodate high-genus solid models with complicated shapes.

(2) **Local refinement and adaptive fitting.** For trivariate splines, both structurally-complicated shape models and feature-enriched models need local refinement. For example, a genus-0 solid bounded by 6 simple four-sided B-spline surfaces has originally $6 \times 1024^2$ control points (DOFs). The size of DOFs increases drastically to $1024^3$ or even larger when we naively convert it to a volumetric spline representation. This exponential increase during volumetric spline conversion poses a great challenge in terms of both storage and fitting costs. Therefore, it is advantageous to use high resolution to approximate boundary surface and low resolution for interior space.

(3) **Singularity free.** A *singular point in a volumetric domain* is a node with valence larger than four along one iso-parametric plane (Fig. 4.1(a)). Handling sin-gularity with tenor-product splines is highly challenging in FEM, thus a singularity-free domain is highly desirable. Unfortunately, singularities commonly exist in many volumetric domains such as hexahedral meshes and cylinder (tube) domains.

(4) **Boundary restriction.** It is a basic requirement for a spline that all blending functions are completely confined within the parametric domain.

(5) **Semi-standardness.** A hierarchical spline is always formulated as Eq. 4.1. Semi-standardness, meaning that $\sum_{i=1}^{B} w_i B_i(u, v, w) \equiv 1$ always holds for all $(u, v, w)$, has a broader appeal to both theoreticians and practitioners.

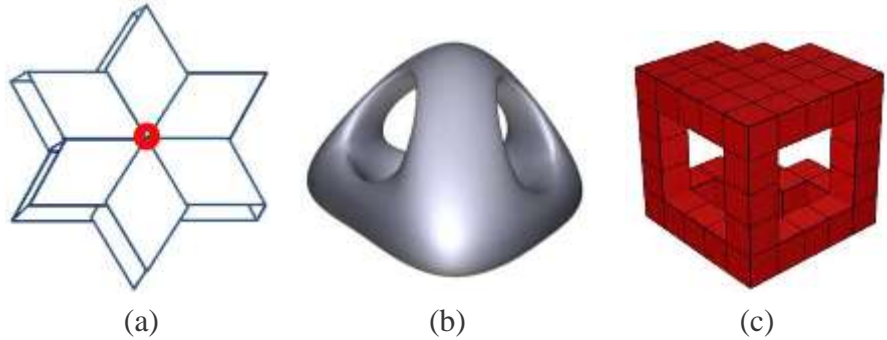

(a)             (b)             (c)

Figure 4.1: (a) The singular point in the volumetric domain. (b-c) A poly-cube domain can mimic the geometry of input and avoid such type of singular point.

Recently, much work has been attempted towards spline modeling of arbitrary topology shape while satisfying the aforementioned requirements, following a top-down fashion like Wang et al.[140]. They have proposed a theoretical trivariate spline scheme, being built upon volumetric poly-cube domains. Poly-cube is a shape composed of cuboids that abut with each other. All cuboids are glued in various merging types like Fig. 4.2, without any singular point (Note that the yellow dots are *not* singular points in the trivariate splines, even though they are singular for surface study). For example, a poly-cube parametric domain like Fig. 4.1(c) is designed to mimic shape geometry Fig. 4.1(b). Although their spline refinement guarantees the features such as semi-standardness and boundary restriction, this theoretical formulation encounters many difficulties. A global one-piece poly-cube domain, together with its 3D embedding, is not versatile enough to handle highly-twisted and high-genus solid datasets. Creating a poly-cube to mimic the input shape requires tedious user work. The boundary restriction procedure in the vicinity of gluing regions (Fig. 4.2, yellow dots/lines) is extremely complicated. Computationally speaking, the global fitting is very time consuming which is completely unsuitable for trivariate splines.

To ameliorate, our framework takes advantage of the bottom-up scheme. The global domain is divided into several components, with a controllable number and

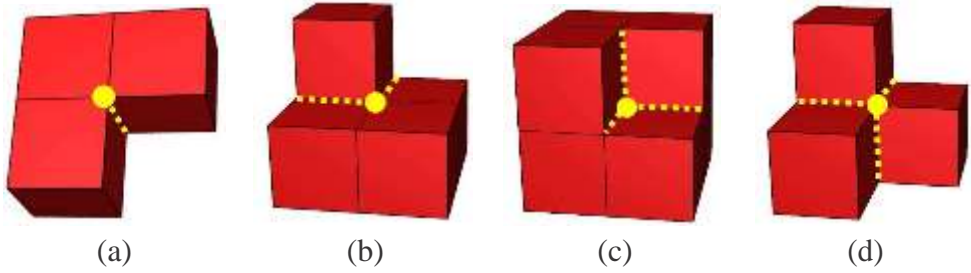

$$(a) \qquad\qquad (b) \qquad\qquad (c) \qquad\qquad (d)$$

Figure 4.2: All possible merging types in a poly-cube ("Type-1" to "Type-4"). To preserve both boundary restriction and semi-standardness, we add extra knots around the control points on the merging boundary (yellow lines and dots).

types of the cuboid merging. We build tensor-product trivariate splines separately for each component, and then glue them together. Compared with the top-down scheme, our divide-and-conquer method is more flexible and powerful to handle high-genus and complex shape. The interior space mapping and remeshing in each component is much easier. Compared with global fitting, our local fitting reduces both the computation time and space consumption significantly.

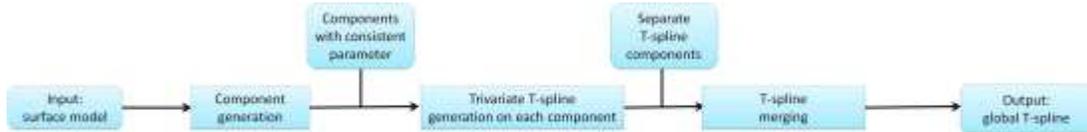

Figure 4.3: The divide-and-conquer scheme.

One key theoretical challenge in our divide-and-conquer scheme lies at designing merging strategies, so that the global spline after merging will still satisfy the semi-standardness and boundary restriction properties, especially around merging regions across adjacent cuboids. All possible cuboid merging types for a poly-cube are shown in Fig. 4.2. The traditional merging technique [27] only handles standard surface T-spline models defined over 2D domains without considering any 3D merging. In our framework, we have to design a new merging strategy, through adding extra knots and modifying weights of blending functions, to handle each merging case in Fig. 4.2, enforcing the semi-standardness and boundary restriction properties everywhere. Fig. 4.3 and 4.4 show the detailed, step-by-step procedure using a high-genus G3 model as an example. Specifically, it includes the following major phases:

(1) Construct a surface poly-cube mapping. To better support our divide-and-conquer scheme, we use the technique [151] to decompose the entire surface model into several components. Each component is a part-aware surface patch and we map it to the boundary surface of a cuboid. We also guarantee in this step that separate cuboid mappings are globally aligned.

(2) Construct a local trivariate tensor-product T-spline on each cuboids (Section 4.3). Adaptive fitting is allowed for a better fitting result.

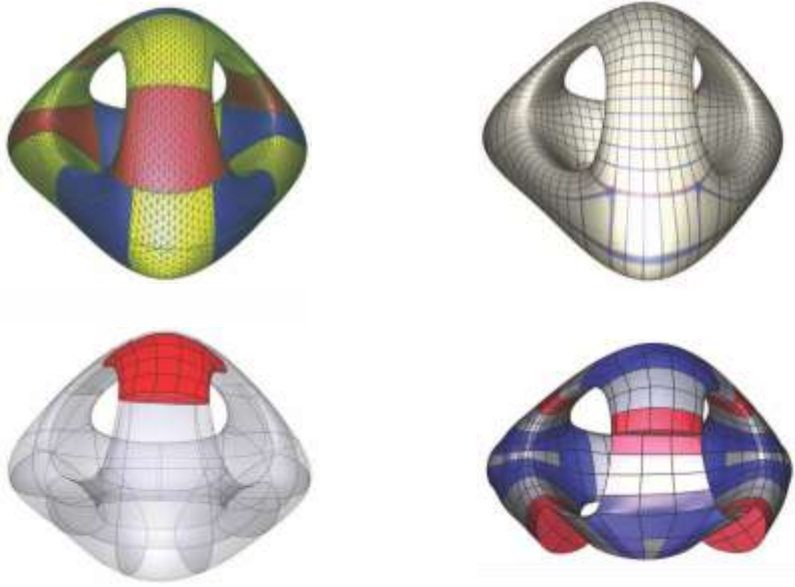

Figure 4.4: Steps to convert the G3 model into a trivariate T-spline solid.

(3) Merge local cuboids into a single global spline (Section 4.4). Note that, the novelty of our merging strategy lies at its comprehensive and complete solution to guarantee the desirable properties: semi-standardness and boundary restriction.

Our new shape modeling framework has the following advantages:

1. Compared with prior top-down strategies, our new divide-and-conquer approach is more flexible and powerful to handle complex solids with arbitrary topology. Each component can be easily converted to a trivariate semi-standard regular spline, which is embraced by industry-standard CAD kernels and facilitates GPU computing like [152].

2. We develop the theory and algorithm to merge adjacent trivariate splines together. Through adding knots and modifying weights, our merging method can enforce semi-standardness and boundary restriction for all possible merging types, even after local adaptive refinement.

3. For solids with homogeneous material, we are capable of generating trivariate splines from poly-cube surface parameterization directly, thus we avoid complicated interior volumetric remeshing. Moreover, our divide-and-conquer strategy makes the modeling and analysis tasks scalable to large-scale volumetric data, in terms of computation time and space consumption during the fitting.

## 4.2  Component Generation and T-splines

This section briefly reviews the required surface poly-cube generation algorithm. We also define the necessary notations for the rest of this chapter. In the interest of understanding, most illustrative figures about knots are simply shown in 2D layout, as their 3D generalizations are straightforward.

### 4.2.1  Component Generation

The starting point of our whole procedure is to decompose an input surface model into several component surface. Each component surface is part-aware and maps to a cuboid surface. The decomposition and mapping must follow the rule that parameters between neighboring components are consistent (i.e., we can glue their parameters together directly as a seamless aligned global poly-cube mapping). We remain agnostic as to which method should be used for such decomposition. However, in order to better promise these requirements, we utilize the algorithm introduced in Chapter 3. Compared with the conventional poly-cube mapping method like [95], our construction is specifically suitable for the divide-and-conquer strategy and spline construction. (1) The conventional poly-cube method always generates an integral poly-cube domain to mimic the whole shape at first. Then we have to decompose this integral domain into small pieces for applying the divide-and-conquer strategy. In contrast, our method directly uses a small set of connected local cuboids, each of which represents a geometrically meaningful patch (e.g., part-aware). This property is particularly suitable for highly-twisted/non-axis aligned/high-genus models (e.g., the g3 model). More importantly, we can use the divide-and-conquer technique directly on our resulting poly-cube without further decomposing anymore. (2) Our method can also reduce the number of cuboids, and control the merging types efficiently: It only generates "Two-cube" and *"Type-1"* (Fig. 4.2(a)) merging, thus it simplifies the merging requirement.

### 4.2.2  Trivariate T-spline

To better prepare readers for the better understanding of the following algorithm, we briefly define the volumetric T-spline representation (The surface T-spline formulation is detailed in [27]). Also we give the detailed explanation of *"Semi-standardness"* and *"Boundary Restriction"* as follows.

We use $\mathsf{T}(\mathsf{V},\mathsf{F},\mathsf{C})$ (or simply $\mathsf{T}$) to denote a control grid domain, where $\mathsf{V},\mathsf{F}$, and $\mathsf{C}$ are sets of vertices, faces and cells, respectively. Given $\mathsf{T}$, a trivariate T-spline can be formulated as:

$$F(u, v, w) = \frac{\sum_{i=1}^{B} w_i p_i B_i(u, v, w)}{\sum_{i=1}^{B} w_i B_i(u, v, w)}, \qquad (4.1)$$

where $(u, v, w)$ denotes parametric coordinates, $p_i$ is a control point, $\mathsf{W}$ and $\mathsf{B}$ are the weight and blending function sets. Each pair of $< w_i B_i >$ is associated with a control point $p_i$. Each $B_i(u, v, w) \in \mathsf{B}$ is a blending function:

$$B_i(u, v, w) = N_{i0}^3(u) N_{i1}^3(v) N_{i2}^3(w), \qquad (4.2)$$

where $N_{i0}^3(u)$, $N_{i1}^3(v)$ and $N_{i2}^3(w)$ are cubic B-spline basis functions along $u, v, w$, respectively.

In the case of cubic T-spline blending functions in Eq. 4.1, the univariate function $N_j^3$ for each blending function $B_i$ is constructed upon knot vector $R^j$, where $R^j$ is a tracing ray parallel to the control grid (See Fig. 4.5(b)): Starting from a knot $k = r_0^0, r_0^1, r_0^2$, we can trace to $r_1^0$ and $r_{-1}^0$, which are the very first intersections when the ray $R(t) = (r_0^0 \pm t, r_0^1, r_0^2)$ comes across one cell face. Naturally, we define the parameter of a control point as the central knot of the knot sequence for the control point.

To support downstream CAE applications, our spline framework has the following requirements:

**Semi-standardness.** $\mathbf{P}_{i=1}^{B} w_i B_i(u, v, w) \equiv 1$ holds for all $(u, v, w)$ in Eq. 4.1, so that the evaluation of spline functions and their derivatives is both efficient and stable. Eq. 4.1 can be rewritten as:

$$F(u, v, w) = \sum_{i=1}^{B} w_i p_i B_i(u, v, w), \qquad (4.3)$$

**Boundary restriction.** We require that blending functions of all control points are strictly confined within parametric domain boundaries. Unfortunately, achieving this requirement is not trivial, especially around the cuboid merging regions. Fig. 4.5 shows a counter-example. A standard control point's blending function (green box), without confinement procedure, tends to intersect with the boundary. In CAE-based force analysis, it means the strain energy "escapes the border", which might lead to an abrupt bend, twist, and flip-over phenomena in experiments. In the follow sections, we usually use *"central points"* for the control point/knot with an unconfined blending function, since the confinement procedure is mainly through adding extra knots/control points around the central point. However, even we de-sign the additional knots carefully and successfully confine the blending function, we still have to recompute all control points' weights around the knots-adding region, otherwise we will break the semi-standardness around this local region.

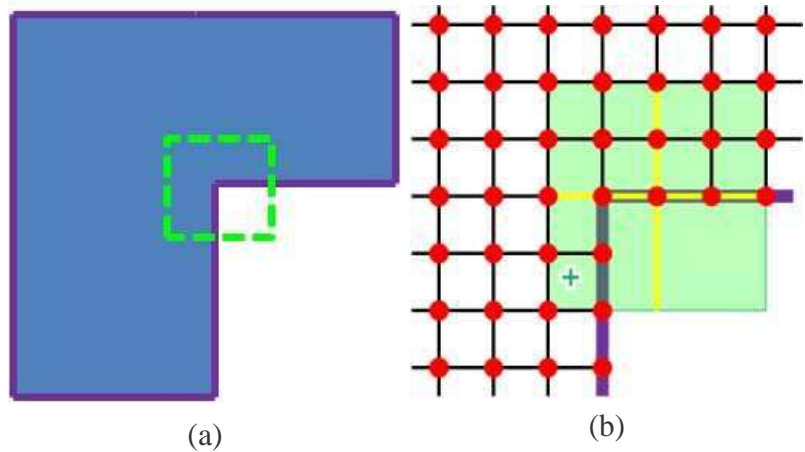

(a)                                          (b)

Figure 4.5: Counter-example of boundary restriction. (a) A "Type-1" merging in a 2D layout. (b) The blending function's supporting region (green box) crosses the boundary. The supporting region is determined by tracing rays (yellow lines).

## 4.3   T-spline Construction for Each Component

The construction of trivariate splines on each component is very critical in our divide-and-conquer method. Two major goals are involved in this step. Besides constructing T-splines preserving desirable features, we have to satisfy the necessary requirement in each component in anticipation for merging. We propose the following procedure to satisfy both goals:

1. Construct a boundary restricted control grid.

2. Perform the meshless fitting to determine locations of all control points.

3. Subdivide the control grid via local refinement iteratively. Perform fitting again in each iteration for a better fitting result.

4. Modify the control grid around merging boundary after each subdivision iteration in anticipation for merging.

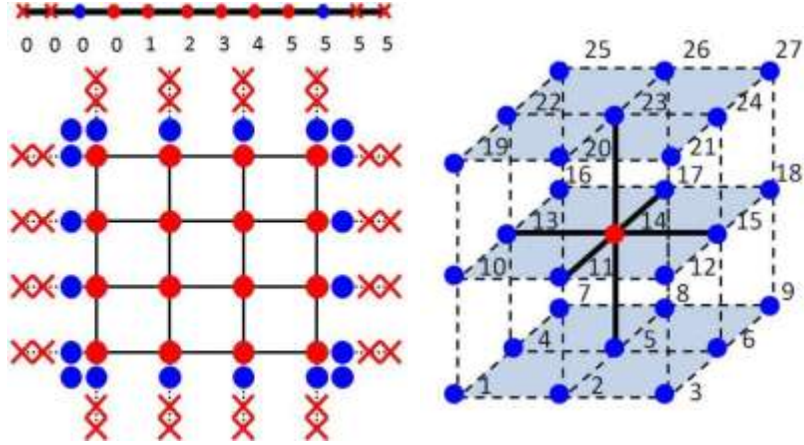

Figure 4.6: (a) Top: Boundary restriction is illustrated on a 1D domain, with 6 "boundary knots" (or called "bd-knots", $[0, 0, 0]$ and $[5, 5, 5]$) and two "boundary control points" (or called "bd-control-points", blue dots) inserted. (a) Bottom: Boundary restricted control grid in a 2D layout. (b) All possible bd-control-points around one central point.

Table 4.1: Refining $N_R$ by inserting $k$ into knot vector $[r_0, r_1, r_2, r_3, r_4]$ generates two basis functions $N_{R_1}$ and $N_{R_2}$.

| $k$ | $R_1$ | $R_2$ |
|---|---|---|
| $r_0 \leq k < r_1$ | $[r_0, k, r_1, r_2, r_3]$ | $[k, r_1, r_2, r_3, r_4]$ |
| $r_1 \leq k < r_2$ | $[r_0, r_1, k, r_2, r_3]$ | $[r_1, k, r_2, r_3, r_4]$ |
| $r_2 \leq k < r_3$ | $[r_0, r_1, r_2, k, r_3]$ | $[r_1, r_2, k, r_3, r_4]$ |
| $r_3 \leq k \leq r_4$ | $[r_0, r_1, r_2, r_3, k]$ | $[r_1, r_2, r_3, k, r_4]$ |

## 4.3.1 Boundary Restricted Control Grid

In order to construct a control grid, we first divide the cuboid block into cells by grid coordinates. The grid coordinates along $k$-axis are denoted as:

$$S_k = [s_1^k, s_2^k, \ldots, s_{n_k}^k], k = 1, 2, 3,$$

where $n_k$ is the resolution of rectilinear grid along $k$-axis and each value in $S_k$ is the normal subdivision of cuboid parameter along $k$-axis. The tensor product of $S_1, S_2, S_3$ divides the block into $(n_1 - 1) \times (n_2 - 1) \times (n_3 - 1)$ cells and gives rise to a point-based spline on $n_1 \times n_2 \times n_3$ control points.

However, this naive spline construction is open boundary and violates the requirement of boundary restriction. To improve, we replicate the non-uniform knots at both ends of $S_k$ to restrict the blending functions within the domain (See Fig. 4.6(a)Top):

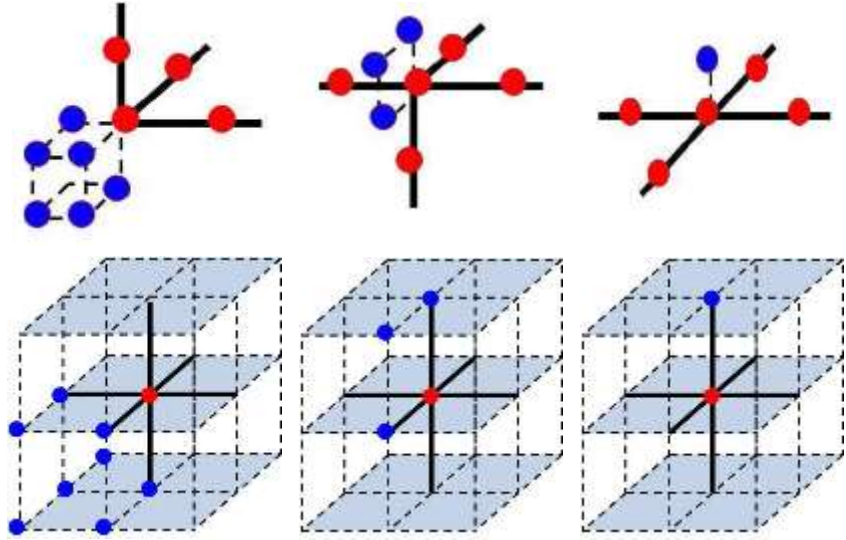

Figure 4.7: The bd-control-point distributions around a central point on the corner/edge/face vertex, respectively.

We add 3 extra knots, called *boundary knots (bd-knots)*, at the end of domain to restrict the boundary. The knot set is expanded:

$$S_k = [s_1^k, s_1^k, s_1^k, s_1^k, s_2^k, \ldots, s_{n_k}^k, s_{n_k}^k, s_{n_k}^k, s_{n_k}^k].$$

We also add 1 extra *boundary control point (bd-control-point)* (blue dots), on the bd-knot outside of the last control point on the boundary. Fig. 4.6(a)Bottom extends it to a 2D domain, and its extension to the 3D domain is in the same pattern. Our spline definition achieves: (1) Every blending function in each domain is confined within the domain boundary; (2) Only bd-control-points' blending functions influence the cuboid boundary, so our following fitting method can rely on this usable property.

In order to represent the bd-control-points conveniently, we can arrange them into a $3 \times 3 \times 3$ grid around the central point as Fig. 4.6(b) (Recall that the central point is the control point with an unconfined blending function). These 27 possible knots share the same parameters as the central point. It is only designed to explicitly record topological relations of these control points in preparation for efficient spline merging. After adding bd-control-points to the 3D control grid, each central point on the corner/edge/face has 8/4/2 control points, respectively (Fig. 4.7). This special bd-control-point representation is uniquely suitable for merging processing as shown in Section 4.4.

### 4.3.2 Meshless Fitting

Our input only includes a control grid and a group of surface sample points extracted from the surface patch (already mapped to a cuboid domain surface). The challenge consists in designing a fitting method for solids without interior volumetric parameterization or remeshing.

1. **Boundary fitting.** We first determine the positions of bd-control-points only. Recall that only bd-control-points $p_i^b$ influence the cuboid surface sample points. Therefore, we can determine their positions by minimizing Eq. 4.4 w.r.t. to surface sample point $v_j^b$:

$$\mathrm{argmin}(\sum_{j=1} ||\mathbf{F}(\mathbf{f}^{-1}(v_j^b)) - v_j^b||) \tag{4.4}$$

$$\Rightarrow \frac{\partial}{\partial \mathbf{p}_r^b} \sum_{j=1} (\mathbf{F}(\mathbf{f}^{-1}(v_j^b)) - v_j^b)^2$$

where $\mathbf{F}$ denotes the spline function as Eq. 4.1 and $\mathbf{f}^{-1}(v_j^b)$ the parameters of $v_j^b$ in the cuboid. The above equation can be rewritten in matrix format as in the least square method:

$$\frac{1}{2}\mathbf{P}^T\mathbf{B}^T\mathbf{B}\mathbf{P} - \mathbf{V}^T\mathbf{B}\mathbf{P} = 0, \tag{4.5}$$

where B is the matrix of blending functions $\mathbf{B}_{ij} = \mathbf{I}_{3\times3}\mathbf{B}_i(\mathbf{f}^{-1}(v_j^b))$, V and P denote the vectors of surface sample points $v_j^b$ and bd-control-points $p_i^b$, respectively. This equation determines bd-control-points and they serve as the constraint in the next interior fitting step.

2. **Interior fitting.** Let u in the set U be the interior parametric value. Each $\mathbf{u}_i = (u, v, w)$ is the interior parameter triplet in the tensor-product parametric grid $(u_0, u_1, \ldots, u_{n_0}) \times (v_0, v_1, \ldots, v_{n_1}) \times (w_0, w_1, \ldots, w_{n_1})$. Theoretically, we have the following harmonic equation w.r.t. interior control points $p_j^{in}$:

$$\mathrm{argmin}(\sum_{i=1}\int_{\Omega_i} ||\nabla \cdot \nabla \mathbf{F}(\mathbf{u}_i)||d\mathbf{u}) \tag{4.6}$$

$$\Rightarrow \frac{\partial}{\partial \mathbf{p}_j^{in}} \sum_{i=1}\int_{\Omega_i} (4\mathbf{F}(\mathbf{u}_i))^2 d\mathbf{u} = 0,$$

where $\Omega_i$ is an infinitesimal parametric volume around $\mathbf{u}_i$. Similar as [74], the above minimized energy $\int_{\Omega_i} \|\triangle F(\mathbf{u}_i)\|$ can be approximated by the following formulation:

$$\sum_{j=0}^{?} w_{ij} F(\mathbf{u}_j) = 0, w_{ij} = \begin{cases} 1 & i = j, \\ -\frac{1}{6} & \mathbf{u}_j \in \mathbf{Nbr}(\mathbf{u}_i) \\ 0 & \text{others} \end{cases} \qquad (4.7)$$

where $\mathbf{Nbr}$ includes 6 immediate neighbors of $\mathbf{u}_i$ in the tensor-product parametric grid. We substitute Eq. 4.7 into Eq. 4.6, which can be solved by the least square method similar to Eq. 4.4. During computing we set already-known $p_i^b$ as constraints and get all other control point positions.

**Global alignment.** Although we execute volumetric fitting separately on every cuboid, our fitting technique still guarantees global alignment of interior fitting results. Recall that we already obtain the identical surface parameters between cuboids before fitting, since we generate aligned poly-edges (i.e., cuboid edges). Therefore, two cuboids minimize precisely the same energy in Eq. 4.4 and Eq. 4.6 on the boundary, leading to the equivalent fitting results.

### 4.3.3 Cell Subdivision and Local Refinement

If the fitting results do not meet certain criteria on each cuboid, we can always perform subdivision over cells in the control grid with large fitting errors and then conduct the volumetric fitting. Each cell is split along 3-axis and divided into eight sub-cells naturally.

The challenge is how to preserve the semi-standardness during subdivision. Sederberg et al. [27] have proposed a feasible approach to refine blending functions on surface patch. We generalize this technique onto our 3D control grid. Let $R = [r_0, r_1, r_2, r_3, r_4]$ be a ray-tracing knot vector and $N_R(\mathbf{u})$ denotes the corresponding cubic B-spline basis function. If there is an additional knot $k \in [r_0, r_4]$ inserted into $R$, $N$ can be written as a linear combination of two B-spline functions:

$$N_R(\mathbf{u}) = c_1 N_{R_1}(\mathbf{u}) + c_2 N_{R_2}(\mathbf{u}). \qquad (4.8)$$

Two knot vectors $R_1, R_2$ are shown in Table 4.1, $c_1$ and $c_2$ are 2 weights that can not exceed 1:

$$c_1 = \min(\frac{k - r_0}{r_3 - r_0}, 1), c_2 = \min(\frac{r_4 - k}{r_4 - r_1}, 1).$$

Since the blending function of $B$ is the tensor product of $N$ along 3-axis, we can also formulate the refined blending functions along one axis:

$$B_i \equiv c_1 B_{i1} + c_2 B_{i2}. \tag{4.9}$$

The procedure of our 3D subdivision and local refinement consists of following steps. The input is a queue of cell $Q_c$.

1. Subdivide cells in $Q_c$ and insert the new vertices into the domain $T$, and update $T$ to $T^*$

2. For all pairs of blending functions $< w_i B_i >$, $w_i \in W, B_i \in B$, compute its new knot vector $R^*$ (See Section 4.2). Then,

   - If the $R^*$ includes the knot which does not exist in $T^*$, insert a new vertex on that knot into the domain $T^*$.

   - If the $R^*$ is more refined than $R$, compute the refinement $B_i = c_1 \times B_{i1} + c_2 \times B_{i2}$. Insert the new blending functions $< w_i \times c_1 B_{i1} >$ and $< w_i \times c_2 B_{i2} >$ into the control grid. Delete the old pair $< w_i B_i >$.

3. Repeat the last step until no new knot vector in $R^*$. Collect all blending functions on the same control point and use the total weight as its new weight.

The above procedure can handle refinement and knot extraction on a complicated 3D control grid. It also determines new required control points automatically to guarantee the semi-standardness. Note that unlike [27], we perform spline fitting again after each refinement iteration to update control point positions. This is mainly because our goal of refinement is to seek for more accurate fitting result. In contrast, the refinement in [27] aims to keep the shape unchanged.

### 4.3.4 Boundary Modification

Boundary modification is necessary for our semi-standard T-spline component, because of the fundamental difference between standard B-spline and our merging strategies. Fig. 4.12 intends to visually show the difference between them. It illustrates the 1D merging method introduced in [26] on our boundary restricted grid. For a $C^2$ merging, 3 control points on one component will be merged with 3 control points on the other component to form a joint new spline. However, the procedure does not take the associated weights into consideration. In standard B-spline, all the weights are uniform. However, in semi-standard T-spline, it is possible that two corresponding soon-to-be-joined control points have different weights. As a result, the semi-standardness around the merged regions will break down. Therefore, we have to add extra requirement about weights to make these control points be capable of merging.

**To-be-merged control point (Definition 1).** *For a control point, if its blending function includes bd-knots around merging boundary, we say this control point is a "To-be-merged" control point (For example, Fig. 4.11(a-b) in a 2D layout).*
**Modification zone (Definition 2).** *For any cell in the control grid, if one of its 8 vertices is "To-be-merged" control point for one boundary, we say this cell is in the "Modification zone".*

A merging-ready spline must have the following properties:
**Boundary requirement (Proposition 1).** *The weights of all "To-be-merged" control points on this boundary must equal to one, such that we can merge two splines and the resulting spline still preserves semi-standardness.*

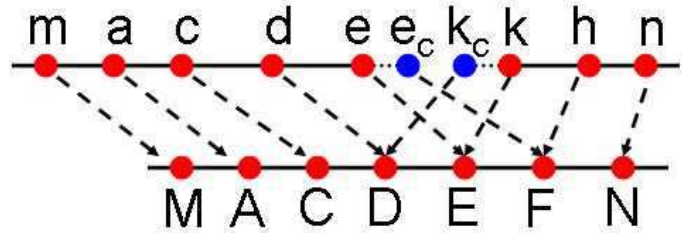

Figure 4.8: Proof of Observation 1.

**Proof:** As shown in Figure 4.8, two 1D splines are merged. Because of the symmetry, we only analyze the left side of merging boundary (line segment [M, E]). During merging, only control points D, E, F update blending functions and weights. The influenced region thus narrows down to the line segment [A, E].

1) Line segment $[A, C]$: The semi-standardness preserves before merging, thus $w_m B_m + w_a B_a + w_c B_c + w_d B_d \equiv 1$ on line segment $[a, c]$, $B_i$ is the blending function at i. After merging, $w_m = w_M$, $w_a = w_A$, $w_c = w_C$, $w_d = w_D = 1$, $B_m = B_M$, $B_a = B_A$, $B_c = B_C$, thus we only need to prove that $B_d = B_D$ between $[A, C]$. The knot vector of $B_d$ is $[a, c, d, e, e]$. The knot vector of $B_D$ is $[a, c, d, e, \mathbf{f}]$. According to Eq. 8,

$$B_D = B_d + \frac{\mathbf{f} - e}{\mathbf{f} - c} B^\lozenge$$

The knot vector of $B^\lozenge$ is $[c, d, e, e, \mathbf{f}]$, which does not influence the line segment $[A, C]$. Therefore, we get $B_D = B_d$ on $[A, C]$.

2) Line segment $[C, E]$: Similar to $[A, E]$, we only need to prove $B_d + B_e + B_{e_c} = B_D + B_E + B_F$. Our subdivision procedure under "boundary requirement" generates a local trivariate B-spline on line segment $[D, F]$ along this direction. According to B-spline merging, $B_d + B_e + B_{e_c} = B_D + B_E + B_F$ on $[C, E]$.

To guarantee that all "To-be-merged" control points' weights equal to one, we need to be able to recognize if a local subdivision breaks the above rule or not:

**Proposition 2.** *Any knot insertion outside of "Modification zone" does not violate boundary requirement (i.e., weights of bd-control-points equal to one).*

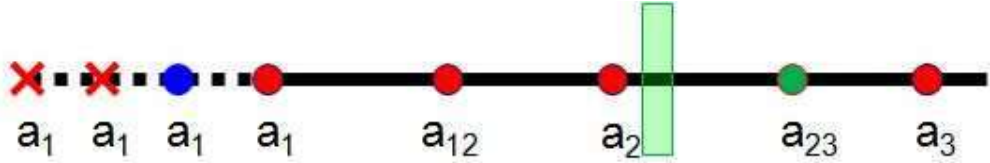

Figure 4.9: Proof of Proposition 2. "Modification zone" is the left of "Green Bar". Three nodes on $a_1$, $a_{12}$ represent "to-be-merged" control points.

**Proof:** The point on $a_{23}$ is the nearest newly-inserted control point outside "Modification Zone". According to Table 1, none of refined blending functions takes $a_1$ as the center of knot vector. Thus the weights of two "to-be-merged" control points on $a_1$ are unchanged. For "to-be-merged" control point $a_{12}$, its refined blending function is only subdivided from the original blending function $B_{a_{12}}$, located at $a_{12}$. According to Eq. 8, $B_{a_{12}} = c_1 \times B^{\parallel}_{a_{12}} + c_2 \times B_{a_2}$. The new weight of $a_{12}$ is $c_1 = \min(\frac{a_{23} - a_1}{a_2 - a_1}, 1) = 1$. Therefore, the new weight on $a_{12}$ still equals to one.

After detecting the potential violation, we can properly handle it using the following proposition:

**Proposition 3.** *If we subdivide all boundary cells around merging region at the same time, the new "To-be-merged" control points still guarantee "Boundary requirement" and their wights all equal to one.*

**Proof:** After subdivision, each blending function is subdivided to several sub-blending functions pairs $< w_i B_i >$. These pairs are distributed to other knots: For example, subdivision of blending function located at D generates new pairs on C, M, D, N, E and the weights on each node can be computed by Eq. 8:

$$D = \frac{(a_{12} - a_1)(a_4 - a_{23})(a_4 - a_{34})}{(a_4 - a_1)(a_4 - a_{12})(a_4 - a_2)}$$

$$+ \frac{(a_{23} - a_{12})(a_4 - a_{34})}{(a_4 - a_{12})(a_4 - a_2)} + \frac{(a_5 - a_{23})(a_{34} - a_2)}{(a_5 - a_2)(a_4 - a_2)}$$

$$M = \frac{(a_{12} - a_1)(a_4 - a_{23})}{(a_4 - a_1)(a_4 - a_{12})} + \frac{a_{23} - a_{12}}{a_4 - a_{12}}, N = \frac{a_5 - a_{34}}{a_5 - a_2}$$

$$C = \frac{(a_{12} - a_1)(a_{23} - a_1)}{(a_4 - a_1)(a_3 - a_1)}, E = \frac{(a_5 - a_{34})(a_5 - a_{45})}{(a_5 - a_2)(a_5 - a_3)}$$

The weight of refined blending function is the summation of subdivided weights.

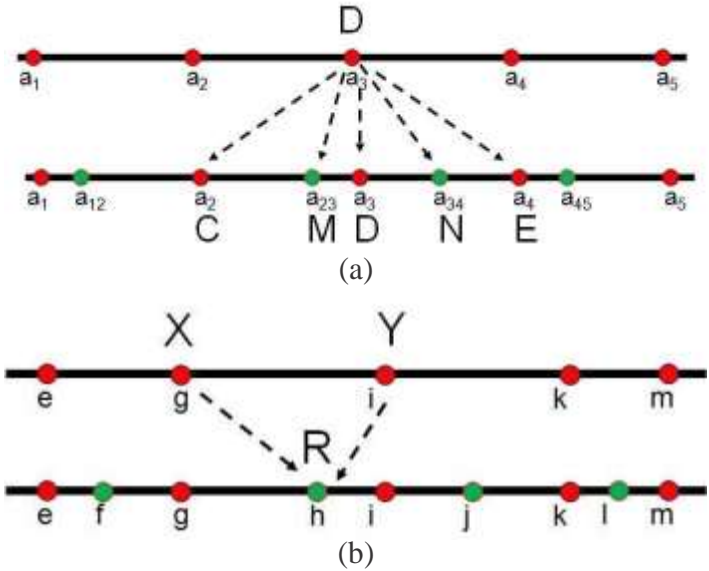

Figure 4.10: Proof of Proposition 3. (a) blending function on D subdivides and generates new blending functions at C, M, D, N, E. (b) updated blending functions on R only result from subdivision of X, Y.

Considering R as an example, the refined blending function on R is only derived from unrefined blending functions on X and Y. According to the above equations, we can compute the weights from X and Y:

$$X \Rightarrow R : \frac{k-h}{k-e}, Y \Rightarrow R : \frac{(f-e)(k-h)}{(k-e)(k-f)} + \frac{h-f}{k-f}.$$

The summation of weights on R is

$$\frac{f-e}{k-e} \times \frac{k-h}{k-f} + \frac{h-f}{k-f} + \frac{k-h}{k-e} \equiv 1.$$

Based on the above propositions, we propose our modification procedure as follows. The input is the newly refined control grid with new subdivided cell set $C_{new}$.

1. For each boundary, assign the cell set $C_T$ as "Modification zone". For any cell with one vertex as a "To-be-merged" control point, we add this cell into $C_T$.

2. For each boundary, detect if there is any new subdivided cell in the "Modification zone":

- $\mathbf{C}_{\text{new}} \overset{\mathbf{T}}{\cap} \mathbf{C}_T = \varnothing$. According to Proposition 2, the refined grid preserves the standardness on the boundary, so no further processing.
- $\mathbf{C}_{\text{new}} \overset{\mathbf{T}}{\cap} \mathbf{C}_T = \varnothing$. Modify the boundary according to Proposition 3: Subdivide all cells on the boundary to satisfy "Boundary requirement".

3. Update control point positions. Instead of fitting again like in Section 4.3.3, we use the same method as in [27] because we seek for keeping spline shape unchanged in this step.

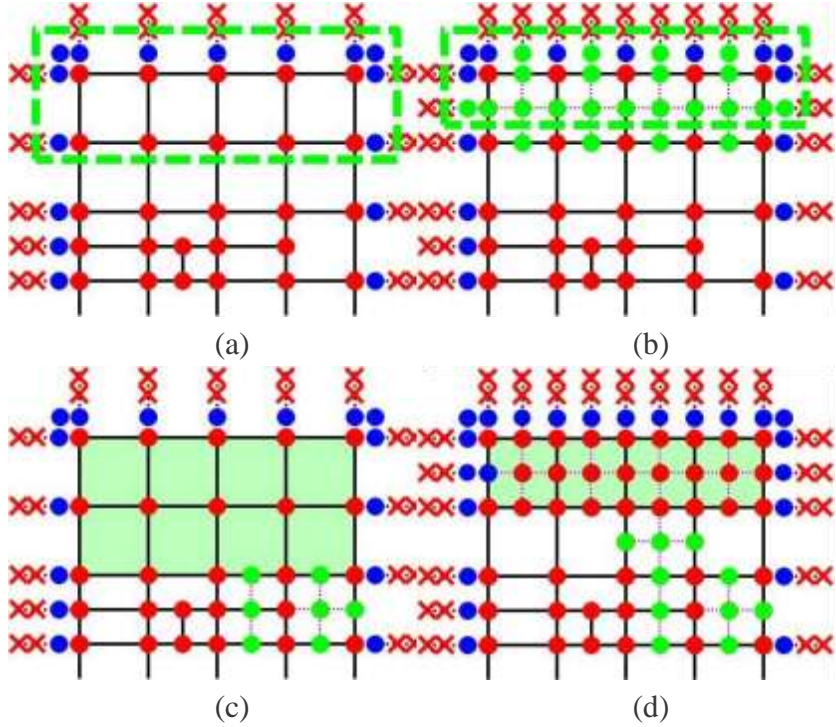

Figure 4.11: Boundary modification. (a) Original "To-be-merged" control points (in the green box). (b) Subdivision all cells along the boundary, according to Proposition 3. The green box covers updated "To-be-merged" control points. (c) and (d) "Modification zone" (green box) of (a) and (b). According to Proposition 2, cell subdivision (by green dots) outside "Modification zone" does not violate "Boundary requirement" (Proposition 1).

## 4.4 Global Merging Strategies

In our framework, the decomposed components can be merged in various different merging types. We develop algorithms to handle different types of merging in this section. As we discussed in Section 4.2.1, our domain only includes "Two-cube" merging (Section 4.4.1) and "Type-1" merging (Section 4.4.2). Also, we seek to handle more complicated conventional poly-cube domains, including all other types of merging in Fig. 4.2 (Section 4.4.3).

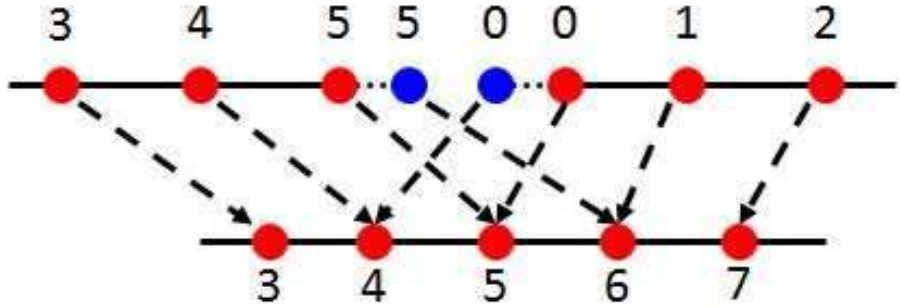

Figure 4.12: Two cube merging in 1D layout. Two control points are combined to form one new control point (4, 5 and 6).

### 4.4.1 "Two-cube" Merging

Merging of 3D components can be simply illustrated by 1D merging. In 1D merging, each boundary parameter corresponds to a new position after merging. For example, in Fig. 4.12, the bd-control-point with parameter 5 corresponds to a new parameter 6. The control point corresponding to $n(n \geq 2)$ original control points simply takes the average position as its new position. Similarly, the merging of two cuboids includes the following steps.

1. Boundary modification. If bd-knot intervals of two components are different, subdivide the cube boundary using the procedure in Section 4.3.4 iteratively until they share the same knot interval (Fig. 4.13(a)).

2. Merging control points. Correspond the original control point to the new control grid. As shown in Fig. 4.13(a)Right, we merge each column along the merging direction as 1D case.

3. Computing control point positions. Each new control point $p^\emptyset$ corresponds to $n(n \leq 2)$ original control points $p_i$. The new control point position is computed by $p^\emptyset = \frac{\sum_i^n p_i}{n}$.

### 4.4.2 "Type-1" Merging

The goal of this merging type is to merge 3 cuboids into one control grid, like Fig. 4.2(a). We can still use the "Two-cube" merging technique to treat most merg-ing regions. But we have to design special confinement method to handle the central points on the yellow dot/lines. Fig. 4.14(b) shows the extra bd-control-points we add around the central point on the yellow lines. For the yellow dot, we add additional 8 bd-control-points around it to confine it into the surface boundary, as shown in Fig. 4.14(c).

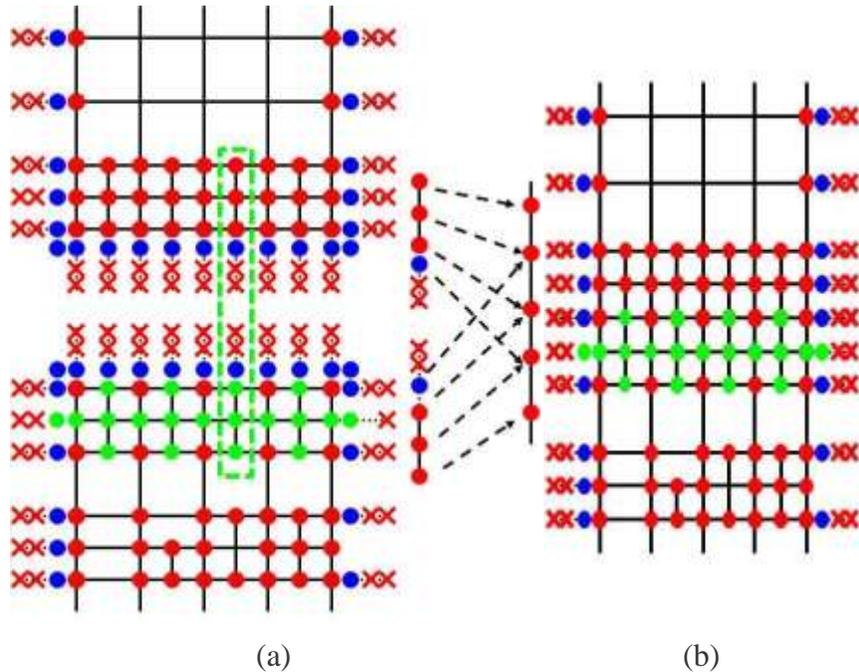

(a)                                (b)

Figure 4.13: "Two-cube" merging. (a) Left: Subdivide the bottom cuboid and insert new control points (green dots) to keep the same knot intervals. Right: Merging along merging direction. (b) The merged control grid.

Fig. 4.15 illustrates the confinement effect Fig. 4.15(a) shows a confined 2D control grid in 2D layout. The extra bd-control-points (blue dots) are inserted around the central point. Fig. 4.15(b-d) showcase its advantage: unlike Fig. 4.5, for any chosen parametric position, none of its control points penetrates the boundary to influence the chosen position.

**Preserving semi-standardness.** Now we still have another challenge. Simply adding these extra control points would violate the semi-standardness property. To preserve semi-standardness, we also modify weights in this newly-merged control grid structure. The weight can be computed as follows (See Fig. 4.14(a)): (1) Before adding bd-knots around the central point, we add an auxiliary control point (green dot) at the corner. Now we locally have a standard rectangular control grid with weights all equal to one initially; (2) Insert the designed bd-knots (blue knots and red crosses in Fig. 4.14(a)) to the grid; (3) Inserting knots triggers the local refinement procedure to recompute the weight of each control points. Note that after refinement, the auxiliary point does not affect inside boundary anymore. Therefore, it is "transparent" and free to be deleted from the spline representation.

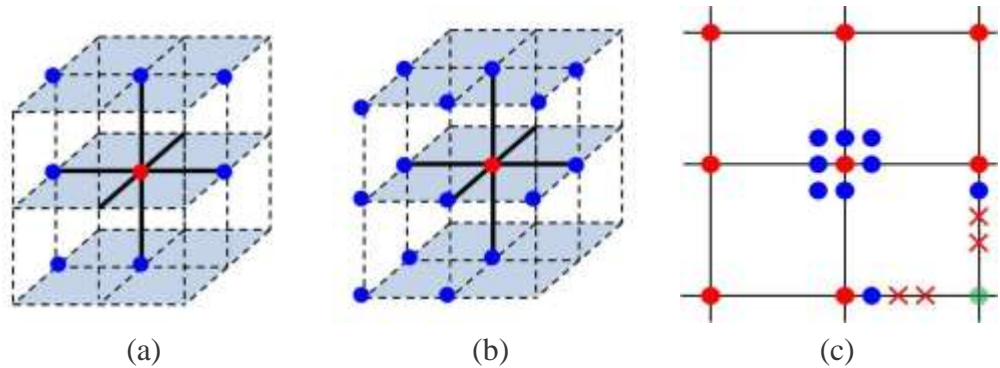

Figure 4.14: "Type-1" merging: (a-b) The 3-D distribution of bd-control-points around the central point on the yellow lines/dot in Fig. 4.2(a). (c) To preserve semi-standardness, bd-knots (blue dots and red crosses) and an auxiliary knot (green) are added. Then we can use local refinement algorithm to compute new control points' weights.

Besides preserving semi-standardness, our weight modification technique also has advantage for pre-computation. The weight computation only depends on the initial knot interval of merged control grid. Thus, we can pre-compute this step and build a look-up table for speedup. Table 4.2(a) shows the indices of control points around the central point (the same as indices in Fig. 4.6(b)). Table 4.2(b) shows the corresponding weights for all control points in Fig. 4.14(a) (Numbers in parentheses correspond to additional control points in Fig. 4.14(b)).

To summarize, "Type-1" merging includes the following steps. The first 3 steps are the same as "Two-cube" merging.

**Step 1** Modify boundary;

**Step 2** Merge control points;

**Step 3** Compute control point positions;

**Step 4** Insert extra bd-control-points as shown in Fig. 4.14(a-b) (We assign the position of the control point on the central point to these new inserted control points);

**Step 5** Modify weight (Change the weight of these bd-control-points by checking the look-up table, as shown in Table 4.2(a)).

### 4.4.3 "Type-2,3,4" Merging

The above two merging algorithms (Sections 4.4.1, 4.4.2) are already functionally sound when handling the merging of all components in our divide-and-conquer framework, because these are the only two merging types in our T-shape based poly-cube. Not just limited to that, Our ambitious goal is to handle any shape of poly-cube domains. Therefore, we offer several more powerful merging operations, which are designed to merge the components like "Type-2,3,4" in Fig. 4.2(b-d). Once again, in order to enforce the boundary restriction, we need to insert extra bd-control-points. For the central points on all yellow line in Fig. 4.2(b-d), they are just "Type-1" merging, so we use the same merging method as as shown Fig. 4.14(a). For the central points on 3 yellow dots, we design the extra bd-control-points, as shown in Fig. 4.22, to preserve boundary restriction.

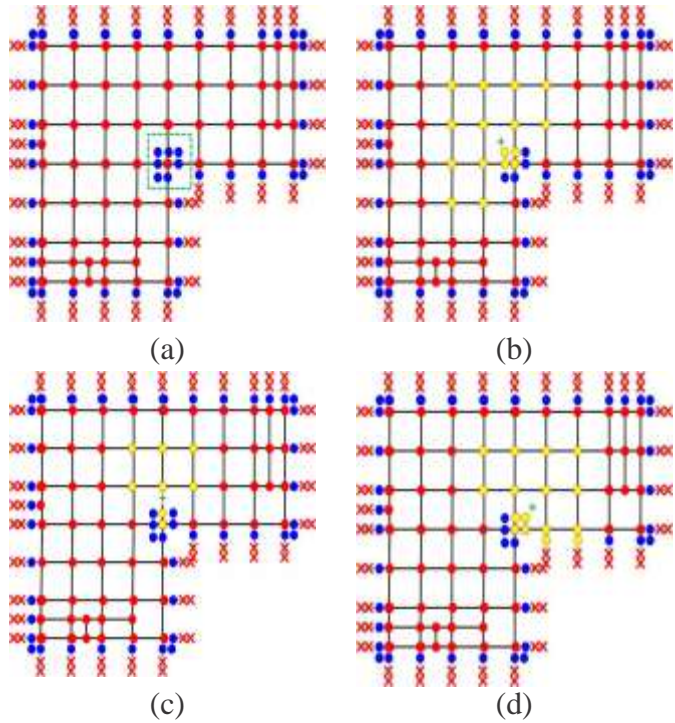

(a)            (b)

(c)            (d)

Figure 4.15: Confinement effect of "Type-1" merging. (a) The 2D layout of a refined control grid, with added bd-control-points (blue dots) around the central point (green box). (b-d) For each parameters (green cross), we highlight all control points (yellow points) that influence this parameter. The violation like Fig. 4.5 is completely eliminated.

To guarantee semi-standardness, we recompute the weight using the same method in Section 4.4.2 as follows. First, we add auxiliary control points, expanding given control grid around the central point to a complete cube-like grid. Second, we insert the designed bd-control-points and perform local refinement to compute the new weight for each control point. Their look-up tables are shown in Table 4.2.

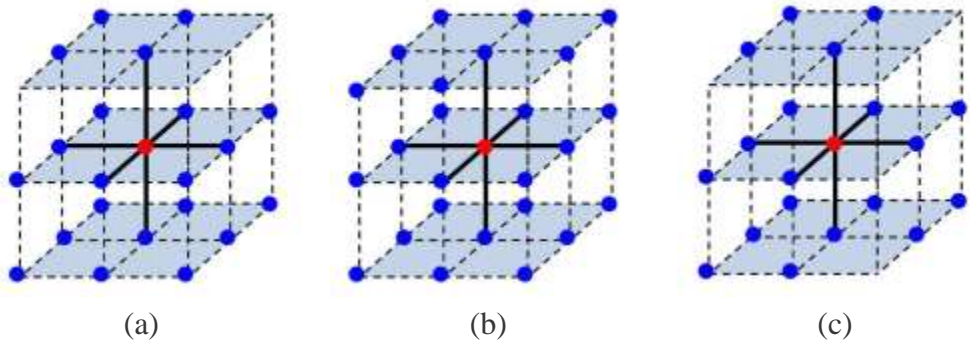

Figure 4.16: The 3D distribution of bd-control-points in "Type-2,3,4" merging. The central points are on the yellow dots in Fig. 4.2(b-d).

Table 4.2: Look-up tables. Row 1: an index table for 27 possible br-control-points in Fig. 4.6. Row 2: weights for "Type-1" merging in Fig. 4.15(b) (weights in parentheses correspond to additional 8 control points in Fig. 4.15(c)). Row (3-5): weights for "Type-2,3,4" merging.

Indices

| 7 | 8 | 9 | 16 | 17 | 18 | 25 | 26 | 27 |
|---|---|---|----|----|----|----|----|----|
| 4 | 5 | 6 | 13 | 14 | 15 | 22 | 23 | 24 |
| 1 | 2 | 3 | 10 | 11 | 12 | 19 | 20 | 21 |

Type-1

| - | - | - | - | - | - | - | - | - |
|---|---|---|---|---|---|---|---|---|
| 1 | 1 | - | $\frac{17}{18}$ | $\frac{35}{36}$ | 1 | $\frac{8}{9}$ | $\frac{17}{18}$ | 1 |
| (1) | (1) | - | $\left(\frac{17}{18}\right)$ | $\left(\frac{35}{36}\right)$ | (1) | $\left(\frac{8}{9}\right)$ | $\left(\frac{17}{18}\right)$ | (1) |

Type-2

| $\frac{26}{27}$ | $\frac{53}{54}$ | 1 | $\frac{53}{54}$ | $\frac{107}{108}$ | 1 | 1 | 1 | - |
|---|---|---|---|---|---|---|---|---|
| $\frac{53}{54}$ | $\frac{107}{108}$ | 1 | $\frac{107}{108}$ | $\frac{209}{216}$ | $\frac{17}{18}$ | 1 | 1 | - |
| 1 | 1 | 1 | 1 | $\frac{17}{18}$ | $\frac{8}{9}$ | - | - | - |

Type-3

| $\frac{20}{27}$ | $\frac{22}{27}$ | $\frac{8}{9}$ | $\frac{22}{27}$ | $\frac{95}{108}$ | $\frac{17}{18}$ | $\frac{8}{9}$ | $\frac{17}{18}$ | 1 |
|---|---|---|---|---|---|---|---|---|
| $\frac{22}{27}$ | $\frac{95}{108}$ | $\frac{17}{18}$ | $\frac{95}{108}$ | $\frac{25}{27}$ | $\frac{35}{36}$ | $\frac{17}{18}$ | $\frac{35}{36}$ | 1 |
| $\frac{8}{9}$ | $\frac{17}{18}$ | 1 | $\frac{17}{18}$ | $\frac{35}{36}$ | 1 | 1 | 1 | - |

Type-4

| $\frac{26}{27}$ | $\frac{53}{54}$ | 1 | $\frac{53}{54}$ | $\frac{107}{108}$ | 1 | 1 | 1 | - |
|---|---|---|---|---|---|---|---|---|
| $\frac{53}{54}$ | $\frac{107}{108}$ | 1 | $\frac{107}{108}$ | $\frac{215}{216}$ | 1 | 1 | 1 | - |
| 1 | 1 | - | 1 | 1 | 1 | - | - | - |

## 4.5 Implementation Issues and Experimental Results

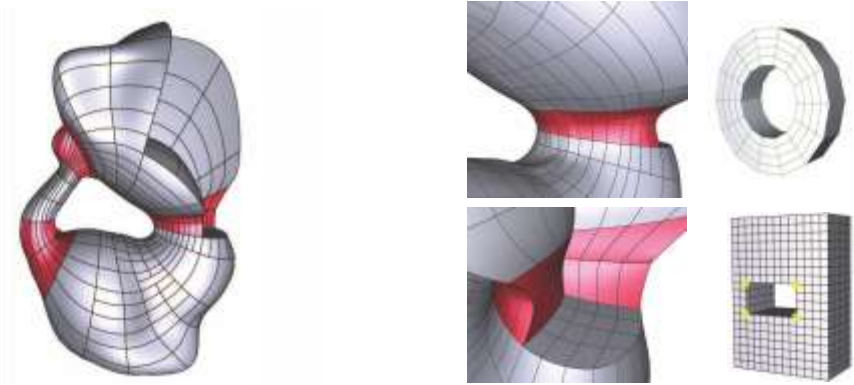

Figure 4.17: "Two-cube" merging for the kitten model.

Our experimental results are implemented on a 3 GHz Pentium-IV PC with 4 Giga RAM. Our first experimental results (Fig. 4.17, and Fig. 4.18) show the application of "Two-cube" merging by considering the kitten and beethoven model as the test datasets. These are the only merging types that exist in our component generation framework. For "Type-2,3,4" merging types that do not exist in our framework, we design a special screwdriver model and domain to demonstrate the power of "Type-2" merging (Fig. 4.19). In terms of poly-cube construction, we recognize that "Type-2" merging is very popular to handle the input with long branches. Yet, "Type-3,4" merging cases rarely exist even in the most conventional poly-cube domains. Geometrically speaking, they are more suitable to mimic highly concave shapes. We use the dark T-junction lines to show control grid knots and use different colors to represent different merging types. Red/Blue/Yellow marks all "to-be-merged" control point knots in 3 merging cases, respectively. We also have a close-up view to show the interior fitting result, demonstrating smoothness around the merging region. The yellow marks on the control grid highlight the ill-points.

In the second group of experimental results (Fig. 4.21, Fig 4.4, Fig. **??**, and Fig. 4.23), we integrate all merging types together to handle the models with high-genus and complex bifurcations, including the eight (genus 2), g3 (genus 3), rockarm, and wrench (genus 1 with bifurcations) model. We first display their component generation results. Then we show a spline model for one local component and the final spline results with a close-up view to highlight the interior fitting and merging regions. Fig. 4.20 also visualizes components' T-shape/poly-cube structures in a more efficient way. We use the same color cuboid to represent one component and the edges to show the cuboid connections. Each green box covers cuboids from the same T-shape. This structure clearly demonstrates that only "Two-cube" and "Type-1" merging are functionally sound in our framework.

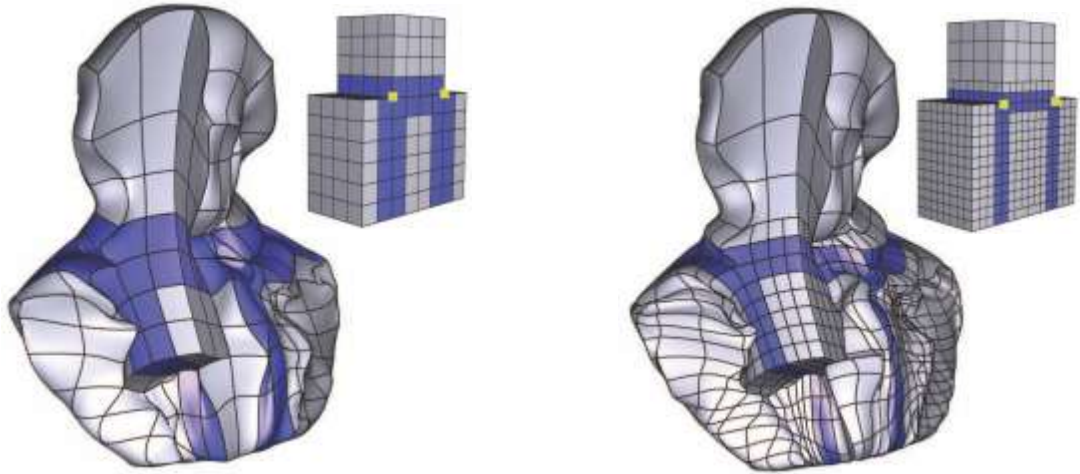

Figure 4.18: "Type-1" merging for the beethoven model.

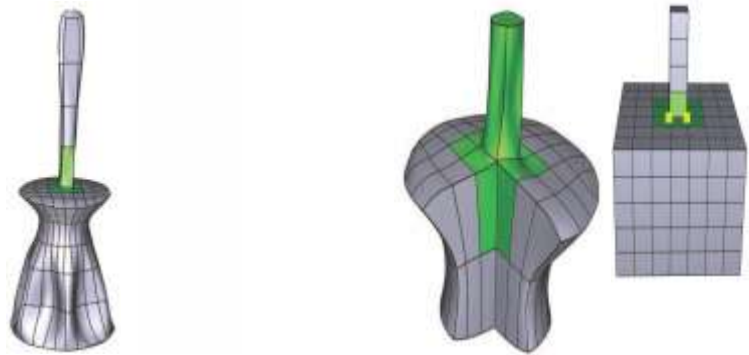

Figure 4.19: "Type-2" merging for the screwdriver model.

In Table 4.3, we document numbers of control points and fitting error. The T-spline scheme can significantly reduce the number of control points. The fitting results are measured by RMS errors which are normalized to the dimension of corresponding solid models. Meanwhile, we demonstrate the interior fitting quality in a close-up view of each model. Also, the table illustrates that adaptive refinement is necessary for trivariate splines, even on a simple surface input model. It is desirable to use high resolution with more DOFs to approximate boundary surface and low resolution with fewer DOFs for volume interior. For example, in the kitten/beethoven model, if we naively use B-spline scheme with hierarchical refinement inside the volume, their control points will increase to 3718/4850, respectively. In the last experiment (Figure 4.24), we apply our technique to convert the fertility model, with the noisy surface, into a trivariate spline and remesh it into a smooth

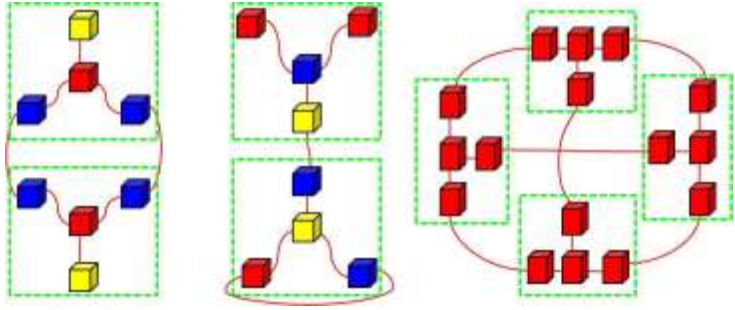

Figure 4.20: The divide-and-conquer structures of the rockarm/wrench/g3 model.

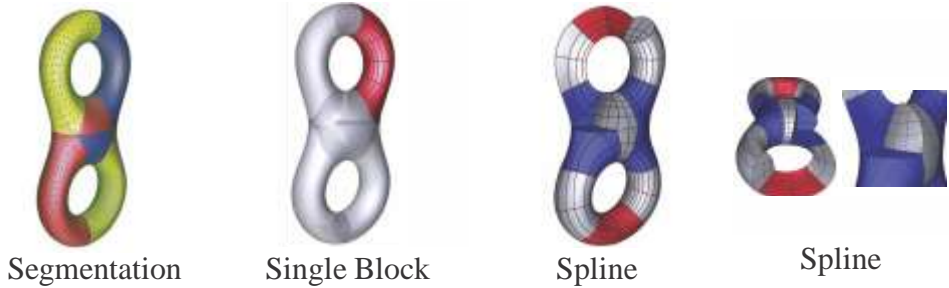

Segmentation     Single Block     Spline          Spline

Figure 4.21: The eight model.

result. The poly-edges (gray-lines) decompose the fertility model into components. Note that poly-edges are aligned everywhere so our local parameters are consistent globally.

**Top-down vs. Divide-and-conquer schemes.** In Table 4.4, we compare the performance between our divide-and-conquer framework with general T-splines using single integral domain in a traditional top-down approach. The most prestigious advantage of divide-and-conquer framework is to easily handle models with bifurcations/highly twisted shape/high-genus. For example, a poly-cube like Fig. 4.1 designed using a top-down scheme is very complicated, with 46 cuboids and they are connected in various types, to mimic the shape of the g3 model. The poly-cube construction also requires tedious manual design. By comparison, its divide-and-conquer domain (Fig. 4.20) includes only 16 cuboids with two certain merging types. Second, we also compare the required spatial consumption between our divide-and-conquer scheme with the top-down scheme. In general, our memory cost is reduced to $\frac{1}{n_s}$, where $n_s$ is the number of cuboids. Third, we compare the computation of $B^0$ between semi-standard T-spline and rational T-spline. We record the computation time on $10^4$ samples for each model. The result shows that our method is at least twice as fast as rational T-splines. This is because the computation avoids division operation completely (See the difference between Eq. 4.1 and 4.3).

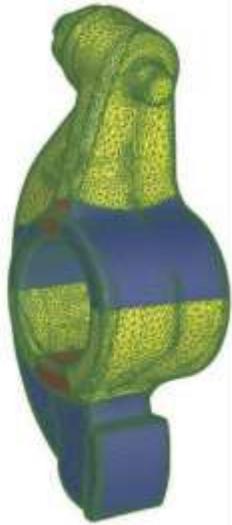

Segmentation

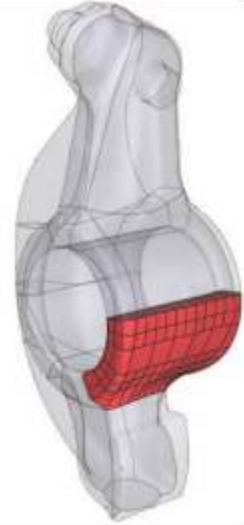

Single Block

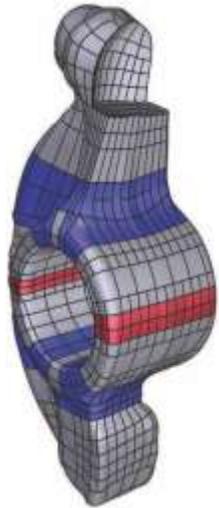

Spline

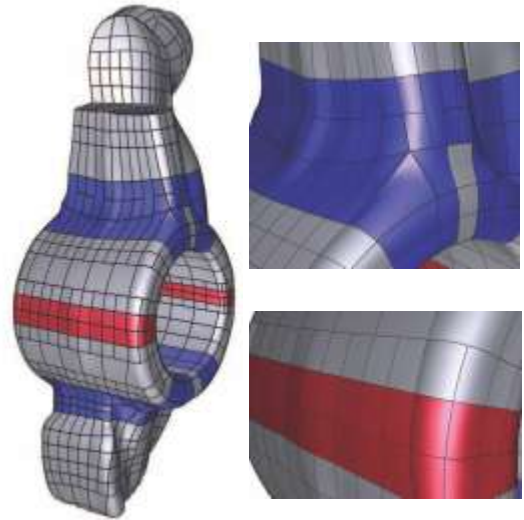

Spline

Figure 4.22: The rockarm model.

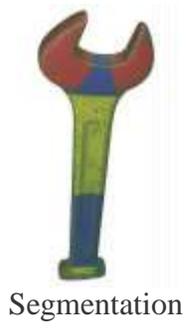 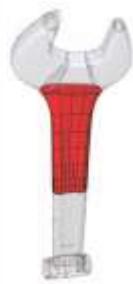 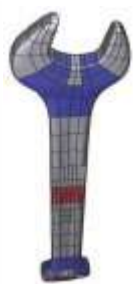 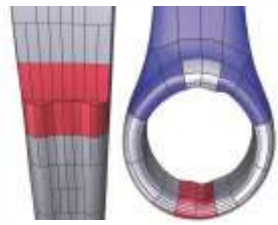

Segmentation        Single Block        Spline        Spline

Figure 4.23: The wrench model.

## 4.6 Chapter Summary

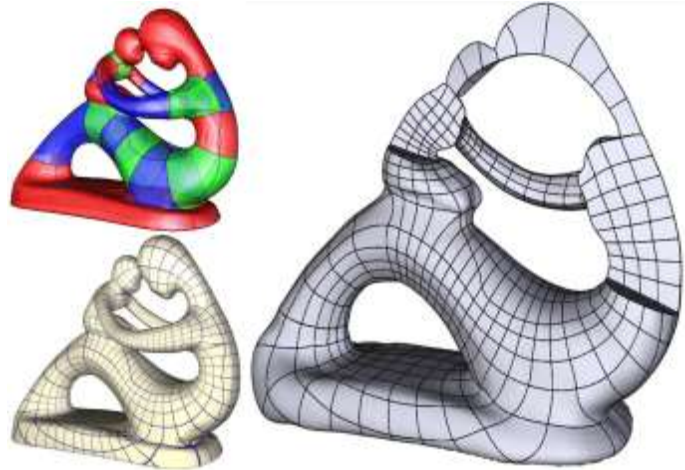

Figure 4.24: Mesh smoothing: We convert the fertility model to a trivariate spline and remesh it into a smooth result. Three figures show the components (with poly-edges), the globally aligned parameters, the remeshing result (with interior cut views), respectively.

Table 4.3: Statistics of various test examples: $N_c$, # of control points; RMS, root-mean-square fitting error ($10^{-3}$). "bv1", "bv2", "ra" and "sd" represent the beethoven (low and high resolution), rockarm, and screwdriver models.

| Model | $N_c$ | RMS | Model | $N_c$ | RMS |
|---|---|---|---|---|---|
| eight | 2058 | 1.63 | wrench | 3756 | 2.3 |
| kitten | 2840 | 3.32 | g3 | 2976 | 1.74 |
| bv1 | 1001 | 1.8 | bv2 | 3273 | 1.36 |
| ra | 4582 | 3.75 | sd | 1261 | 1.65 |

In this chapter, we have presented a novel framework to construct trivariate T-splines with arbitrary topology. Because of the flexible and versatile divide-and-conquer scheme, our framework can naturally handle solid objects with high genus and complex bifurcations. We decompose the input surface model into several part-aware components so that we can fit each component without the need of volumetric parameterization. The proposed spline scheme supports local refinement hierarchically, and the global trivariate T-splines satisfy the attractive properties of semi-standardness and boundary restriction. These novel contributions have a broad appeal to both theoreticians and engineers working in the shape modeling and its application areas.

Table 4.4: Comparison between our splines and general splines: Space required by fitting; Time to compute derivatives of basis functions; $N_c$, and Number of cuboids.

| Model | Our Method | | | General Method | | |
|---|---|---|---|---|---|---|
| | Space | $B^{\Downarrow}$ | $N_c$ | Space | $B^{\Downarrow}$ | $N_c$ |
| kitten | 116802 | 2.38s | 1 | 300688 | 4.53s | 8 |
| eight | 24714 | 2.25s | 6 | 174124 | 4.35s | 15 |
| g3 | 18952 | 2.17s | 16 | 314832 | 4.23s | 46 |

# Chapter 5

# Spline-based Volume Reconstruction

In Chapter 3 and 4, we have introduced the techniques to construct trivariate splines from surface 3D model. In this chapter, we adapt relevant trivariate splines to volumetric data reconstruction: we attempt to apply trivariate splines to represent 3D volume images.

## 5.1 Motivation

For volumetric scalar fields defined over a set of discrete samples, the reconstruction of the data is a fundamental problem with very significant applications. For instance in visualization, the size of volume data we have been dealing with increases dramatically to $1024^3$ voxels commonly or even larger. This trend of ever-increasing data size poses a great challenge in terms of both storage and rendering costs and thus requires reconstruction.

An ideal model would provide an accurate and efficient approximation for huge data sets, as well as the exact evaluation of function values and gradients which are required for high-quality visualization and physical simulation. An appropriate reconstruction involves following common quality requirements:

**Accuracy.** The reconstructed model should faithfully preserve the density function.

**Feature-alignment.** In regions with well-pronounced feature directions, parametric lines should guide and follow the shape feature.

**Compactness.** The number of patch layout as well as the degree of freedom for each patch should be as few as possible.

**Structured regularity.** Locally, each 3D patch is a subdivided cube-structured domain; Globally, the gluing between patches should avoid singularity.

**As-homogenous-as-possible.** The density distribution in one single patch should be narrowed in favor of approximation accuracy.

**Continuity.** A continuous representation supports high-order derivatives for high quality visualization.

An ideal reconstruction framework should optimize the output simultaneously with respect to all above criteria. However, existing techniques typically prefer offering a tradeoff between above conflicting requirements. The major reconstruction strategy is through multi-resolution data hierarchy to compress the data representation. Many algorithms have been developed to support hierarchical data reconstruction, including multi-dimensional trees [153] and octree-based hierarchies [154], [155]. However, these methods tend to produce an extreme large set of sub-blocks and require extra effort to pack them into a single structure, which undoubtedly violates the aforementioned compactness requirement. Moreover, the shape of produced block is limited as the axis-aligned texture/cube (i.e.,"flat block"). In contrast, an ideal candidate for feature-driven applications should utilize feature-aligned texture/cube. Other reconstruction methods seek to generate a continuous spline representation to approximate the data. In general, spline based reconstruction can be divided into non-regular and regular splines. Rossl et al. [43] have developed quadratic super splines to reconstruct and visualize non-discrete models from discrete samples. Finkbeiner et al. [156] have demonstrated that box splines deployed on body-centered cubic lattices in the input data are also feasible models for fast evaluation and GPU-acceleration. Tan et al. [42] have presented a reconstruction algorithm for medical images taking advantages of trivariate simplex splines. Meanwhile, compared to non-regular splines, many types of techniques (e.g., volume rendering [157]) and applications (e.g., iso-geometric analysis [41]) have a preference for regular-structured schemes. However, the major challenge lies at they rely heavily on spatial parameterizations and for arbitrary 3D objects such parameterizations become a rather non-trivial task. The goal of vectorization is to convert a raster object (2D or 3D image) into a vector graphics that is compact, scalable, editable and easy to animate, which is very similar to our research goal. In object-based vectorization [158], the whole image is segmented into a few objects. The color of each object then is approximated by spline patch. Recently, gradient meshes ([159]) serve as very powerful tools on 2D image representations and have been studied in depth. In a gradient mesh, position and pixel vary according to the specified gradients. However, it is not easy to directly update it to 3D volumetric image application because of its inefficiency of handling complex topology.

In order to achieve all above requirements, we propose a novel reconstruction approach that converts the discrete data to a small number of volumetric patch layouts. Each patch is a regular tensor-product cube grid while maintaining shape features. The voxels in every single patch have the almost homogenous density values in favor of accurate approximation for each patch.

In this chapter, we provide a novel framework to help a user to reconstruct a discrete volume data into regular patches and spline representations. Our representation has significant advantages: Each patch has regular structure while maintaining the shape features. The whole data is compactly represented by a very small number of patches. The density in each patch is as-homogenous-as-possible thus both the shape and density function can be accurately approximated by a high-order spline representation.

In order to achieve these advantages, our approach consists of the following major steps:

1. Starting with the computing of local tensors and principal curvatures, we generate an optimized frame field to respect the shape feature.

2. A regular structured parametrization of (u, v, w) is generated, whose gradients align the above field everywhere. Then we produce a set of volumetric patches based on the parametrization result.

3. We construct on each patch a trivariate T-spline to approximate the function F (u, v, w) using as-few-as-possible control points.

The remainder of this chapter is organized as follows. Section 5.2 is the frame field generation stage and Section 5.3 involves the volumetric parametrization and patch remeshing. We discuss the spline approximation, implementation details, and demonstrate experimental results in Section 5.4. We conclude in Section 5.5.

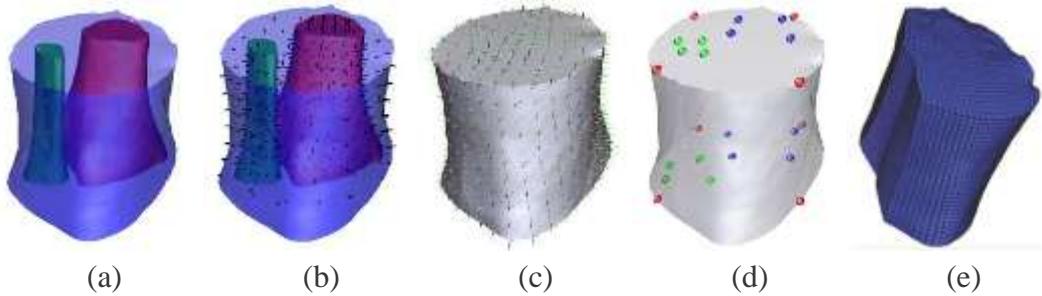

(a)       (b)       (c)       (d)       (e)

Figure 5.1: Main steps of the reconstruction. (a) Input model with material-aware boundary surfaces. (b) The tensor and principal direction field is computed on each voxel in the input data. The major principal directions on the boundary surfaces serve as the constraints for the next step. (c) In a frame field optimization procedure, an as-smooth-as-possible frame field is generated while maintaining the given constraints. (d) Corner points are selected to determine the domain structure. Additional constraints are added into next step of parametrization computing. (e) A volumetric parametrization.

## 5.2  Frame Field

In this section we mainly focus on frame field generation. We start from designing an operator of tensor to describe the local feature ( Section 5.2.1). Then we discuss the optimization of a 3-direction frame field in Section 5.2.2.

### 5.2.1  Tensor and Principal Curvature

Traditionally, curvature has be used as a shape feature descriptor widely. Theoretically this differential property characterizes only an infinitesimal neighborhood. Therefore it is desirable to design a numerically adaptable operator for the discrete data to compute this property. Although much work deals with this task on the surface (see [160] for an overview), we still need a new curvature operator for the discrete 3D hyper-volume data. Our operator captures statistically the shape of a neighborhood around a central point by fitting a continuous function, and thus mimics the 3D differential curvature and encodes anisotropy along 3 orthogonal directions. To summarize this shape, we use a cubic polynomial function $\mathbf{I}^H(u, v, w)$ to approximate the local density function, because they are the simplest form that can sufficiently express the shape variability we need to encode in a continuous manner.

Specifically, the given volumetric data set is represented using a uniform grid $G = (V, E, C)$, where $V = v_0, v_1, \ldots, v_n$ denotes the voxels and $E, C$ denote the set of edges and cubes in the grid, respectively. Each grid voxel $v_i = (x_i, y_i, z_i, \mathbf{I}_i^D)$ includes 4 components: geometric position in the grid $(x_i, y_i, z_i)$ and the discrete density value $\mathbf{I}_i^D$.

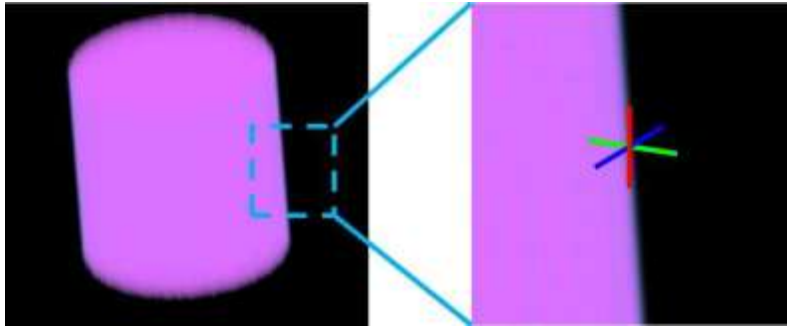

Figure 5.2: Left: The input local data around a voxel. Right: The approximated result and 3 principal directions.

In order to get a local polynomial function $\mathbf{I}^H(u, v, w)$ around center voxel $v_i$, we assign a local parameter value $(u_0, v_0, w_0)$ to $v_i$. For each of its adjacent k-ring neighbor voxels $v_j \in N(v_i)$, the local parameter is $(u_j, v_j, w_j) = (u_0 + x_j - x_i, v_0 + y_j - y_i, w_0 + z_j - z_i)$. Then our fitting cubic polynomial can be formulated as:

$$\mathbf{I}^H(u, v, w) = \sum_{i,j,k \geq 0}^{i+j+k \leq 3} c_m u^i v^j w^k = \mathbf{P}(u, v, w)C^T, \qquad (5.1)$$

where $C$ denotes the vector of unknown coefficients $c_m$. $\mathbf{P}$ is the vector of $u^i v^j w^k$. Similarly, we can also describe derivatives of $u, v, w$. For instance,

$$\mathbf{I}_u^H(u, v, w) = \sum_{i,j,k \geq 0}^{i+j+k \leq 3} c_m \times i \times u^{i-1} v^j w^k = \mathbf{P}_u(u, v, w)C^T, \qquad (5.2)$$

where $\mathbf{P}_u$ is the vector of $i \times u^{i-1} v^j w^k$ (we set $u^m = 0$ if $m < 0$). In the same way, we can also describe other derivatives $\mathbf{I}_v^H, \mathbf{I}_w^H, \mathbf{I}_{uu}^H, \mathbf{I}_{vv}^H, \mathbf{I}_{ww}^H, \mathbf{I}_{uv}^H, \mathbf{I}_{uw}^H, \mathbf{I}_{vw}^H$ by determining $\mathbf{P}_v, \mathbf{P}_w, \mathbf{P}_{uu}, \mathbf{P}_{vv}, \mathbf{P}_{ww}, \mathbf{P}_{uv}, \mathbf{P}_{uw}, \mathbf{P}_{vw}$.

In order to describe the currently unknown coefficients $C$, we construct a fitting equation:

$$QC^T = \mathbf{I}^D, \qquad (5.3)$$

where $Q$ is the fitting matrix. Each row $\mathbf{Q}_{j:}$ in the matrix depends on a voxel $\mathbf{Q}_{j:} = \mathbf{P}(u_j, v_j, w_j), j \in i N(i)$. $\mathbf{I}^D$ is the vector of discrete value $\mathbf{I}_j^D$ on each voxel. Because the size of unknown variables is very small, we can solve this linear least-square problem through multiplying the matrix $Q$ by its transpose:

$$C = (Q^T Q)^{-1} Q^T \mathbf{I}^D. \qquad (5.4)$$

We notice that $(Q^T Q)^{-1} Q^T$ is constant for every local function if we choose the same k for k-ring neighbors of each voxel.

**Tensor and Principal Curvature.** After the above calculations, we now can represent the tensor as the following matrix:

$$T = \begin{bmatrix} \mathbf{I}_{uu}^H & \mathbf{I}_{uv}^H & \mathbf{I}_{uw}^H \\ \mathbf{I}_{uv}^H & \mathbf{I}_{vv}^H & \mathbf{I}_{vw}^H \\ \mathbf{I}_{uw}^H & \mathbf{I}_{vw}^H & \mathbf{I}_{ww}^H \end{bmatrix}. \qquad (5.5)$$

This matrix is equal to the second fundamental form of our hyper-volume representation. Therefore, we can compute 3 eigenvectors of the local tensor matrix $T$ and thus get 3 directions. We use them to describe the feature on each voxel. Compared to the conventional texture-gradient based feature, our tensor feature has very obvious advantages: it produces 3 directions rather than one; all local 3 direction fields follow the shape anisotropy thus global fields are already almost smooth. As a result it simplifies the complexity and time consumption of the following op-timization step.

## 5.2.2 Field Smoothing

Although we can use initial principal directions to compute the parametrization without optimization, it will stuck in a local minimum. To overcome this problem, we propose an optimization method which respects only the most dominant directions. First, we extract iso-surfaces of interest and take them as constraints to respect the shape. Second, the frame field in each iso-surfaces is iteratively optimized.

**Iso-surface extraction.** It is natural to take feature on iso-surfaces as constraints, because the final parametrization result and patch must respect the shape of iso-surfaces. Moreover, each sub-space in an iso-surface always tends to be as-homogenous-as-possible, which is an ideal property for final shape and density approximation.

Frequently, input datasets contain multiple structures and iso-surfaces that need to be differentiated. However, if those features have the same density and gradient values, existing clustering methods are limited at effectively classifying those similar features accurately. Thus, we apply the texture-based classification method for the iso-surface extraction. In the first step, we simply remove the background voxels. It does not influence the information of the feature of interest while significantly decreasing the computational time and operation complexity. After the background elimination, sixteen statistical attributes (angular second moment, contrast, correlation, variance, inverse difference moment, individual entropy, sum average, sum variance, sum entropy, skewness, kurtosis, correlation information measurements, intensity, gradient and second order derivative) can be extracted following the feature equations defined in [161] and [162]. For the sake of fast computation and easy programming, we use k-mean clustering in the high- dimension parameter space to automatically detect various features. One or more features can be selected with respect to the user's requirement. The boundary of each cluster finally becomes one of our iso-surfaces.

The constraints are added towards voxels on iso-surfaces, automatically or manually. In practice, to efficiently describe the feature of iso-surfaces, we set only one of 3 principal directions as the constraint, one of which follows the normal direction of the iso-surface. As shown in the following sections, only-one-direction constraints are functionally sound to preserve the feature and have extra flexibility when handling smoothing and parametrization.

After this preprocessing step, the input is decomposed to an independent subspace $V_i$ bounded by an iso-surface $S_i$. The subspace may also cover several smaller subspaces with iso-surfaces

$$S^{sub} = \{S_0^{sub}, S_1^{sub}, \ldots, S_n^{sub}\}.$$

Each voxel in the subspace $V_i$ has 3 initial directions and each voxel on the iso-surfaces $S_i$ $S^{sub}$ has one direction as the constraint. The following smoothing step will modify the directions on each voxel while maintaining the constrained directions.

**Field smoothing.** The smoothness of a unit frame field can be measured as the integrated rotation differences between every two neighboring voxels. [163] have studied the energy of a 2D cross field and simplified it to a linear representation. In our 3D volume, the challenge lies at smoothing 3 vectors in separate directions while maintaining their orthogonality. Therefore, we take the local rotation matrix as the unknown variable. $F(v_i) = f_0, f_1, f_2$ is a frame with 3 orthogonal vector directions on each voxel $v_i$. We can also uniquely describe this frame by rotating from the origin fame to it. Each row of the rotation matrix $R(v_i)$ is a vector direction $R_{r:} = f_r, r = 0, 1, 2$. Now, the energy turns out to be the sum of all

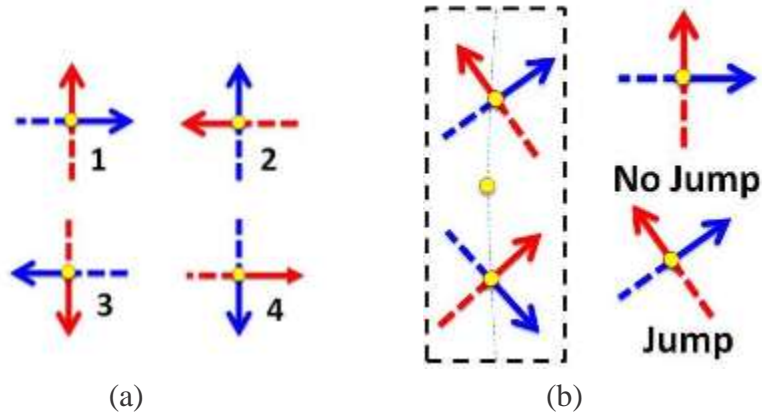

(a)                                        (b)

Figure 5.3: (a) Jump matching: 4 frames has different principal directions along red and blue arrows. The smooth energy between them should be zero ideally. (b): The smoothing results with/without considering period jump.

corresponding vector differences between adjacent voxels

$$E_{smooth} = \sum_{e_{ij} \in E} \sum_{r=0} ||R_{r:}(v_i) - R_{r:}(v_j)||^2. \qquad (5.6)$$

In order to solve unknown rotation matrix, we have to apply nonlinear solver (e.g.,Gaussian-Newton method) to minimize the energy function. Another difficulty is that Equation 5.6 predetermines the one-to-one mapping of 3 directions on two voxels, without considering "jump matching". "Jump matching" means all permutation cases of direction mapping. Fig. 5.3(a) shows all 4 "jump matching" cases for a 2-direction field. Similarly, we can have 24 "jump matching" cases for a 3-direction field. An ideal optimization algorithm should dynamically change direction mapping to get the best result. Fig. 5.3(b) shows a simple frame optimization on one voxel according to two adjacent voxels. Using jump matching we can get the perfect optimization result, while traditional method fails.

To overcome these problems, we design a novel optimization method. The key idea is that we compute the registration energy [164] between one voxel and its neighboring voxels. We extend 3 orthogonal principal directions into a length-normalized frame. Each frame gives 6 end positions $\{\mathsf{P}(\mathsf{v}_i)\} = \{\mathsf{p}_0, \dots, \mathsf{p}_5\}$ at the end of 3 frame lines.

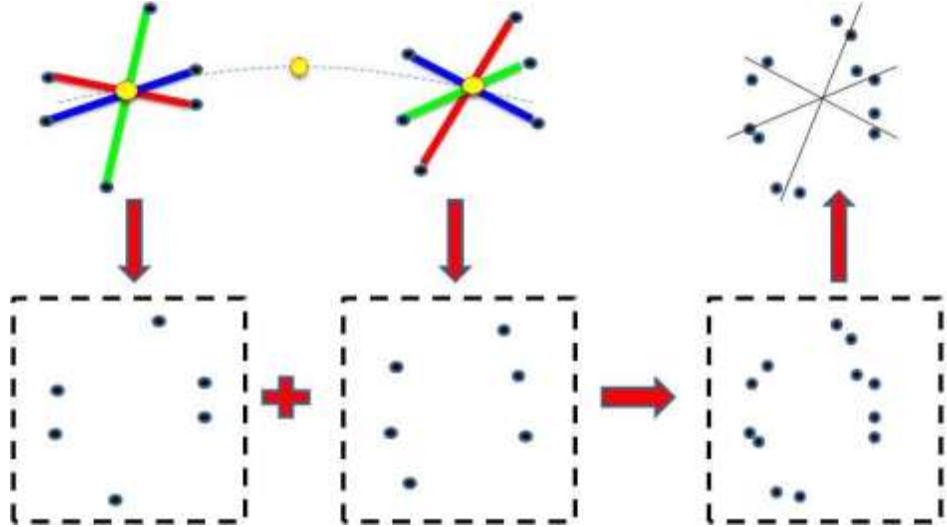

Figure 5.4: Major steps of optimization: (1) Union of ending points. (2) ICP-registration. (3) Compute rotation to get updated frame.

1. Get the union of all frame end positions on neighboring voxels: $\{\mathsf{S}_2\} = \bigcup_{\mathsf{v}_j \in \mathsf{N}(\mathsf{v}_i)} \mathsf{P}(\mathsf{v}_j)$.

2. The original point set $\{\mathsf{S}_1\} = \{\mathsf{P}(\mathsf{v}_i)\}$ is the frame ending positions of $\mathsf{v}_i$. Using the ICP-based registration [164], we compute a matrix T that approximately transforms voxels of $\{\mathsf{S}_1\}$ to those of the approximated set $\{\mathsf{S}_2\}$.

3. Decompose the transformation matrix T into a rotation matrix R and a shear matrix S using polar decomposition. Add the rotation R to the frame of $\mathsf{v}_i$.

For an iso-surface voxel $v_i$ which has a constrained direction, we first apply the above algorithm without considering constraint. Then, in the updated frame, we search for the closest direction and project it to the constrained direction by a rotation. The whole frame is also rotated as the final updated result.

The above algorithm is computed on each voxel iteratively until we get a promisingly smooth field. Starting from initially smooth tensor field will make optimization converge quickly. Our optimization algorithm avoids solving non-linear equations; Moreover, we utilize jump matching to get a much better result.

## 5.3 Volumetric Parametrization

The parametrization should be locally oriented to the frame field from Section 5.2. Therefore, the parametrization is computed as a solution to the following energy minimization problem:

$$E_{param} = \sum_{v_i \in V} ||\nabla u_i - u_i||^2 + ||\nabla v_i - v_i||^2 + ||\nabla w_i - w_i||^2, \quad (5.7)$$

where $u_i$, $v_i$, $w_i$ are the unknown parameters and $u_i$, $v_i$ and $w_i$ are 3 frame field directions on each voxel. In practice, in order to respect the iso-surface and edge features, as well as preserving regularity in the final parametrization result, our parametrization algorithm has following steps:

1. Corner detection and selection: Determine all corner candidates from the frame field. Interactively select corner points from the candidates, serving as corners of the final parameter domain. These corner points directly determine the structure of the final parameter domain.

2. Energy minimization with constraints: Add parameter constraints on corner points and other points if necessary. Add these parameter constraints into the energy minimization equation. Compute the minimization again to get the final parametrization result.

3. Remeshing: Guided by the generated parameter, trace and generate a small set of volumetric patches.

### 5.3.1 Corner Points

Intuitively, in a parameter domain as shown in Fig. 5.5, a corner candidate $v_i^c$ is the intersection point of 3 iso-parametric surfaces on u, v, w respectively. Consequently, some of its neighboring voxels should separately distribute on 3 iso-parametric surfaces and their normal vector follows 3 different parameter gradients.

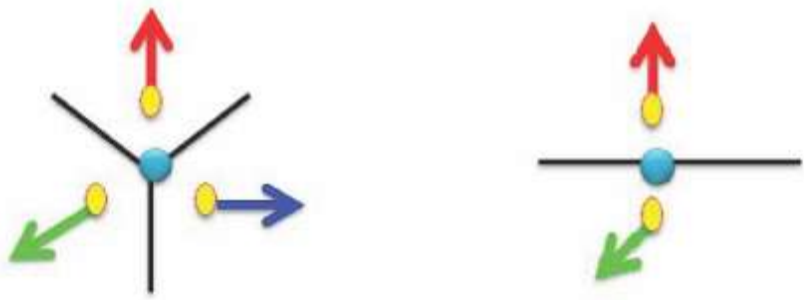

Figure 5.5: Corner point and edge point. Each vector is a constrained direction following the gradient of different scalar field u, v, w (red, green, blue) separately.

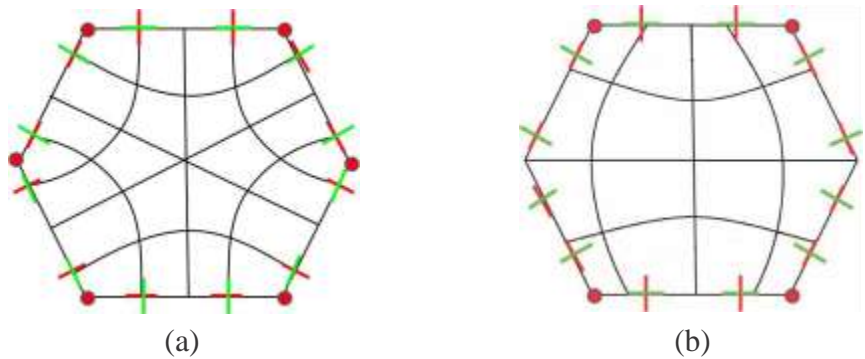

| (a) | (b) |

Figure 5.6: (a) 2D layout of a frame field. It has 6 corners (red nodes) and one singularity. (b) Recomputed frame field. 4 nodes are selected as corners. The jump-match of the frame on the boundary is limited.

In practice, we define **Corner point** as the voxel that has neighboring voxels with constrained normal directions along 3 different gradients $\nabla u$, $\nabla v$ and $\nabla w$ separately. Similarly, we define **Edge point** in a similar way, but its neighboring voxels' constrained normal directions only follows 2 different gradients. For example in Fig. 5.6(a) 2D layout, 6 nodes are detected as the corners according to our definition.

From these corner candidates, we interactively choose several corners as the final corner points. These corners will be mapped to the corners of the parameter domain. Consequently, the edge points connecting a pair of corners will be mapped to the iso-parametric lines on the parameter domain. These edge points also partition the boundary surface into several patches. Intuitively, each patch should be mapped to an iso-parametric surface on the parameter domain.

**Frame field recomputing.** We notice that the original frame field tend to produce unnecessary singularities (Fig. 5.6(a)), making the parameter result and patch structure complicated [165]. To eliminate this problem, we can re-compute the frame field with new constraints: The normal direction of a voxel on an iso-parametric patch must be aligned to the parametric normal direction. As shown in Fig. 5.6(b), the normal directions on the left and right boundaries are forced to be aligned to the green direction, leading to a singularity-free frame field.

**Additional constraints.** Parameter constraints must be added into the energy minimization to make sure that any corner point we select will locate on a corner of the parameter domain. Thus, we associate each corner with a known parameter before solving the energy equation. However, this constraint may cause serious distortion in the solved parametrization. Therefore, we need to add more constraints to get a better parametrization. We observe that the distortion always happens on the geometric-complicated boundary surface patch which maps to an iso-parametric surface in the domain. Therefore, we can avoid this distortion by adding the additional constraints on the boundary surface patch if necessary.

### 5.3.2 Energy Minimization

In order to minimize Equation 5.7, we have to design a linear formulation of the gradient operator $\nabla$ for any scalar field (i.e., U, V or W) on each voxel $v_i$. We notice that the gradient computing is invariant to the choice of parameter. Therefore, we again use the density function (Equation 5.1) and its derivatives to numerically describe the gradient operator $\nabla$. Equation 5.2 and 5.4 together describe the gradient operator on a voxel. For instance, we represent $\nabla u_i$ as :

$$\nabla u_i = (P_u C, P_v C, P_w C) = (P_u, P_v, P_w)(Q^T Q)^{-1} Q^T U^D, \qquad (5.8)$$

where $U^D$ represents the vector of unknown scalar value u on $v_i$ and its neighboring voxels. Then, we substitute them into the energy equation, for example:

$$\sum_{v_i \in V} ||\nabla u_i - u||^2 = \sum_{v_i \in V} ||(P_u, P_v, P_w)(Q^T Q)^{-1} Q^T U^D - u_i||^2. \qquad (5.9)$$

Equation 5.9 is a typical fitting problem, which can be converted into a linear system $AU^T = B$ through computing $\frac{\partial E}{\partial u} = 0$, where $U^T$ is the vector of unknown value u on all voxels. We can simply solve it by least square method.

**Modified norm.** It is obvious that feature orientation is more important than exact edge length. The orientation can be improved by less penalizing stretch which is in the direction of the desired iso-lines. In order to achieve this, [78] have introduced an anisotropic norm and we extend it to 3D vector computing:

$$||(u, v, w)||_{(\alpha, \beta, \gamma)} = \alpha u^2 + \beta v^2 + \gamma w^2.$$

This norm penalizes the deviation along the major directions with different weights.

Then we modify the energy equation to the new form:

$$\sum_{v_i \in V} ||\nabla u_i - u_i||_{(,1,1)} + ||\nabla v_i - v_i||_{(1,,1)} + ||\nabla w_i - w_i||_{(1,1,)},  \quad (5.10)$$

with $\leq 1$.

## 5.4  Spline Approximation and Experimental Results

The previous steps generate a set of regular structured parametric patches thus it is very straight forward to define a regular high-order representation to approximate the shape and the density function of each patch. In our framework, we utilize T-splines for final approximation. A trivariate T-spline [140] can be formulated as:

$$F(u, v, w) = \frac{\sum w_i p_i B_i(u, v, w)}{\sum w_i B_i(u, v, w)}, \quad (5.11)$$

where $(u, v, w)$ denotes parameter coordinates, $p_i = (X_i, Y_i, Z_i, I_i)$ denotes each control point, $w_i$ and $B_i$ are the weight and blending function sets. Each pair of $< w_i B_i >$ is associated with a control point $p_i$. Each $B_i(u, v, w)$ is a blending function given by $B_i(u, v, w) = N^3_{i0}(u) N^3_{i1}(v) N^3_{i2}(w)$, where $N^3_{i0}(u)$, $N^3_{i1}(v)$ and $N^3_{i2}(w)$ are cubic B-spline basis functions along $u, v, w$, respectively. We choose T-spline because it has two significant advantages: First, the refinement of control mesh is subdivided locally to reduce a large percentage of superfluous points and thus enhances the simplicity and accelerates the potential visualization applications; Second, T-spline scheme guarantees $\sum_i w_i B_i(u, v, w) \equiv 1$ across the entire space. Thus the computing of $F(u, v, w)$ and its derivatives can be much more efficient. We notice that, although our domain is globally consistent, each patch is treated as a single object and an independent T-spline in order to better approximate sharp feature.

### 5.4.1  Experimental Results

We introduce our experimental results in this section. A prototype system is implemented on a PC with 3.5GHz P4 CPU and 4GB RAM. We consider the Atom, Fuel, Ankle and Tooth as the test models, and use T-splines to approximate the density function based on our domain. Fig. 5.7 shows the continuous representation results. Compared with the original discrete data, reconstructed models perfectly preserve the shape and density information of the object. They also completely remove the background noise and simplify the procedure of transfer function design for the user. Fig. 5.8 shows more details about our parameterization: the corner points, parameter domain, surface parametrization and volumetric parametrization respectively. Table 5.1 summarizes the statistics of the performance of our processing on four models. These figures and tables showcase that our system effectively reconstruct the model with lower number of control points without sacrificing visual quality.

Table 5.1: Statistics of various test examples: $N_d$, # of voxels; RMS, root-mean-square fitting error (density only, $10^{-2}$); $N_c$, # of corners; $N_c^{()}$, # of control points.

| Model | $N_d$ | RMS | $N_c$ | $N_c^{()}$ |
|-------|-------|-----|-------|------------|
| Atom | $256^3$ | 0.122 | 12 | 1.5 $*$ |
| Fuel | $64^3$ | 0.877 | 16 | 7.2 $*$ |
| Ankle | $128^3$ | 0.422 | 12 | 1.6 $*$ |
| Tooth | $256^2 \times 161$ | 0.393 | 24 | 5.1 $*$ |

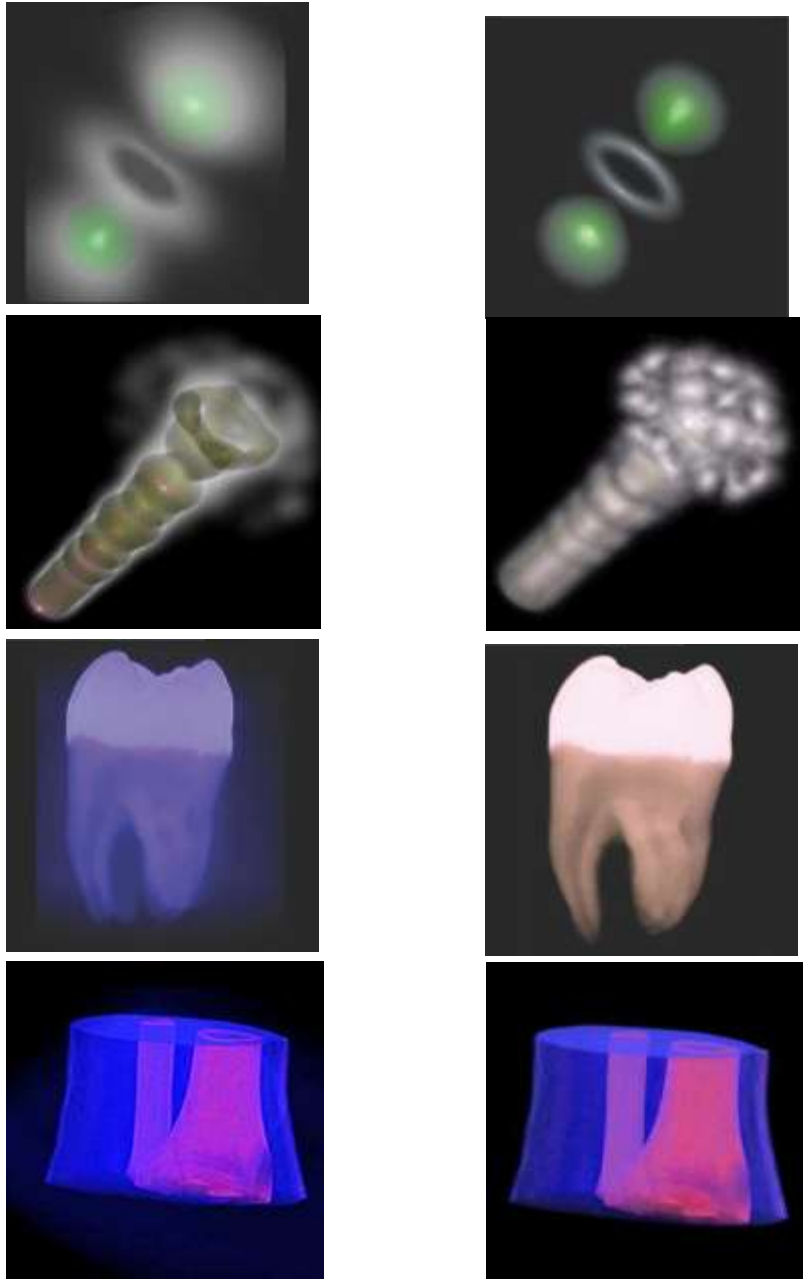

Figure 5.7: Left column: Volume visualization using input discrete models; Right column: Reconstructed models.

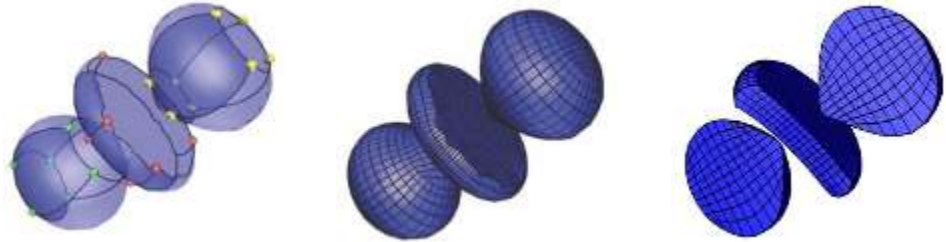

Figure 5.8: The atom model. Left column: Corner points and parameter domain. Middle column: Surface parametrization. Right column Interior parametrization.

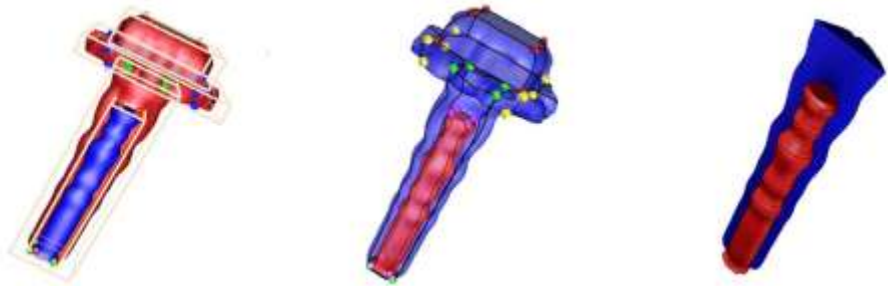

Figure 5.9: The fuel model. Left column: Corner points and parameter domain. Middle column: Surface parametrization. Right column Interior parametrization.

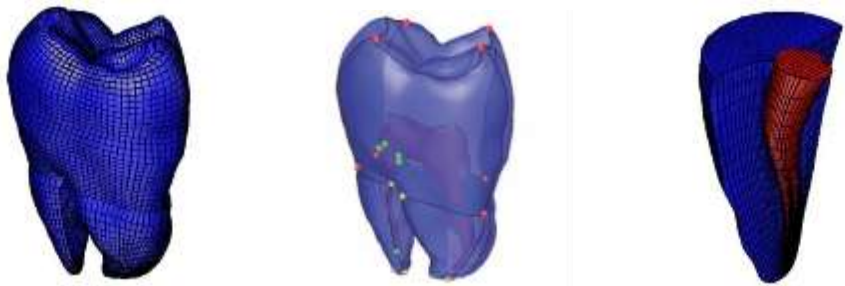

Figure 5.10: The tooth model. Left column: Corner points and parameter domain. Middle column: Surface parametrization. Right column Interior parametrization.

## 5.5   Chapter Summary

In this chapter, we have proposed a novel method that reconstructs the discrete volumetric data into the regular continuous representation. We start with the computing of principal curvatures on a hyper-volume and then find reliable feature-aligned constraints. Then we compute a smooth field respecting the most dominant shape features. Corner points are then computed and placed at geometrically meaningful locations. Based on the frame field, we can generate a regular parametrization which takes material feature-alignment constraints into account, producing a small number of regular patches. We construct trivariate T-splines on all patches to approximate geometry and density functions together. Our test results clearly verify our design.

Our framework perfectly promises a lot requirements in visualization such as feature-alignment, compactness, regular structure, high-order representation and as-homogenous-as-possible, etc. These modeling advantages naturally prompt us to explore its uncharted potential in the near future. We anticipate further novel GPU-accelerated isosurface direct visualization techniques based on our high-order regular representations. Meanwhile, the conjunctions between material-based physical analysis/simulation and our continuous hyper-volume shape functions are of great interest for potential physics-based applications.

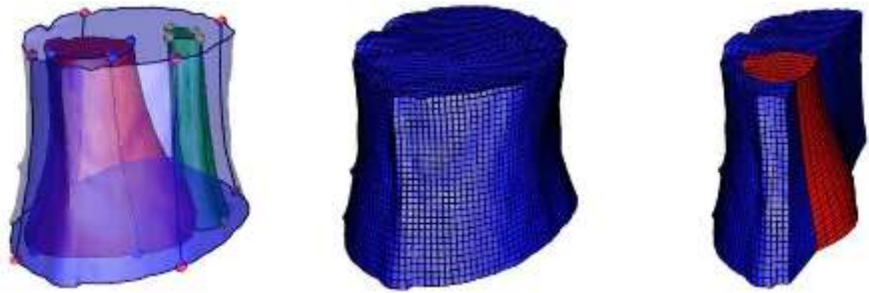

Figure 5.11: The ankle model. Left column: Corner points and parameter domain. Middle column: Surface parametrization. Right column Interior parametrization.

# Chapter 6

# Metrics-based Focus+Context Lens

In all previous chapters, we have discussed the techniques about "how to construct volumetric parameterization, spline construction and representation". Here, we argue that our volumetric parameterizations techniques also have various applications on computer graphics and visualization research. Therefore in this chapter, we attempt to study "how to apply the developed volumetric modeling techniques in other possible research areas".

As we introduced in Chapter 2, there is a stronger-than-ever need for visualizing large-scale datasets in various science/engineering applications. Meanwhile, with the explosive emergence of various types of portable devices (e.g., iPad), the industry frequently pursues as-large-as-possible data visualization on physically-small-sized screen of mobile device in recent years. Therefore, a careful tradeoff is required to deal with the potentially conflicting requirement of the inherent screen size limitation and ever-increasing data size. Focus+Context visualization offers a good strategy when tackling this problem.

Our ultimate goal is to design a flexible F+C methodology on 3D volume image. Therefore, we attempt to design a practical algorithm framework to support this idea. In this chapter, we first apply this framework onto 2D image data as the first step to 3D application. This choice is natural and necessary, because our idea is based on geometric modeling techniques and all relevant numerical computations on 2D manifolds are more mature, stable and robust than on 3D manifolds. Therefore, we decide to adapt it on image operations to test its efficiency.

In essence, we can view our core framework as a "reverse-parameterization" process. Instead of mapping a high-dimension object into a low-dimension space, we attempt to reversely map a low-dimension object into a high-dimension space, such that the visual information is enlarged. In the following sections we will discuss the algorithm in details.

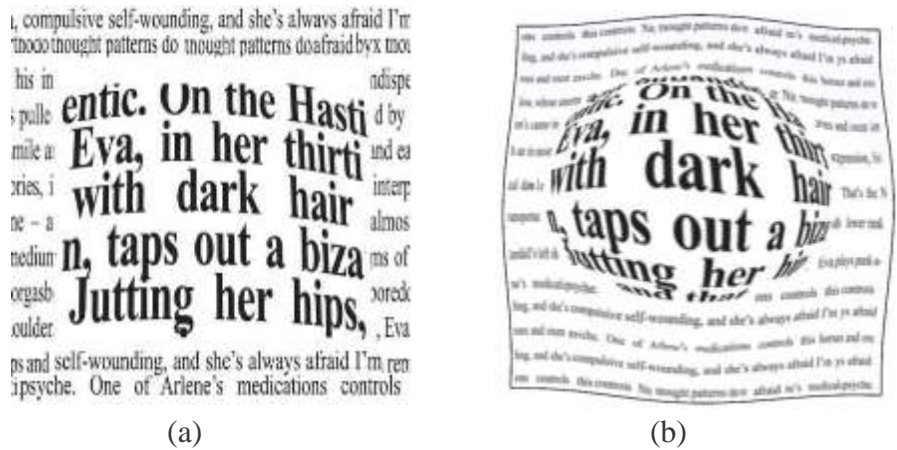

(a)                                         (b)

Figure 6.1: (a): Direct zoom-in. (b): Our geometric approach to simulate magnification lens

# 6.1 Motivation

The traditional method is direct zoom-in, as shown in Fig. 6.1(a). Focus+Context (F+C) visualization, as a natural solution, has gained much research momentum recently. In order to display regions of interest (ROIs) with high resolution, F+C allows the user to access and address the detail of interest ("Focus") while still keeping the overall content of the whole data to accommodate human cognitive custom ("Context"). Attractive F+C visualization should consider the following quality-centric aspects:

(1) **Shape-preserving.** Shape (such as angle, rigidity) plays a crucial role during magnification when improving the visual cognition. The improper magnification distortion may cause serious cognitive confusion.

(2) **Smooth transition.** Any visual gain from unifying the detail with the surrounding context may easily be lost if the transition between the focus and context regions is difficult to understand.

(3) **Flexibility.** For data with complex and multiple ROIs, the user may have preference for using different magnification methods or focusing on different shapes on the same input.

It is a tremendous challenge to optimize the output simultaneously with respect to all of these criteria. For example, many recent methods attempted to simulate optical lenses in depth (e.g., fish-eyes, bifocal lens) for magnification. The most challenging side effect is that, it rarely considers shape-preserving and smooth transition, thus lens distortions are intolerable when features become sufficiently intricate.

Inspired by recent image manipulation techniques such as resizing [166] [96], our new idea is to address the lens design and simulation problem using novel ge-ometric modeling methods. The F+C visualization is then solved by a deformation metric design and optimization solution. This way, we examine this conventional 2D deformation task from a completely innovative perspective of 3D geometric processing. Rather than minimizing deformation energy on a 2D image/grid, we transform the 2D input to 3D mesh, and then conduct 3D deformations which minimize the shape distortions and magnify the ROIs. To achieve our goal, we design a novel deformation framework that functionally acts as a "lens". We first build a special 3D mesh (*"Lens-Mesh"*) that magnifies any area of interest while keeping the rest of area with little distortion. Then, we automatically deform the lens-mesh back into 2D space for viewing. Both steps require us to find distortion minimization for each individual mesh element with an appropriate family of geometric metrics.

In this chapter we present a general theoretical and computational framework, in which 3D geometric modeling techniques can be systematically applied to the 2D lens simulation. The main contributions of our lens design and simulation include: (1) Our algorithm minimizes the geometric deformation metric distortion thus it is particulary suitable to satisfy the shape preserving property. Moreover, our deformation scheme lets the deformable mesh locally confine the resulting distortion with great flexibility rather than letting the distortion uniformly spread throughout the nearby spaces; The resulting transition between the focus and context regions is also smooth and seamless; (2) Instead of only using lenses with a regular circle or square shape, it is very easy to design an arbitrary shape of magnifiers using our lens-mesh to adapt various shapes; (3) The user can iteratively specify the geometric metrics, which allows easy production of visually pleasing effects. The whole algorithm is shown to be of high efficiency, because of the computation of a linear system with pre-processing.

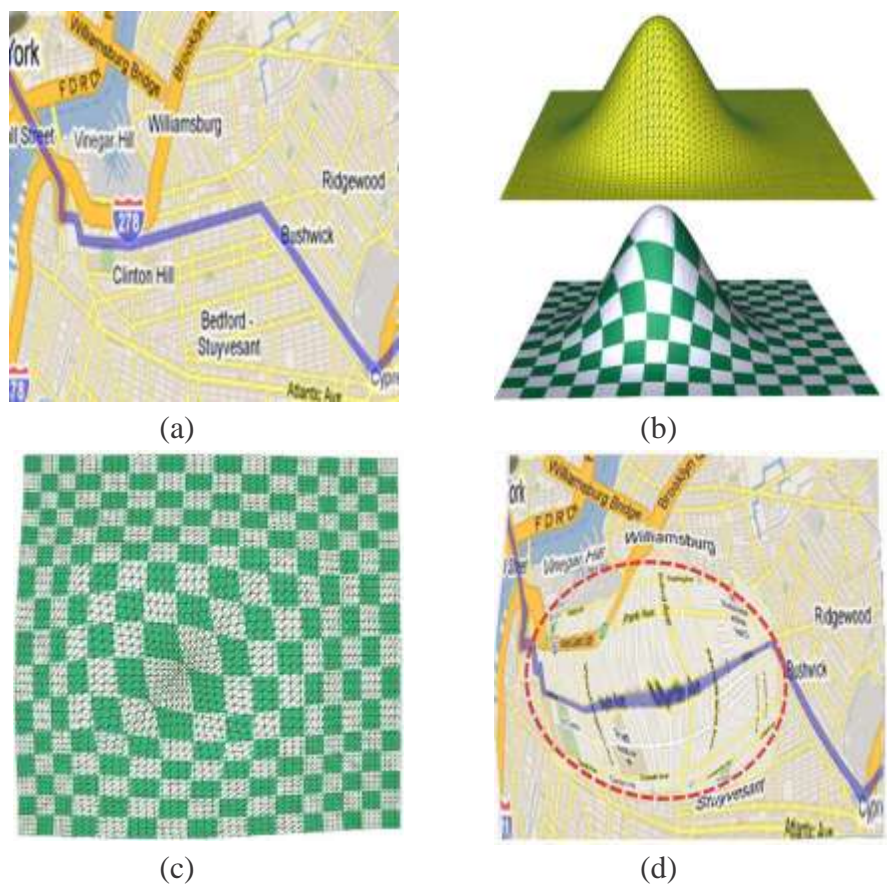

(a)

(b)

(c)

(d)

Figure 6.2: An example of our entire framework: (a) The input image. (b) We generate a 3D lens-mesh to magnify the area of ROI. Then we transfer the texture from the input to the lens-mesh. (c) We deform the lens-mesh back into a 2D plane with minimized distortion. (d) Finally we get a new 2D image with area of ROI magnified.

## 6.2   Framework

This section gives a high level overview of our proposed framework. Our system takes as input a ready-to-display 2D image. For 3D dataset (e.g., volume datasets and 3D scanning models), we can generate the 2D format image through volume rendering. In geometric deformation, we can consider our input as a 2D regular triangle mesh $\mathsf{M} = \{\mathsf{V}, \mathsf{E}, \mathsf{T}\}$. $\mathsf{T} = \mathsf{t}_1, \mathsf{t}_2, \ldots, \mathsf{t}_n$ denotes every individual triangle, and $\{\mathsf{E}, \mathsf{V}\}$ denotes the sets of edges and vertices. Each vertex $\mathsf{v}_i = (\mathsf{p}_i, \varphi_i)$ includes the vertex 2D position $\mathsf{p}_i = (\mathsf{x}_i, \mathsf{y}_i)$ and texture mapping coordinate $\varphi_i = (\mathsf{x}_i, \mathsf{y}_i)$. Note that in the input mesh the vertex position and mapping coordinates have the same value. The output is also a 2D triangle $\mathsf{M}^{\mathrm{out}}$ which has the same structure as $\mathsf{M}$, but every vertex's position and mapping coordinate are updated. Fig. 6.2 illustrates our framework step-by-step using a google map as the example. Our framework mainly includes the following steps.

**Step 1.** The user makes an initial choice about regions of interest (ROIs). We can use a simple user sketch (e.g., drawing a circle) as the ROI boundary to enclose each ROI, or use the exact shape/boundary of every ROI. The boundary can be determined by an automatic feature segmentation operation such as [167] or simple heuristic methods.

**Step 2.** Generate a 3D mesh $\mathsf{M}^{\mathrm{3D}}$ based on the initial mesh $\mathsf{M}$ in order to magnify the area of mesh on ROI.

- (2.1) For each ROI, we deform the original 2D surface patch in ROI into a specified 3D surface, with the ROI boundary as constraints (no shape changes outside the boundary). Every triangle's area in the boundary is therefore magnified.

- (2.2) We transfer the texture from $\mathsf{M}$ to $\mathsf{M}^{\mathrm{3D}}$ while satisfying the shape preserving property. To achieve this, for each vertex inside ROI boundary
  we compute texturing mapping coordinates $[\mathsf{u}, \mathsf{v}]$ on $\mathsf{M}^{\mathrm{3D}}$ by solving the harmonic equation $\nabla^2 \mathsf{u} = 0$ and $\nabla^2 \mathsf{v} = 0$.

**Step 3.** We deform $\mathsf{M}^{\mathrm{3D}}$ back into a 2D plane with distortion minimization. We flatten each triangle $\mathsf{t}_i^{\mathrm{3D}}$ in $\mathsf{M}^{\mathrm{3D}}$ back to 2D by rotation, and we denote this 2D triangle as standard triangle $\mathsf{t}_i^{\mathrm{std}}$. To make each triangle in the final output $\mathsf{M}^{\mathrm{out}}$ approximate to its standard triangle, we design an iterative-executed algorithm with two phases: For each iteration k, we have a starting 2D triangle mesh $\tilde{\mathsf{M}}^k$ which is the result from $(\mathrm{k} - 1)$th iteration ($\mathrm{M}^0$ is initialized by projecting $\mathsf{M}^{\mathrm{3D}}$ to 2D).

- (3.1) For each triangle $\mathsf{t}_i$ in $\mathsf{M}^k$, we compute a deformation metric $\mathsf{M}_i$ (for- mulated as a $2 \times 2$ matrix) using the standard triangle $\mathsf{t}_i^{\mathrm{std}}$.

- (3.2) We determine the updated position of every vertex by solving the linear equation to approximate the deformation metric $\mathbf{M_i}$ for each triangle.

## 6.3   Mesh Generation

The input of our framework is the uniform 2D dataset. Aiming to effectively generate the 2D rendered image from the mesh model/volumetric dataset, we adapt the fragment program (initially proposed by Stemaier et al. [168]) for rendering, considering many parameters including depth, view angle, and camera position. The steps include: cast the ray into the mesh model/volume dataset and composite the color based on the surface/volume data and transfer functions, and render the result into the frame buffer for display.

In most practical focus+context visualization applications, the user only chooses a general approximate region via simple user sketch and/or basic geometric primitives (like the region within a drawn circle), enclosing both mesh segment and nearby context space as a reasonable proxy. The choice of circle lens is natural and humans are more accustomed to it with better visual understanding compared with other geometric primitives. In practice, we first visually choose a general/approximate region, then we pick the center c of this region as the center of sphere associated with a radius $\mathbf{r}$. $\mathbf{r}$ must be large enough to enclose the entire ROI.

After we setting the lens, we magnify it by moving each vertex to a 3D position. we use gaussian function to compute $z_i$ for each vertex: $z_i = g(1 - d_i)h_0$, where $d_i$ denotes the distance to the circle center c, g(x) denotes a standard gaussian function $e^{x^2}$ and $h_0$ is a user input to scale the magnification; As an alternative solution, we can also use a standard sphere instead of gaussian function to accommodate user's visual preference: $z_i = \frac{r}{r} \sqrt{r^2 - d^2}$.

**Arbitrary ROI boundary design.** Our system also allows an exact boundary of an object in the image as the ROI boundary. We denote the triangle mesh patch inside this object as $\mathbf{M_p}$ and $\partial\mathbf{M_p}$ as the patch boundary. We first conduct the medial axis transform for $\mathbf{M_p}$, generating a central curved path $\mathbf{C}$ and each vertex $v_i$ in $\mathbf{M_p}$ has a distance $d_i$ as the shortest distance to the path. The user decides the height $h_0$ of curved path $\mathbf{C}$. For each vertex $v_i$, we have its new position $(x_i, y_i, z_i)$, $z_i = g(1 - \frac{d_i}{d_m})h_0$, where $d_m$ is the maximum distance. We need to subdivide the triangle if it is scaled or sheared too much after magnification. Then we interpolate the locations, colors, distances and heights linearly for newly-inserted vertices.

The automatic algorithm can handle versatile models very well, sometimes users still prefer to use special shapes as the desirable lenses for ROI. Fig. 6.3 shows different visual effects with different meshes.

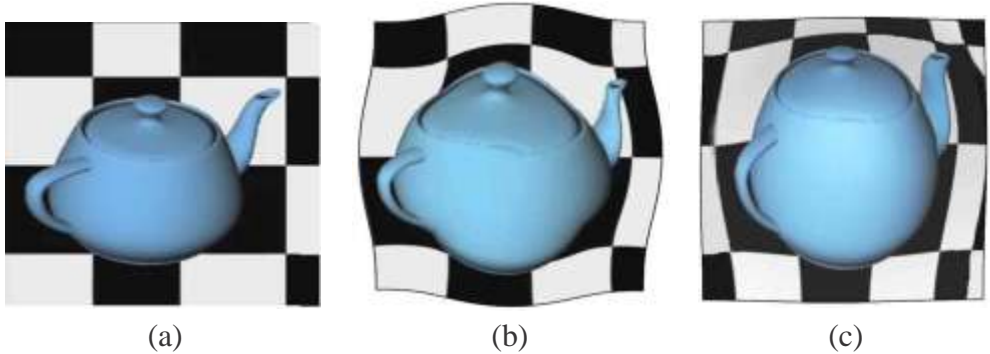

<div align="center">(a)              (b)              (c)</div>

Figure 6.3: Magnification results using different shapes of lenses for the 3D teapot mesh model. (a) Original teapot mesh model. (b-c) Magnification results using the square-shaped and our automatically-generated ROI-guided meshes, respectively.

### 6.3.1 Texture Mapping

The objective of this step is to assign the texture to the magnified 3D triangle mesh, otherwise the texture will be distorted after changing every triangle's shape inside ROIs.

Since both the input mesh $\mathbf{M}$ and magnified mesh $\mathbf{M}^{3D}$ have squared boundary, we treat this problem as the energy minimization problem. We shall map the mesh $\mathbf{M}^{3D}$ to a uniform 2D domain by solving the harmonic functions $\nabla^2 u = 0$ and $\nabla^2 v = 0$, where $\nabla^2 = \frac{\partial^2}{\partial x^2} + \frac{\partial^2}{\partial y^2}$. In practice, solving equations for any but the simplest geometries must resort to an efficient approximation due to the lack of closed-form analytical solutions in the general setting, we shall use mean value coordinates [62] to solve it numerically.

- We assign each vertex an initial coordinate. In practice we initialize it with its original 2D position $(u_i, v_i) = (x_i, y_i)$.

- We iteratively update the coordinates for each vertex $(u_i, v_i) = \sum_{Ng(v_i)} w_j(u_j, v_j)$, and $Ng(v_i)$ is the one-ring neighbor of $v_i$, $(u_j, v_j)$ is a neighbor's coordinate, $w_j$ is the local mean value coordinate [62] computed on $\mathbf{M}^{3D}$. Two types of vertices serve as the Dirichlet boundary conditions (i.e., we avoid changing their coordinates): (1) The squared boundary only; (2) All regions outside any ROI.

## 6.4 Flattening

We search for a flattened mesh so that we can display the result on the popular flat screen (Note that, our algorithm also supports curved screen like "IMAX"). The key challenge in this problem is to preserve the important geometric deformation metric for each triangle. The shape distortion can be measured as the total differences between the resulting triangles and the original triangles. We use the following algorithm to minimize the differences.

**Step 1.** For each triangle $t_i$ in 3D space, we reformulate it into a standard 2D triangle $t_i^{std}$ which keeps its original shape. Suppose $v_1, v_2, v_3$ are 3 vertices of $t_i^{3D}$ in 3D space, $e_1 = v_1 - v_2, e_2 = v_2 - v_3, e_3 = v_1 - v_3$ are 3 edge vectors. We recompute 2D positions of 3 vertices as $v_1 = (0,0), v_2 = (||e_1||, 0)$ and $v_3 = (||e_2||\cos\theta, ||e_2||\sin\theta)$ (Fig. 6.4). $\theta$ is the angle between $e_1$ and $e_2$. Note that, we flatten the triangle separately so a vertex in $\mathsf{M}$ has different 2D positions in different $t_i^{std}$.

**Step 2.** Now we flatten the mesh back to 2D. This step includes 2 iteratively computed phases. The output mesh $\mathsf{M}^{out}$ has the same triangle mesh structure as $\mathsf{M}$ while every vertex has only a 2D position. Initially, we guess $\mathsf{M}^0 = \mathsf{M}^{3D}$ and we reduce the dimension of vertices to 2D by projecting along axis-z: $v_i = (x_i, y_i)$.

(2.1) In this phase we compute the deformation metric for each triangle. The metric represents the transformation from the localized standard $t_i^{std}$ to its k-th iteration counterpart $t_i^k$. We represent this transformation as a $2 \times 2$ matrix $M_i$ and we want to approximate this metric in the output $\mathsf{M}^{out}$. The computation of $M_i$ is detailed in Section 6.4.1.

(2.2) In this phase, we compute the position of each vertex from the following equation.

$$E^k = \sum_{i=0} \sum_{j=1} w_{ij} ||e_{ij}^k - M_i^k e_{ij}^{std}||^2, \qquad (6.1)$$

where $e_{ij}^k, e_{ij}^{std}$ are edge vectors on the triangle $t_i^k$ and standard triangle $t_i^{std}$. We rewrite the function in terms of every edge vector:

$$E^t = \sum_{\substack{M_t \\ i,j}} w_{ij} ||(v_i^k - v_j^k) - (v_{m}^{std}_i - v^{std}_j)||^2, \qquad (6.2)$$

where each pair of $(v_i, v_j)$ belongs to the triangle $t_m$ (Note that $(v_i, v_j)$ and $(v_j, v_i)$ are 2 different vectors that belong to different triangles). $w_{ij}$ is the weight for each edge (see Paragraph "Weights" for details). Setting the gradient to zero, we obtain the following linear equation:

$$LV^{kT} = MLV^{stdT}, \qquad (6.3)$$

where the matrix $L$ represents the edge relationship of vertices (weighted by $w_{ij}$). The matrix $M$ includes all local matrix $M_{t_m}$, $V^k$ and $V^{std}$ are vectors including all vertices' positions on $M^k$ and standard triangles. $V^k$ is the only unknown vector here and solving this equation gives rise to the positions of all vertices in $V^k$.

**Pre-factorization.** We observe that the above matrix $L$ depends only on the geometry of $M$. Thus this sparse matrix is fixed during iterations, allowing us to pre-factorize it with Cholesky decomposition and we can reuse the factorization many times throughout the algorithm in order to accelerate the process, which has a significant impact on algorithm efficiency. The total distortion error $E^k$ converges and we end the iteration when $||E^k - E^{k-1}||$ is smaller than the threshold $\alpha$ (we set $\alpha = 0.1\%$).

**Weights.** The choice of weight $w_{ij}$ in Eq.(7.11) depends on the importance of the triangle. The triangles around the ROI center are more sensitive to distortion. Meanwhile, the distortion on a large triangle is more visually confusing than that on the tiny ones. Therefore, we design the weight as $w_{ij} = (1 + h_m)A_m \cot(\theta)$, where $A_m$ is the area of the triangle $t_m$, $h_m$ is the averaged height (z-values) of the triangle, and $\theta$ is the opposite angle of the edge vector $(v_i, v_j)$ in $t_m$.

### 6.4.1 Computing Metrics

The vertex postion in $M$ is determined by our designed metric $MI$. In our system, we want to achieve a flexible metric such that the user can generate variable visual effects with easy interaction. We notice that each transformation matrix includes two factors: one rotation matrix and two scaling values along two orthogonal directions. Inspired by [55], which blended the angle-only metric and rigid-only metric, we provide a new method that allows the user to specify a "mixed" metric that actually blends between two factors.

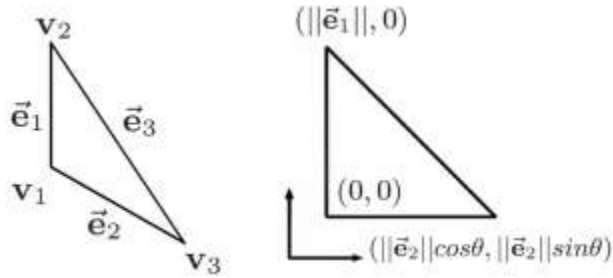

Figure 6.4: Generating a 2D standard triangle. Left: Original 3D triangle. Right: 2D standard triangle $t_i^{std}$.

We start first by computing the transformation matrix between a triangle $\mathbf{t}_i^k$ in $\mathbf{M}^k$ and the standard triangle $\mathbf{t}_i^{std}$. Equivalent to [57] and [169], we compute the Jacobian matrix $\mathbf{J}$ between two triangles.

$$\mathbf{J}(\mathbf{t}_i^k) = \sum_{i=1} \mathbf{e}_i^k (\mathbf{e}_i^{std})^T \qquad (6.4)$$

This matrix measures two tetrahedra's deformation on two factors: rotation and scaling. We can decompose two factors by singular value decomposition.

$$\mathbf{J} = \mathbf{U}\Sigma\mathbf{V}^T, \mathbf{M}_r = \mathbf{U}\mathbf{V}^T. \qquad (6.5)$$

Here $\mathbf{M}_r$ is a rotation-only matrix. and $\Sigma$ includes two scaling values $\sigma_1$ and $\sigma_2$.

$$\Sigma = \begin{array}{cc} \sigma_1 & 0 \\ 0 & \sigma_2 \end{array}$$

To compute a flexible matrix, we can change this $2 \times 2$ diagonal matrix $\Sigma$ with blended scaling values. We allow the user to input a blending parameter $\alpha (0 \leq \alpha \leq 0.5)$. Then the resulting matrix is formulated as:

$$\mathbf{M} = \mathbf{U} \begin{array}{cc} \sigma_1^b & 0 \\ 0 & \sigma_2^b \end{array} \mathbf{V}, \qquad (6.6)$$

where $\sigma_1^b = \alpha(\sigma_1 - 1) + 1, \sigma_2^b = \alpha(\sigma_2 - 1) + 1$.

**Algorithm 2** The flattening algorithm.

```
Input: triangle mesh M³ᴰ,
        Blending parameter α ∈ [0, 0.5]
        Fitting error threshold
Output: 2D mesh Mᵒᵘᵗ
```
$L = \text{BuildM atrix}(\mathsf{M})$ // See Eq.(3)
$\text{Cholesky} - \text{Decomposition}(L)$
**for all** $\mathbf{t}_i^{3D} \in \mathsf{M}^{3D}$ **do**
   //Compute the 2D standard triangle
   $\mathbf{t}_i^{std} = 2D - \text{Standard}(\mathbf{t}_i^{3D})$
**end for**
```
Initial guess
```
$\mathsf{M}^0 = \text{P rojection}(\mathsf{M}^{3D})$
**while** $\|E^k - E^{(k-1)}\| > \quad$ **do**
   **for all** $\mathbf{t}_i^k \in \mathsf{M}^k$ **do**
      //Compute metrics. See Eq. (4)
      $M_{\mathbf{t}_i} = \text{Compute}(\mathbf{t}_i^{3D}, \mathbf{t}_i, \alpha)$
   **end for**
   // Build and solve Eq.(3) to get $\mathsf{M}^k$
   $E^t = \text{F ittingError}(\mathsf{M}^k, \mathsf{M}^{3D})$
   $k = k + 1$
**end while**
$\mathsf{M}^{out} = \mathsf{M}^k$

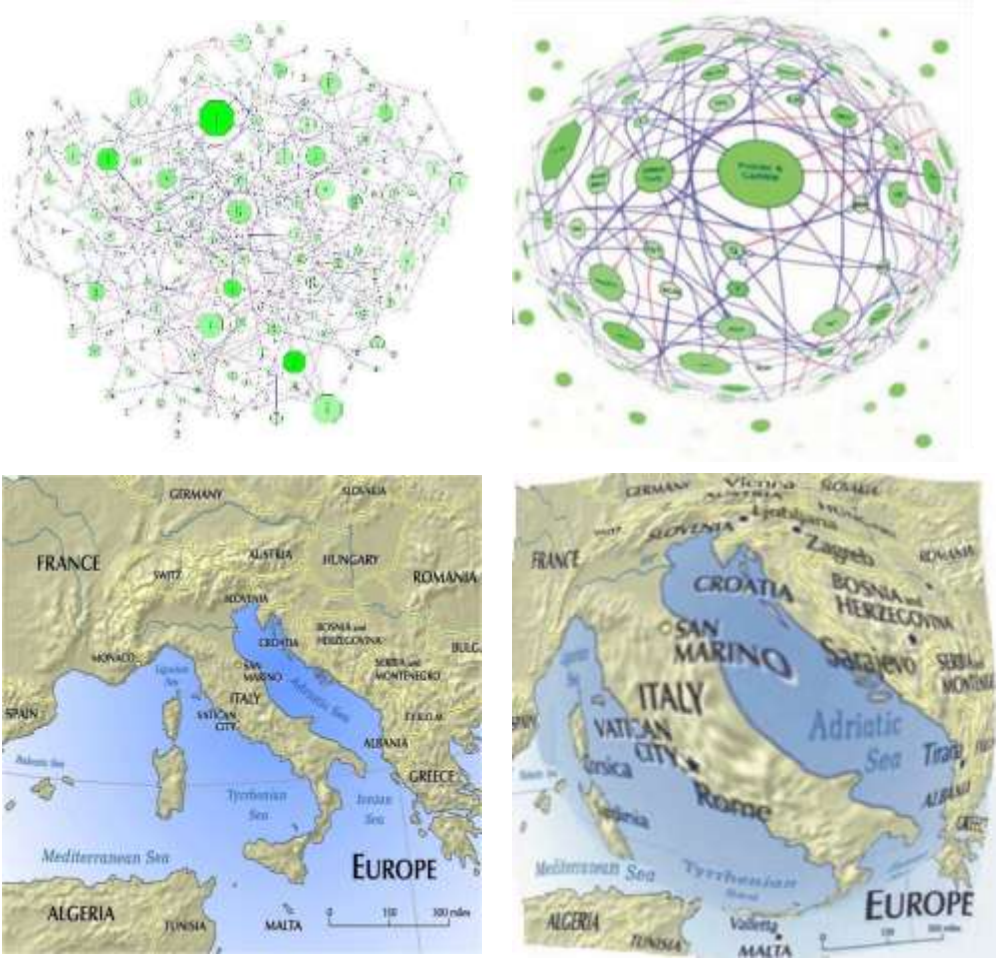

Figure 6.5: Applications of our lens simulation. Left row: Inputs. Right row: Graph of company relations, the connecting edges are revealed by the magnification; European map, major cities of Italy are revealed now.

## 6.5 Experimental Results

Our system can effectively provide F+C information to the user, allowing the user to get detailed focal region while maintaining the integral perception of the model. The results shown in the following figures demonstrate the power of our technique. Our experimental results are implemented on a 3GHz Pentium-IV PC with 4Giga RAM. In Fig. 6.5, we test our lens using several popular data structures such as graph, city, map, and text for information visualization: Graph is an abstract data structure representing relationships or connections. For access to relative nodes or to the particularly important nodes, our lens makes it easy to find and navigate toward these nodes; Our framework also improves the magnification functions with results of multi-scale map/satellite magnification, which reveal and magnify the additional details (e.g., additional country names); Our lens provides the efficient scanning function for the text reading as well. We can place the magnifier to zoom in the focus region while the remaining regions are evenly distributed to the context area (as shown in Fig. 6.1).

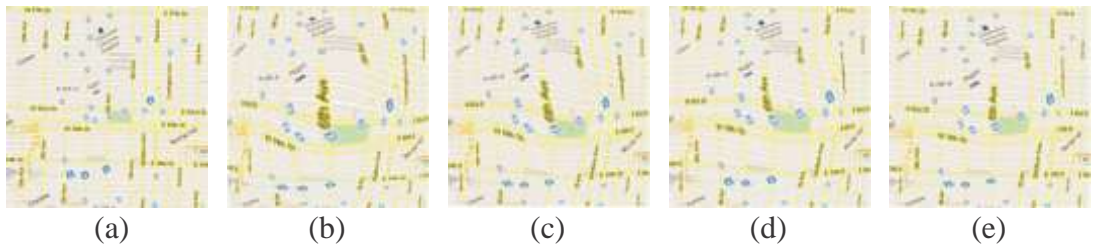

| (a) | (b) | (c) | (d) | (e) |

Figure 6.6: A group of different metrics with modified blending parameter α (α =0, 0.01 , 0.1, 0.5).

Fig. 6.2(d) is another excellent example to demonstrate that our technique offers a powerful lens for the route magnification. Using our lens, the user can see the additional route information and easily panning or zooming to achieve their requirements. Meanwhile, there is no any obviously visual distortion in both focus or context areas (the transition area with two view scales merges using linear interpolation). The global road distributions and orientations are preserved, and detailed streets are displayed around ROI.

As a general rule, a good F+C method should be able to maximally support the shape/feature preservation of objects of interest, such as conformal (angle) preservation or/and authalic (area) preservation, while minimizing context distortions. Instead of only minimizing angle distortion in [170, 179-181], Fig. 6.6 shows a group of lenses with the same input but different metrics, with the blending parameter α = 0, 0.01, 0.1, 0.5. This blend metrics enrich the result and thus the user can modify the blending parameter to interactively change the visual effect until one result is satisfactory from the user's perspective.

**Performance.** Unlike other methods, the performance of our framework does not depend on the input image but the size of our triangle mesh. So a conventional performance table ("model-by-model") is not necessary for the analysis purpose. The sample images we tested are all between $512 \times 512$ and $1024 \times 1024$. We provide two meshes with sizes of $100 \times 100$ and $200 \times 200$ to handle small and large images separately. The smaller mesh (10k vertices) uses only 0.3 second for one iteration and it always converges in 2 iterations. We use the larger mesh (40k vertices) to handle very high-detailed application and it uses 1.3 seconds for one iteration. The pre-processing (matrix assembling and pre-factorization) requires only about 1.0 second.

**Distortion.** Similar to Eq.(7.11), we apply the following term to measure the shape distortion on every triangle $T_i$.

$$E_i = \sum_{j=1}^{\times} w_{ij} \|e_j - M_i e_j^d\|^2, \tag{6.7}$$

Fig. 6.7 compares the distortion between our lens and poly-focal lens [114] (We consider the input image of poly-focal lens as a regular grid mesh. The deformation equation is defined in [114]). Although poly-focal lens or fisheye lens can have similar continuous magnification F+C view as our lens, it creates noticeable distortions towards its edges and has no method to formally control the focus region as well as to preserve local features in the context region. The comparison is meaningful because both methods allow "free-boundary" to obtain better shape-preserving effects. To measure the distortion of poly-focal lens, we also consider their resulting image as a deformed mesh with each vertex/color moving to the new position. Thus we can also use the same criteria to measure the shape distortion. The color indicates that our method can reduce the shape distortion in a much better way. We use blue color to represent zero distortion and red the maximum (0.45 in our result).

**Comparison for Magnification Results.** We apply our method to a volumetric colon dataset to verify the advantages of our lens and compare with others as shown in Fig. 6.8. Local shape preservation and smooth transition have important applications in the clinic education, diagnose, and even virtual surgery. In the normal clinic exam, the colonoscopy needle navigates along the colon axis and the lens is added along the same direction such that the clinicians are able to recognize polyps on the folds (the wrinkles on the colon wall, red circle). The folds in Fig. 6.8(b-c) are seriously distorted which may sabotage the clinicians' expertise on polyps detection. No matter how we modify their lenses in (b-c), the distorted folds always exist along the lens boundary. In sharp contrast, the fold details in (d) are better preserved and easy for recognition.

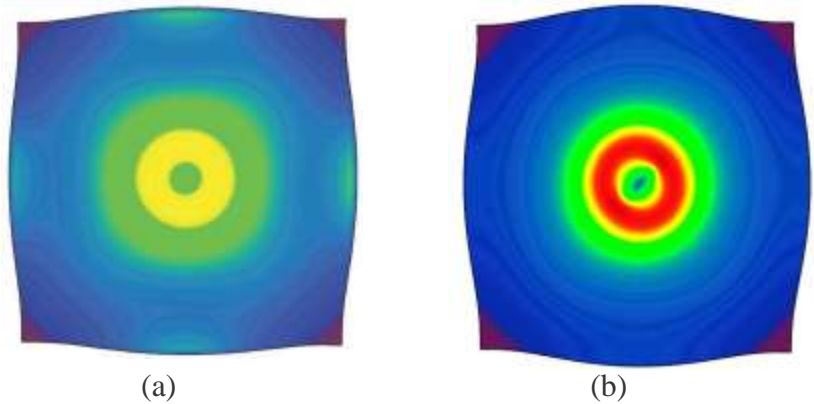

(a)                              (b)

Figure 6.7: (a-b) The distortion of our mesh and poly-focal lens. The distortion is color-coded from blue (minimum) to red (maximum).

We compare our method with other approaches, like zoom-in, fish-eye, bi-focal, perspective wall, poly-focal [114] and cube deformation [119] in Table 6.1. Our method has advantages in the following aspects. First, our solution works well particularly with the complex shape, because it can flexibly design arbitrary shapes for lenses. Our method emphasizes angle and rigidity metrics for the shape-preserving purpose. Moreover, it allows the user to interactively design and blend various metrics.

**Limitations.** Our system flattens the mesh to achieve F+C visualization, but potentially it may result in flip-over phenomenon (i.e., the resulting triangle covers another one or its orientation is reversed). Fortunately, this phenomenon always happens especially on a highly curved surface with complex topology. In contrast, our 3D mesh is relatively very simple compared with common models used in geometric modeling study and there are no flip-over triangles in all examples during our experiments. The texturing step (Section 6.3.1) also produces a fine mapping as a good initial guess. Meanwhile, we can always solve the flip-over problem using the existing algorithm [171].

Compared with the direct zoom-in and bi-focal methods, our method can not authentically keep exactly the same feature of a local region as the original input. Also, our metric lacks of the measurement to preserve the global structure, shape symmetry, or long straight lines. However, our human cognitive system for recognition is accustomed to automatically compensating these slight variations of a local region and thus it relieves possible disturbing experience for the user.

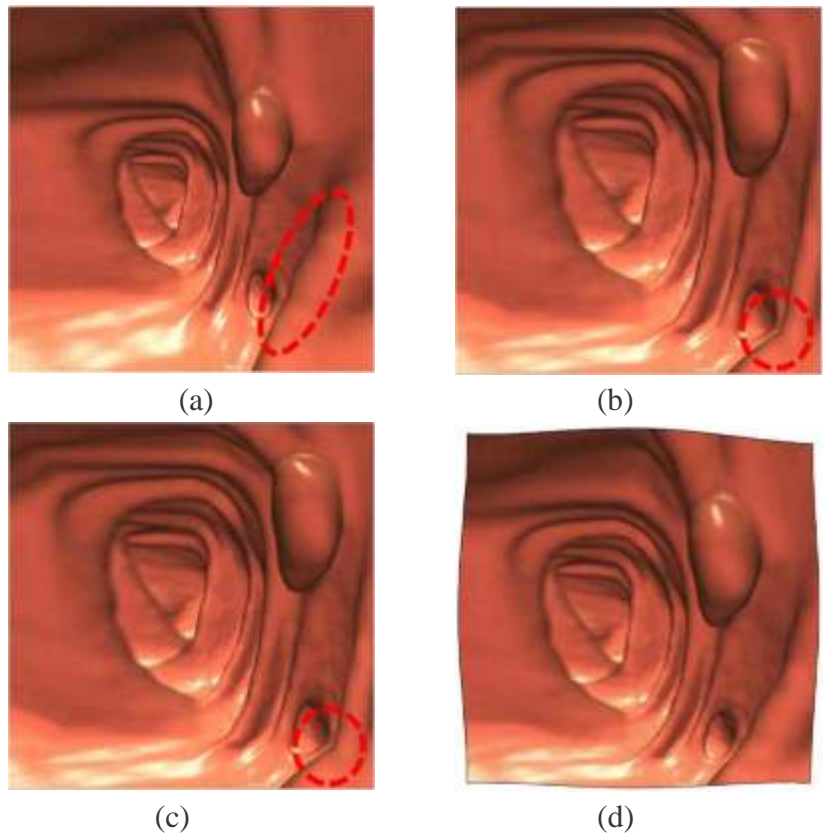

<div align="center">(a)    (b)</div>

<div align="center">(c)    (d)</div>

Figure 6.8: Magnification results using different lenses for volumetric colon dataset. (a) Original colon dataset. (b-d) Magnification results using bifocal, polyfocal, and our lenses. By comparison, the folds on the interior colon surface are seriously distorted by all the other lenses because of the sharp transition between the focus and context regions, while our lens shows the accurate shapes/features of the interior colon surface without any obvious distortion.

Table 6.1: Comparison with the existing approaches.

| Method | zooming | fisheye | bifocal | perspective wall | poly-focal | mesh editing | our method |
|---|---|---|---|---|---|---|---|
| Shape preserving | yes (focus) | no | yes(focus) no(transition) | no | no(focus) yes(transition) | yes | angle+ rigidity |
| Smooth transition | no | yes | no | no | yes | yes | yes |
| Arbitrary lens shape | no | no | no | no | no | no | yes |
| Interactive metric design | no | no | no | no | no | no | yes |

## 6.6    Chapter Summary

We have developed a novel and interactive technique to achieve Focus+Context visualization based on geometric deformations. Specifically, we develop from the input a 3D lens-mesh and magnify the ROIs through deformation on the lens-mesh. Our lens design methodology and the prototype system manifest that the geometric deformation metrics greatly enhance the F+C visualization, and our approach is expected to transcend the traditional boundary of geometric modeling and will benefit data visualization and visual analytics.

The important features of our framework can be summarized as: (1) **Shape-preserving.** The geometric deformation metrics are minimized so that the resulting details appear similar to their original counterparts. Geometric deformation also generates a continuous transition region where the user can get a smooth viewing transition from the highly-magnified interior region to the non-magnified exterior region; (2) **Robustness.** It enables the user to design arbitrary number/shape of magnifiers to effectively display the entire ROIs for visualization of multiple and complex features. It also allows the user to interactively specify geometric metrics for various visual effects; (3) **Efficiency.** The computation is very efficient because of our pre-factorization processing. Our experimental results have demonstrated that our lens, as a novel F+C technique, has great potentials in many visualization applications.

# Chapter 7

# Four Dimensional Magnification Lens

In the last chapter we introduce a novel geometry-based method for image focus+context visualization. The success inspires us to extend this pipeline to volume visualization. The rapid advances in 3D scanning, acquisition, and modeling techniques have given rise to the explosive increase of volumetric digital models with extra density information like MRI, textured solid models [172, 173] or CAD models containing materials. The great progresses in GPU rendering, and internet bandwidth push forward a stronger-than-ever need for visualizing large scale volume datasets in various science/engineering applications. Meanwhile, the explosive emergence of various types of potable mobile devices (e.g., smart phone) pursues the visualization technique to display large scale models on a physically limited device screen. It requires us to non-homogeneously rescale different regions while keeping the global shape of models within the screen space.

The traditional method is through the use of 2D screen region-of-interest (ROI) magnification techniques, which functions as "lens" and offers a good strategy to magnify a local region only. However, compared with magnification on the image projected on the screen, it is more preferable to locally magnify the 3D volume datasets directly. For example, the user can translate, rotate, cut and visualize the dataset from different angles without computing magnification again and again. Magnifying datasets directly is also necessary for many virtual reality applications (e.g., cultural heritage and walkthrough).

From practitioners' perspective, an attractive magnification should address the following quality-centric aspects: **Shape-preserving.** Shape (such as angle, rigidity) plays a crucial role during magnification when improving the visual cognition. The improper magnification distortion may cause serious cognitive confusion. We should preserve the shape of both focus region and surrounding context region and global shape simultaneously; **Smooth transition.** Any visual gain from unifying the local detail with the surrounding context may easily be lost if the transition between the focus and context regions is difficult to understand; **Simple interaction.** In most practical applications, the user only prefers to use simple user sketch (e.g., draw a circle) to enclose the focus region. An ideal system should support such simple interaction.

However, it is a tremendous challenge to optimize the output simultaneously with respect to all of the aforementioned aspects. The most challenging side effect is that: in a 3D world, a local region's magnification inevitably compresses the rest region and leads to distortion. More severely, the conventional methods are more likely to spread the distortion throughout the 3D space. Any optimization technique only moderates but never eliminates distortion. Meanwhile, the existing techniques consider neither shape-preserving nor smooth transition from the rigorous geometry's point of view, thus lens distortions are intolerable when features become sufficiently intricate.

To tackle the above challenges, we are inspired by the following idea: Rather than magnifying ROIs and shrinking the rest region in the 3D world, we could increase ROIs' volume in the additional dimension without changing the rest region. Also, it is a well-known knowledge that the differential geometry theory and its practical techniques (e.g., surface parameterization) can handle angle distortion rigorously and quantitatively. In this way, we examine this conventional magnification task from a completely innovative perspective of 3D/4D geometry processing.

To achieve this goal, we propose a framework to simulate 4D lens in order to achieve local magnification while minimizing global angle distortion. This framework starts from transforming the 3D input into a 4D mesh with an initial fourth dimension for every vertex. Then we conduct 4D deformation which enlarges ROI's volume while keeping the rest unchanged. Then, we automatically deform the mesh back into 3D space for other applications. Both steps require us to seek distortion minimization for each individual mesh element during deformation. Specifically, our contributions in this work include:

1. A framework to address the 3D volume dataset magnification. In contrast to other possible deformation solutions, our method lets the additional dimension's space absorb the volume magnification rather than spreading throughout the nearby space in the original dimensions. Therefore, our result can resemble the original interior texture and the resulting transition between ROIs and the rest is also smooth and seamless.

2. Techniques for distortion minimization with high dimensions. To achieve this, we propose a piece-wise method to solve the harmonic function on nD tetrahedral mesh. Meanwhile, we develop a flattening method to model the 4D shape flattening back into 3D and preserve the shape.

Our system has the very unique feature that we can preserve the shape around both focus region and context region/global shape. Our geometry-based method can also achieve distortion control and quantifying. Therefore our system can effectively magnify and visualize volume datasets while keeping distortion unnoticeable. Theoretically, our research first demonstrates that 4D geometry is a powerful tool for volume visualization and modeling, and has great potential for 3D graphics-relevant tasks.

After discussing related literatures, a framework overview is given in Section 7.1. On a global view, modeling the 4D magnification in Section 7.2 is the first stage in our framework, followed by flattening techniques in Section 7.3. In Section 7.4, we demonstrate our experimental results and document more comprehensive discussion, respectively.

## 7.1   Framework

This section gives a high level overview of our proposed framework. Our system takes as input a wide range of 3D textured solid models (Fig. 7.1). For a tetrahedral mesh without texture, Takayama et al. [173] proposed a method for interior solid texturing modeling. For volumetric datasets (like CT and MRI) with texture information only, we partition the given volumetric dataset using a uniform grid. Each vertex in the grid is associated with a 3D parameter (u, v, w). The original volume dataset now becomes the volume texture of the uniform grid. We further decompose each grid into several tetrahedra and convert the input to a 3D textured tetrahedral mesh, as shown in Fig. 7.1(Bottom).

Now we can describe an arbitrary input by a uniform format. We define the input as a tetrahedral mesh $M = \{T, E, V\}$. $T = \{t_1, t_2, \ldots, t_n\}$ denotes the set of tetrahedra, and $\{E, V\}$ denotes the set of edges and vertices. A mapping function $\varphi$ maps vertices to the texture. In a discrete setting, each vertex $v_i = (p_i, \varphi_i)$ includes two items: p denotes vertex's position (we use $p^{3D} = (x, y, z)$ in 3D and $p^{4D} = (x, y, z, h)$ in 4D). $\varphi_i = (u, v, w)$ denotes a volumetric parameter corresponding to the volume texture. Our output is a new tetrahedral mesh $M^{out}$ with updated p and $\varphi$ for each vertex. Our framework includes the following steps.

**Step 1: Choosing ROI.** The user makes an initial choice about regions of interest (ROIs). The shape/boundary of a ROI can be determined by a bounding sphere that encloses user's interested region, or, by a more accurate ROI's boundary. We could detect an accurate ROI's boundary through automatic boundary extraction operations (e.g., marching cube) or simple heuristic methods.

**Step 2: Magnification.** In order to magnify the total volume in ROI, we generate a new 4D mesh $M^{4D}$ based on the initial mesh.

- (2.1) For each ROI, we deform the original 3D tetrahedral patch inside the ROI in the 4D space, with the ROI boundary as constraints (so that no shape changes outside ROI's boundary). The total volume within the boundary is magnified after this operation.

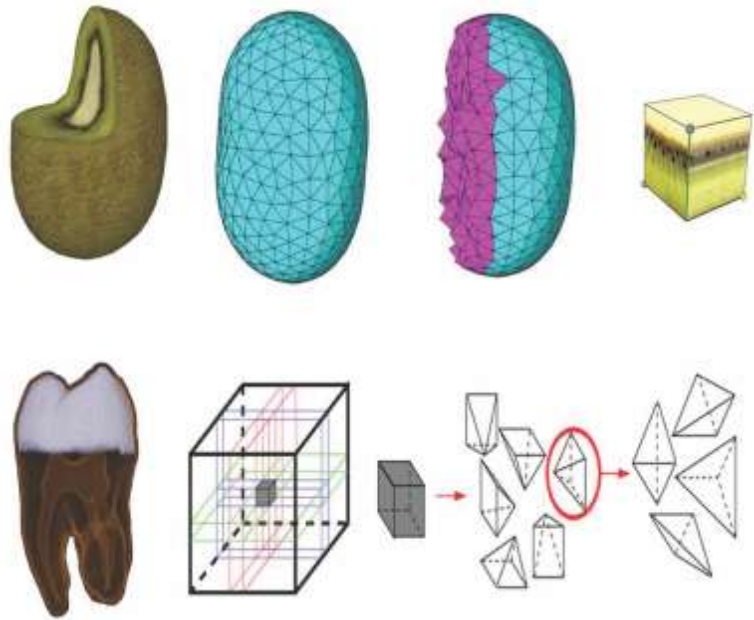

Figure 7.1: Inputs of our framework. Top: A 3D solid textured model is a tetra-hedral mesh mapped by the color texture. Bottom: For a volumetric dataset, we partition the space into grids and each grid is uniformly subdivided into tetrahedra.

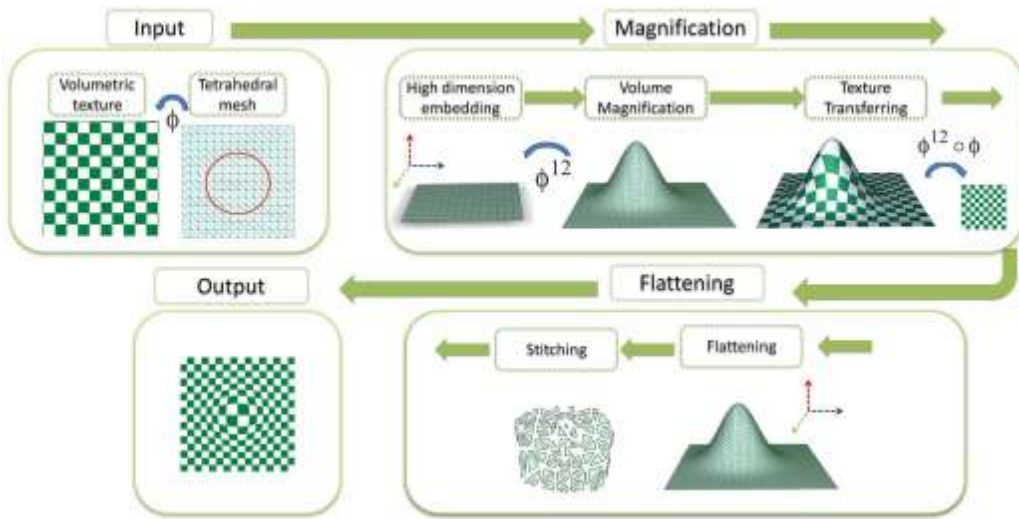

Figure 7.2: Illustration of the framework. Because it is impossible to visualize 4D space, we use an image, a planar triangle mesh and a 3D triangle mesh to represent a volumetric dataset, a 3D tetrahedral mesh and a 4D tetrahedral mesh. After preprocessing, the input is a tetrahedral mesh with a volumetric dataset as the texture. The tetrahedral mesh is first embedded into a high dimensional space and we magnify the total volume in a ROI through the additional dimension. We solve the harmonic function to recompute the mapping and transfer the texture to the new 4D tetrahedral mesh. Finally, we flatten the 4D tetrahedral mesh back into 3D for flexible visualization.

- (2.2) We recompute each vertex's parameter to remedy the shape distortion during magnification. To achieve this, we solve the volumetric harmonic function: $\Delta\varphi^{12} = 0$, where $\varphi^{12}$ is a texture transfer function $\mathbf{M}^{4D} \rightarrow \mathbf{M}$. Then for a vertex $\mathbf{v}_i = (\mathbf{p}_i^{4D}, \varphi_i)$ in $\mathbf{M}^{4D}$, we update its parameter as: $\varphi_i = \varphi(\varphi^{12}(\mathbf{p}_i^{4D}))$, where $\varphi$ is the parameter on the original 3D mesh $\mathbf{M}$.

**Step 3: Flattening.** In Step 2 we have already magnified $\mathbf{M}$ to $\mathbf{M}^{4D}$. In order to visualize $\mathbf{M}^{4D}$, it is necessary to flatten $\mathbf{M}^{4D}$ back into a 3D mesh as the final output $\mathbf{M}^{out}$ and preserve the magnification effect. We use a $4 \times 3$ rotation matrix to rotate each 4D tetrahedron $\mathbf{t}_i^{4D}$ back to a "flattened" 3D tetrahedron $\mathbf{t}_i^{F}$. Then we stitch all separate tetrahedra together as the sole mesh $\mathbf{M}^{out}$, and keep each tetrahedron's shape to roughly approximate to $\mathbf{t}_i^{3D}$ after stitching. We can execute this step iteratively until getting a visually promising result.

- (3.1) We initially guess a 3D tetrahedral mesh (e.g., from the last iteration's result, or by simple projection from $\mathbf{M}^{4D}$ in the first iteration). By comparing between the "guess" tetrahedron $\mathbf{t}_i^{3D}$ in $\mathbf{M}^{3D}$ and rotation-generated "flattened" tetrahedron $\mathbf{t}_i^{F}$, we can compute a $3 \times 3$ Jacobian matrix $\mathbf{J}_i$ between two corresponding tetrahedra. Then we can extract from $\mathbf{J}_i$ a stretching-free/rotation-only matrix $\mathbf{R}_i$.

- (3.2) We solve the linear optimization equation to determine every vertex's position in $\mathbf{M}^{out}$ such that, in the resulting mesh, the Jacobian matrix between the resulting tetrahedron and the "guess" tetrahedron approximates $\mathbf{R}_i$.

Fig. 7.2 shows our framework in a step-by-step fashion. Since it is extremely difficult to visualize the 3D-to-4D deformation in an intuitive way, we utilize 2D-to-3D deformation to simply illustrate the entire framework: 2D image or triangle mesh to mimic volume dataset / tetrahedral mesh, and deformed 3D triangle mesh to mimic a 4D tetrahedral mesh.

## 7.2 3D-to-4D Magnification

In order to magnify in 4D space, we first extend the input $\mathbf{M}$ by embedding it into 4D space. For each vertex with a 3D position $\mathbf{p}^{3D} = (x, y, z)$, we expand it to $\mathbf{p}^{4D} = (x, y, z, h)$ where the additional height $h = 0$. We can imagine this operation in the 2D layout as pulling a 2D plane from 2D to a real 3D world with shape unchanged (still a 2D plane but embedded in a 3D world after pulling).

### 7.2.1 ROI Magnification

Now we start to magnify ROIs. ROI is a region in the volume. Each ROI encloses a mesh patch $M_p$ and we use $\partial M_p$ to represent the boundary of patch $M_p$. To magnify the ROI's volume, we seek a solution that could stretch all vertices inside $M_p$ to new positions while keeping other vertices unchanged.

In most practical focus+context visualization applications, the user only chooses a general approximate region via simple user sketch and/or basic geometric primitives (like the region within a drawn sphere), enclosing both mesh segment and nearby context space as a reasonable proxy. In our system, we use a sphere to enclose the focus region and simulate lens in most applications. The choice of sphere lens is natural and humans are more accustomed to it with better visual understanding compared with other geometric primitives. In practice, we first visually choose a general/approximate region, then we pick the center c of this region as the center of sphere associated with radius r. It may be noted that, $r$ must be large enough to enclose the entire ROI.

After setting the lens, we magnify its volume by moving each vertex to a new position along the fourth dimension. As shown in Fig. 7.2, we use a gaussian function to compute $h_i$ in each $p_i^{4D}$ because the shape changing in such case is not severe but smooth. For each vertex we compute $h_i = g(1 - \frac{d_i}{r})h_0$, where $d_i$ denotes the distance to the sphere center c, $g(x)$ denotes a standard gaussian function $e^{x^2}$ and $h_0$ is a user input to scale the magnification. As an alternative solution, we can also use a standard 4D sphere instead of gaussian function to accommodate user's visual preference: $h_i = \sqrt{r^2 - d_i^2}$.

In some applications, the user may seek for a lens with an arbitrary shape. For example, a focus object extracted from the volume may have complex shape or high genus boundary and the user prefers to use this exact boundary to be the lens (like Fig. 7.3(b)). To achieve this, we can generate a central skeleton-like curved path $C$ (e.g., [137]) and get the medial axis transform for every point on the object boundary. Each vertex $v_i$ inside the lens associates the shortest distance $d_i$ with the axis path $C$. Now again we can use gaussian function to compute $h_i$ for each vertex: $h_i = g(1 - \frac{d_i}{d_m})h_0$, where $d_m$ is the maximum distance value.

Large scale magnification may stretch/shear the tetrahedron and sabotage the mesh quality. To solve this, we need to subdivide the highly-stretched tetrahedron and compute the locations and parameters $(p, \varphi)$ for newly-inserted vertices. We utilize barycentric coordinates and linear interpolation to interpolate new positions and parameters. For a point $p_c$ inside a tetrahedron, its barycentric coordinate is:

$$f_c = \sum_i \lambda_i f_i, \quad \lambda_i = \frac{1 < p_c, s_i >}{3 \quad V}, \tag{7.1}$$

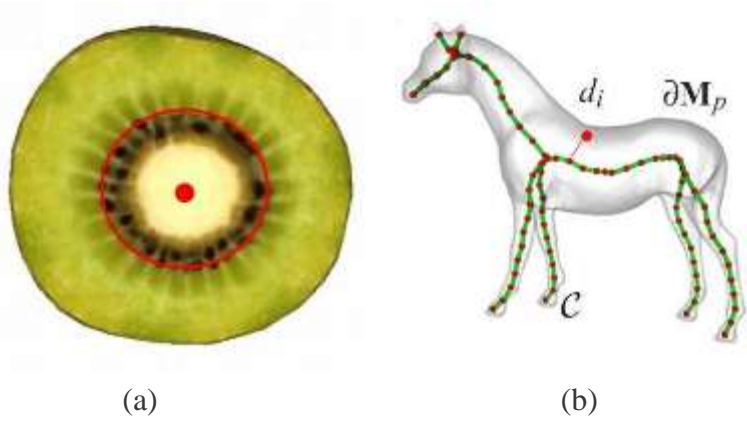

(a)                                    (b)

Figure 7.3: Two ways of lens shape design: (a) We can use a 3D sphere, with a center c (red point), to enclose the entire ROI. The radius is r. (b) For an arbitrary shape lens like an extracted object's boundary (horse) from the volume, its medial axis can assist us to generate the lens. Each vertex inside the ROI associates a distance value $d_i$ with the axis.

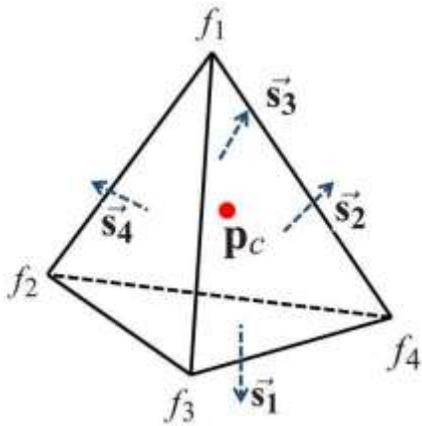

Figure 7.4: A tetrahedron and face normal vectors.

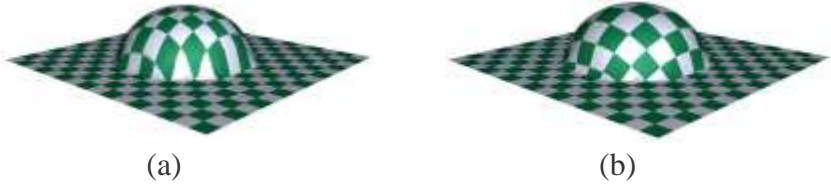

<center>(a)                    (b)</center>

Figure 7.5: Texture transfer. We use 2D layout to illustrate the effectiveness of texture transferring. (a) Direct magnification without recomputing texture transfer. (b) The result after recomputing texture transfer.

where V is the volume and $s_i = A_i n_i$. $A_i$ indicates the area of one tetrahedron's face triangle (and each tetrahedron has four face triangles). Using the barycentric coordinates, we can keep the shape unchanged before and after adding vertices. Although the texture interpolation may not be optimal under this strategy, we compensate it by modifying the texture coordinates in the following texture transfer step.

## 7.2.2 Texture Transfer

After the magnification step, the tetrahedral mesh in the focus region has already been magnified. Now we need to recompute the texture mapping to minimize distortion around both focus and context regions. The texture transfer is necessary because after the above magnification step, the tetrahedron in the focus region has already been significantly deformed to a different shape, thus still using the unchanged coordinates to map and interpolate the texture will inevitably cause angle distortion. Therefore we need to recompute and modify the texture mapping to preserve the original texture shape after deformation. Fig. 7.5 uses a 2D example to illustrate the necessity of texture transfer. In the left figure, direct magnification without texture transfer produces severe distortion effect for the context region, which will be significantly improved after texture transfer as shown in the right figure.

The objective of this step is to texture the new mesh using the original texture, while preserving the interior texture shape. We have the tetrahedral mesh $M \in R^3$ and $M^{4D} \in R^4$ before and after the magnification. To transfer the texture information from $M$ (with the texture function $\varphi$) to $M^{4D}$, it is desirable to construct a function $\varphi^{12} : M^{4D} \rightarrow M$, that maps the entire space of $M^{4D}$ onto M. Then we can describe the transferred texture mapping function on $M^{4D}$ as $\varphi \circ \varphi^{12}$.

We solve the following harmonic function by computing $\varphi^{12}$ and minimizing the mapping distortion:

$$\Delta\varphi^{12} = 0, \tag{7.2}$$

where $\Delta = \frac{\partial^2}{\partial x^2} + \frac{\partial^2}{\partial y^2} + \frac{\partial^2}{\partial z^2}$.

In practice, solving equations for any but the simplest geometries must resort to an efficient approximation due to the lack of closed-form analytical solutions in the general setting. In our system we use discrete piece-wise coordinates to solve it numerically.

1. In $M^{4D}$, we shall use each vertex's original 3D position as the initial parameter $\varphi_i = (u_i, v_i, w_i) = (x_i, y_i, z_i)$.

2. To solve the harmonic function $\Delta \varphi^{12} = 0$, we iteratively update the parameter for each vertex $(u_i, v_i, w_i) = \sum_{Ng(v_i)} \omega_{ij}(u_j, v_j, w_j)$, where $Ng(v_i)$ is the one-ring neighbor of $v_i$, $(u_j, v_j, w_j)$ is every neighbor's parameter, $\omega_{ij}$ is the local coordinate associated with each neighbor. Meanwhile, vertices on the boundary of volume $M^{4D}$ serve as Dirichlet boundary conditions (i.e., we avoid changing their parameters).

3. $\varphi^{12}$ now maps vertex $v_i$ to one point location $(u_i, v_i, w_i)$ on $M$. Now we assign the texture parameter on this point in $M$ to $v_i$. This point must locate inside one tetrahedron $t_i$ in $M$, and the parameter of $v_i$ can be represented as the weighted average of four vertices' parameters on $t_i$. We again use Eq. 7.1 to compute the weight for four vertices on $t_i$.

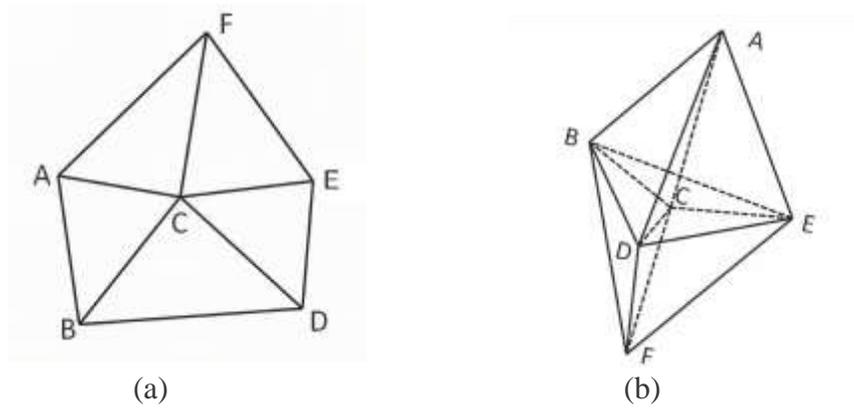

(a)                                              (b)

Figure 7.6: (a) Cotangent coordinates on a triangle mesh. (b) Cotangent coordinates on a tetrahedral mesh.

**Local Coordinates.** In our system solving Eq. 7.2 requires an affine combination as local coordinates $\omega_{ij}$. We require that $\sum_{Ng(v_i)} \omega_{ij} = 1$, and this partition of unity property allows us to use every vertex of a polygon as a basis to interpolate any function.

Cotangent coordinate is a robust coordinate system and widely used on triangle mesh processing. We use two angles opposite to one edge to compute its cotangent coordinate: $k_{C,D} = \cot \angle C\,BD + \cot \angle C\,ED$ for edge $E_{C\,D}$ (Fig. 7.6(a)). In our volume-based system, we generalize the formula from the triangle mesh onto the tetrahedral mesh, using cotangents of dihedral angles opposite to the edge. Note that, there are generally more than two tetrahedra sharing the same edge. Suppose for edge $E_{uv}$, it is shared by n tetrahedra thus it is corresponding to n dihedral angles, $\theta_i, i = 1, \ldots, n$, we define the string energy:

$$k_{u,v} = \sum_{i=1}^{\mathbf{X}} \cot \theta_i. \tag{7.3}$$

Then for a vertex $v_i$, we express its one-ring neighbor's local coordinates as:

$$\omega_{ij} = \frac{k_{i,j}}{\sum_{j \in N_{g(i)}} k_{i,j}}. \tag{7.4}$$

As shown in Eq. 7.3, determining local coordinates involves computing the dihedral angles between two faces. We compute a dihedral angle in 3D as follows. In Fig. 7.6(b), we can compute the cosine of the dihedral angle between two opposite faces $\triangle ABD$ and $\triangle CDB$ as the following multiplicative term (up to the product of the norm of these vectors):

$$(AB \wedge AD) \cdot (C\,D \wedge C\,B). \tag{7.5}$$

However, in our 4D space $\mathbf{M}^{4D}$ this formula is not suitable for computing. It turns out that in 4D space, cross product operator "$\wedge$" requires 3 vectors rather than just 2. To avoid using $\wedge$, we can use Lagrange's identity to compute the above formula:

$$(s \cdot u)(t \cdot v) - (s \cdot v)(t \cdot u) = (s \wedge t) \cdot (u \wedge v). \tag{7.6}$$

Now we can compute the cosine of the dihedral angle with the following updated formula:

$$(AB \cdot C\,D)(AD \cdot C\,B) - (AB \cdot C\,B)(AD \cdot C\,D). \tag{7.7}$$

## 7.3 Flattening

After the above step, we have already magnified the volume of ROI in a 4D mesh $\mathbf{M}^{4D}$. However, we have to flatten it back to 3D space for visualization and other typical applications. The key challenge in this step is to preserve every magnified tetrahedron's volume/shape during flattening. Inspired by 3D techniques like [55, 57], we devise a two-step algorithm to handle 4D flattening. We first rotate each 4D tetrahedron $\mathbf{t}_i^{4D}$ individually back to 3D space as the 3D tetrahedron (without changing shape except rotation). We denote this "flattened" 3D tetrahedron $\mathbf{t}_i^F$. Note that every tetrahedron is rotated back to 3D independently thus all $\mathbf{t}_i^{3D}$ are separate from each other without being glued together. The second stage includes stitching them together into one piece as the original tetrahedral mesh structure. During stitching we minimize the shape distortion such that the final tetrahedron in $\mathbf{M}_i^{out}$ preserves the shape of $\mathbf{t}_i^F$.

Rotating a 4D tetrahedron $\mathbf{t}_i^{4D}$ back to a 3D tetrahedron $\mathbf{t}_i^{3D}$ is simple. The challenge lies at keeping its shape close to $\mathbf{t}_i^F$ in the resulting mesh $\mathbf{M}_i^{out}$. Our system affords two iteratively computed phases to achieve this goal. To clearly describe the algorithm, we denote k as the current iteration, then $\mathbf{M}^k$, $v_i^k$, $\mathbf{t}_i^k$ as the tetrahedral mesh, a vertex and a tetrahedron in the k-th iteration, respectively. Note that $\mathbf{M}^k$ always keeps the same mesh structure as the input mesh M. Initially, we generate the mesh $\mathbf{M}^0$ in the first iteration by removing the fourth dimension from every vertex in $\mathbf{M}^{4D}$: For a vertex with $\mathbf{p}_i^{4D} = (x_i, y_i, z_i, h_i)$ in $\mathbf{M}^{4D}$, we initialize its position in $\mathbf{M}^0$ as $\mathbf{p}_i^{3D} = (x_i, y_i, z_i)$.

In the first phase we compute the Jacobian deformation matrix for each tetrahedron $\mathbf{t}_i^k$. The matrix represents the transformation from the localized flattened tetrahedron $\mathbf{t}_i^F$ to its counterpart $\mathbf{t}_i^k$. We represent this transformation as a $3 \times 3$ matrix $\mathbf{J}_i$. Generalized from [169], we can compute this Jacobian matrix as:

$$\mathbf{J}(\mathbf{t}_i^k) = \sum_{i=1} e_i^k (e_i^F)^T,\tag{7.8}$$

where $e_i^k$ and $(e_i^F)^T$ are the corresponding edges between $\mathbf{t}^k$ and $\mathbf{t}^F$ (Totally there are six pairs of edges for every tetrahedron). This matrix measures two tetrahedral deformation on two factors: rotation and scaling. Our goal is to preserve the shape of each tetrahedron thus we allow a rotation-only matrix, which can be decomposed separately by singular value decomposition of $\mathbf{J}$.

$$\mathbf{J}(\mathbf{t}_i^k) = \mathbf{U}\Sigma\mathbf{V}^T, \mathbf{R}_i = \mathbf{U}\mathbf{V}^T,\tag{7.9}$$

where $\mathbf{R}_i$ is the rotation-only matrix.

Now in the second phase, we can update the position of each vertex by minimizing the following energy:

$$E^k = \sum_{i} \sum_{j=1} \kappa_{ij} ||e_{ij}^k - \mathbf{R}_i e_{ij}^F||^2,\tag{7.10}$$

**Algorithm 3** The flattening algorithm.

```
Input: Initial 4D mesh M⁴ᴰ,
           threshold
Output: 3D mesh Mᵒᵘᵗ
```

**for all** $t_i^{4D} \in M^{4D}$ **do**

    `//Compute a flattened tetrahedron`

    $t_i^F = Flatten(t_i)$

**end for**

$M^0 = Initialize(M^{4D})$

$k = 0, d = $ `INF_MAX`

**while** $d > $ **do**

    **for all** $t_i^k \in M_i^k$ **do**

        `//Compute Jacobian matrix`

        $J_i = Jacobian(t_i^k, t_i^F)$

        `//Rotation-only matrix`

        $R_i = SVD(J_i)$

    **end for**

    `// Build and solve Eq. 12`

    $Assemble(L, R, V^F)$

    $V^k = SolveEquation(L, R, V^F)$

    `//Compute moving distance`

    $d = MaxDistance(M^{k-1}, M^k)$

    $k = k + 1$

**end while**

$M^{out} = M^k$

```
Output: Mᵒᵘᵗ
```

where $|\mathbf{T}|$ is the set of all tetrahedra, $e_{ij}^k$, $e_{ij}^F$ are 6 edges on the tetrahedron $\mathbf{t}_i^k$ and $\mathbf{t}_i^F$, $\kappa_{ij}$ is the weight associated with the edge. Now we rewrite the function in terms of every edge vector:

$$E^k = \sum_{m,n} \kappa_{mn} ||(v_m^k - v_n^k) - R_l(v_m^F - v_n^F)||^2, \qquad (7.11)$$

where we use $v_m^k - v_n^k$ to represent every edge in Eq. 7.10, $R_l$ and $\kappa_{mn}$ are the rotation-only matrix and weight of the tetrahedron $\mathbf{t}_l$ which the edge $(v_m^k, v_n^k)$ belongs to. Note that an edge $v_m^k - v_n^k$ may appear multiple times if it is shared by more than one tetrahedron, and thus we use different $R_l$ when the edge appears more than once. Setting the gradient to zero, we obtain the following linear equation:

$$L(V^k)^T = RL(V^F)^T, \qquad (7.12)$$

where the matrix $L$ represents the edge relationship of vertices (weighted by $\kappa_{mn}$) in Eq. 7.11. The matrix $R$ includes all local matrix $R_l$, $V^k$ and $V^F$ are vectors including all vertices' positions on $M^k$ and $M^F$. $V^k$ is the only unknown vector here and solving this equation gives rise to the positions of all vertices in $V^k$.

After updating the positions, we compute the moving distance for each vertex between $M^{k-1}$ and $M^k$. The distance is normalized to the diagonal length of the volume. We record the maximum moving distance among all vertices, and the iteration loop stops if this distance is smaller than the threshold. We set the threshold to be $1e^{-4}$. In practice for all experimental results our algorithm converges in at most 2 iterations.

**Weights.** The choice of weight $\kappa_{mn}$ in Eq. 7.11 depends on the importance of a tetrahedron. From the cognitive perspective, tetrahedra around the ROI center are more sensitive. Also a tetrahedron with large volume should have a higher weight than the one with small volume, because the distortion on a large tetrahedron is more visually confusing. For each edge, we design the weight as $(1 + h)V \, k_{u,v}$, where $V$ is the average volume of connected tetrahedra, h is the averaged height (h-values), and $k_{u,v}$ is computed from Eq. 7.3.

**Boundary Constraints.** For a solid textured model, it is necessary to keep the boundary shape. For a volumetric dataset, the user also prefers to get a resulting shape with an original square boundary. Therefore, we keep the position of every boundary vertex unchanged during all iterations.

# 7.4 Experimental Results and Discussions

Our system can effectively provide magnification information to the user, allowing the user to get detailed focal region while maintaining the integral perception of the model. The results shown in the following figures demonstrate the power of our techniques. Our experimental results are implemented on a 3GHz Pentium-IV PC with 4 Giga RAM.

We test our system on both solid textured models and volumetric datasets. From Fig. 7.7 to Fig. 7.8, we test various solid textured models such as watermelon, and kiwi and visualize their original/magnification results. Fig. 7.7 demonstrates one important application using our focus+context magnification. The figure shows that more seeds appear after magnification. Also, the distribution of seeds (i.e., their relative positions between seeds) is preserved. Preserving particle distribution and relative positions during magnifying has many potential applications in experiment-driven science and engineering (e.g., structural biology, game design, etc.). Our focus+context magnification provides an effective magnification lens for this category of applications.

Fig. 7.8 shows another example. Compared with [120], in which the sphere-like shape is severely distorted (e.g., the brain model is severely distorted to an irregular heart-like model), our lens successfully keeps the structure of kiwi core still as the spherical shape, and the shape of context region is also unchanged.

From Fig. 7.9 to Fig. 7.13, we test several volumetric dataset examples: aneurism, nucleon,lobb, bonsai and fuel. In these tests, we magnified different shapes like tumor in Fig. 7.9, oxygen atomic nucleus in Fig. 7.10, 3D wave in Fig. 7.11, trunk in Fig. 7.12, irregular air head in Fig. 7.13. All experimental results clearly demonstrate that our framework can keep the prominent global shape and the context region unchanged for viewers' easy recognition. Meanwhile, in Fig. 7.9 we demonstrate an application on structure-aware visualization using a model with many branches (note that a model's geometric structure typically has many branches). We magnify the tumor model while long branches (thinner vessels) are preserved without occlusion or relative position distortion. This example shows that our method could be of great value to structure-critical applications (e.g., oil pipeline optimization and detection, indoor routing and planning, etc.).

Fig. 7.12 demonstrates the application of arbitrary shape lens. In most of our examples we use the standard sphere shape lens. However, as we discussed in Section 7.2, we can generate arbitrary shape of lens from medial axis to preserve features. For example in Fig. 7.12, we utilize the medial axis of the trunk to generate a special lens for the trunk part. In the result, the trunk is magnified and the shape is well preserved.

In Fig. 7.14 and Fig. 7.15 we demonstrate more applications such as medical and physics experiment visualization with complicated models. We magnify the bladder part in Fig. 7.14 and the resulting model preserves the context region very well. The user (doctor) can easily recognize each surrounding part (pelvis, artery, etc.) without any difficulty. This advantage enables doctors to obtain the accurate information and avoid misdiagnose. In Fig. 7.15 , we magnify the smoke obstacle while we can still recognize the shape and number of surrounding flows.

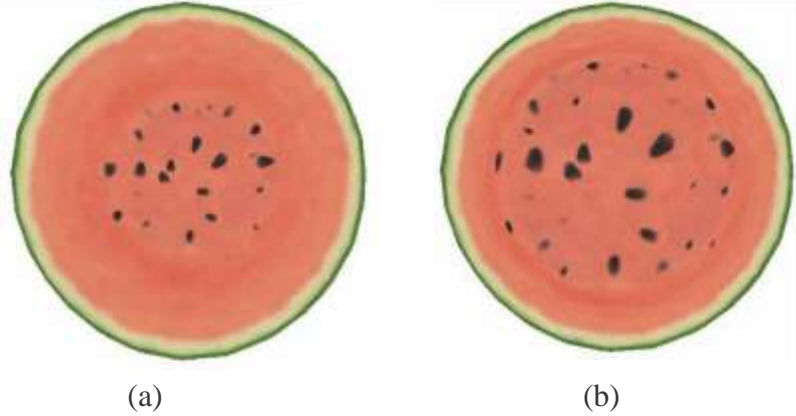

    (a)                  (b)

Figure 7.7: The tetrahedral mesh of watermelon.

**Performance.** The performance of our framework does not depend on the size of texture/volumetric datasets but the size of vertices in the input tetrahedral mesh. The statistics of examples are shown in Table 7.1. About computational time, in practice, we can interactively use a sparse low resolution cube/tetrahedral volume, like in [120], to accelerate the computation and get a fast result. Furthermore, we can pre-compute magnification and flattening on pre-designed mesh and later use it on different volumes by just changing textures of the mesh.

Compared with other optical/voxel/resizing based methods, our geometry-based method has the advantage that we can quantify the local distortion by computing mesh angle distortion, instead of just displaying visual effects. In the conventional lens design techniques, the user can only recognize the distortion through observation because of lacking an accurate measurement method. By comparison, our focus+context lens defines two categories of distortions: The local distortion and global distortion. We define the local distortion as the angle distortion in each tetrahedron. This metric can be quantified by computing the ratio of the single values $\sigma_1$ and $\sigma_2$ from the Jacobian matrix $\mathbf{J}$ (The metric is normalized by the diagonal length of the whole cube grid volume).

During our flattening step, one robustness issue involves avoiding self-intersection. This question is related to our flattening step. To theoretically illustrate its robustness on how to avoid self-intersection, we shall notice that our flattening algorithm is a 3D generalization from the surface method [55, 57], which is originally designed to handle very complex and/or high genus surface model input with no self-intersecting triangle in the output. In practice this method can effectively handle a model with very complex shape without self-intersection. Compared with these complex surface models, our model's geometry and topology structure is rather sim-ple: a flattened $R^3$ plane with a simple gaussian function in the middle. That means that, the deformation is rather slight from this simple input to a flattened output. In our experimental results, the self-intersection does not visually appear. Consequently, degeneration prevention is not practically necessary for our mesh, thus our system does not need to provide more mechanism to prevent self-intersection.

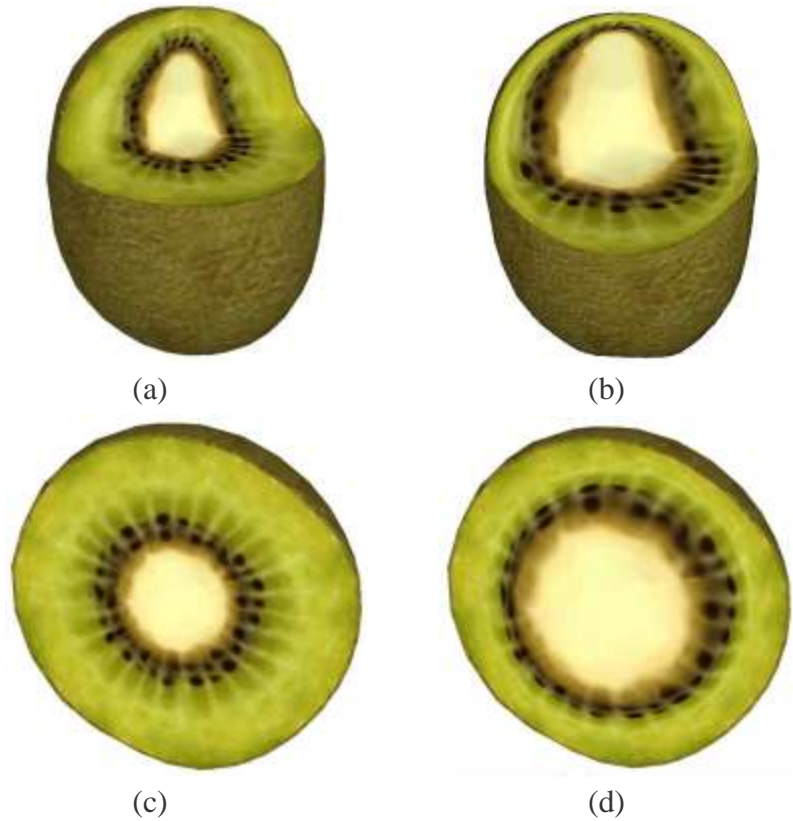

(a)                   (b)

(c)                   (d)

Figure 7.8: The tetrahedral mesh of kiwi.

Our flattening is computed iteratively. The convergence depends on the moving distance of every vertex between two iterations. We set a small number ($10^{-4}$) as the threshold. In each iteration, we compute this moving distance for every vertex (normalized by the diagonal length of the cube grid volume). The iteration stops if the maximum moving distance is smaller than the threshold. Our model converges in one or two iterations in all of our experiments. The reason of the fast convergence is that our tetrahedral mesh is very simple (just a volume as a $R^3$ plane with a simple gaussian function in the middle).

**Comparisons.** Currently most of magnification lens design focuses on 2D image visualization only. Recently, Wang et al. in [120] introduced a data reduction method which can achieve magnification effect. Compared with [120], our method's most important advantage is that: we can preserve both the focus region's shape and the nearby/global shape. Although the method in [120] can preserve the shape surrounding the focus region, it is incapable of preserving the nearby transition region (e.g., context), especially the global shape. These phenomena appear in the examples of [120] and show their method's major limitation. For example in [120] Fig.1 column 2, in order to magnify the focus region, the entire brain model (i.e., the global shape) is distorted significantly: from an original sphere-like shape to an irregular heart-like shape. In another focus+context visualization example in [120] Fig. 8, the contour of skull is severely deformed. Such severe distortion of the global shape may cause misunderstanding/misdiagnose ([174]). By comparison, our technique preserves the context region and global shape much better than [120]. For example, our method can keep the sphere boundary of watermelon and kiwi unchanged after magnification (Fig. 7.7, Fig. 7.8). Therefore, our method with an improved context region/global shape preserving capability could be more useful in the relevant applications.

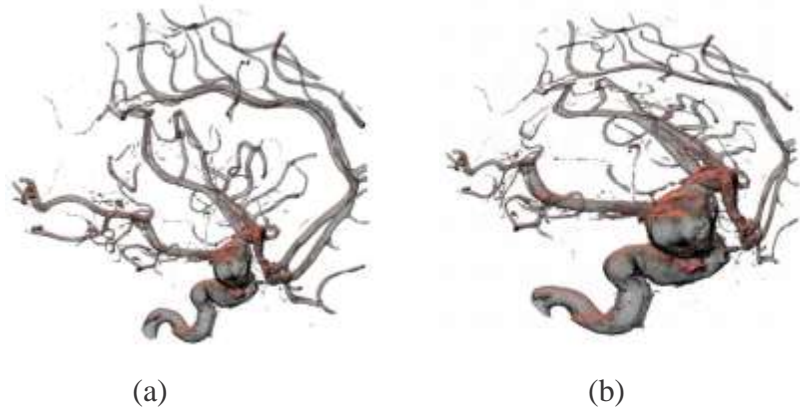

(a)             (b)

Figure 7.9: The volumetric aneurism dataset.

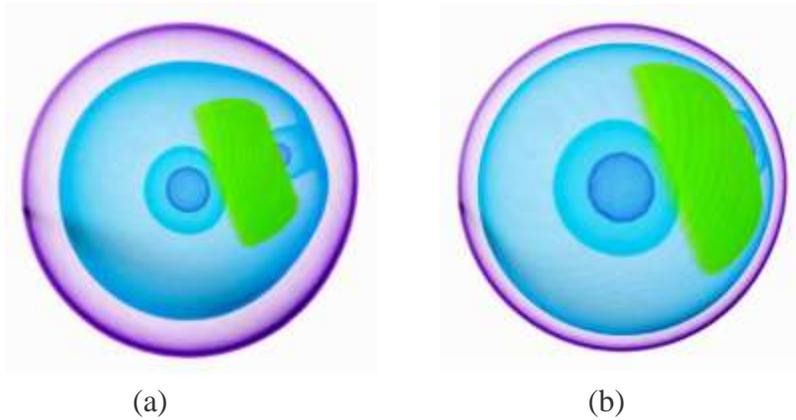

(a)             (b)

Figure 7.10: The volumetric nucleon dataset.

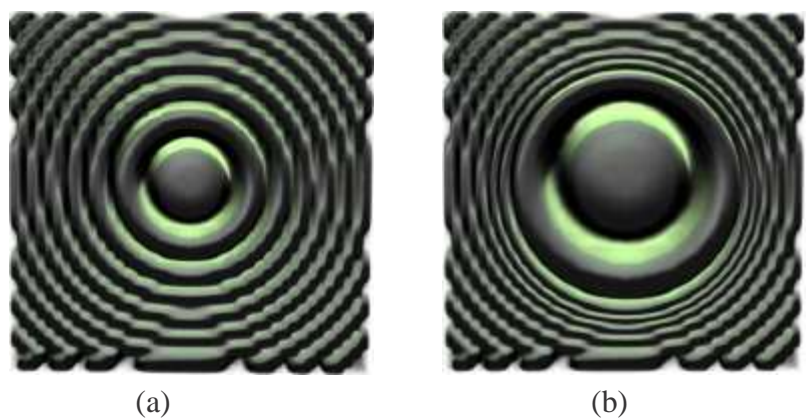

(a)                                    (b)

Figure 7.11: The volumetric marschner/lobb dataset.

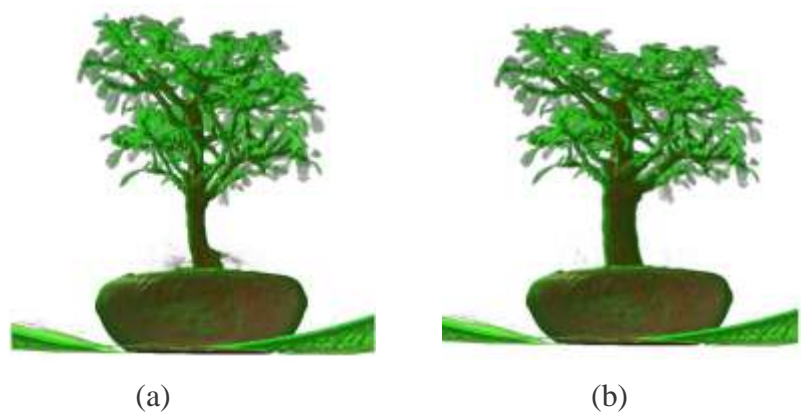

(a)                                    (b)

Figure 7.12: The volumetric bonsai tree dataset with magnified trunk.

Another comparison is on distortion controlling and quantifying. The distortion mechanism in [120] is highly arbitrary, determined by weighted cube grid magnification. Our method is geometry based and generalized from the surface conformal parameterization technique, thus we can control the local angle distortion much better from the perception's point of view. Angle-oriented shape perseveration and distortion minimization are more perceptually pleasing than using cube grid in [120]. Their cube resolution is very coarse with hundreds or voxels inside each cube. The linear interpolation of these voxels after cube grid deformation will cause additional angle distortion. Therefore, the cube grid distortion metric is always inaccurate. Our system can visualize the distortion not only through visual display but also quantifying such effect by computing angle distortion in a more

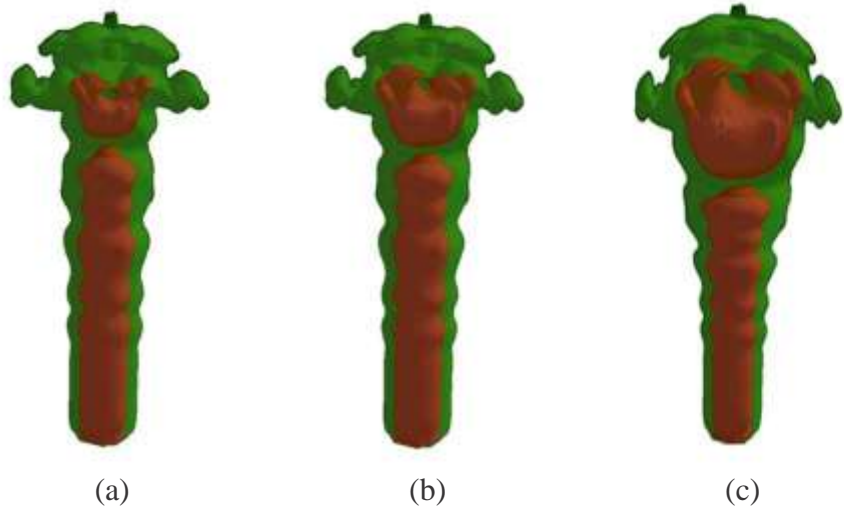

<div align="center">

(a)                    (b)                    (c)

</div>

Figure 7.13: The volumetric fuel dataset with magnified head.

accurate way (which is the ratio of two singular values from Jacobian matrix).

We also compare our method with another focus+context technique [119]. We shall notice that our system handles much more complicated scenarios than those in [119]. The input in [119] is only surface boundary model, so it has no interior or nearby information to display or magnify (all nearby context regions are empty 3D space). Consequently, [119]'s system can hide severe distortions in the empty context region without any visual information (since it is invisible). By comparison, our input is 3D solid model or volume with multiple materials/tissues, both inside the focus region and outside such region. When we magnify a focus region inside our model, all nearby context regions should avoid distortion because they also contain important tissue, material and shape information. By comparison, our geometry-based method can accommodate more complicated models with well-preserved magnification results for interior and exterior regions.

Since our lens is geometry-based, it can effectively obtain a better global distortion minimization even on surface mesh when only compared with [119]. We can simply modify our framework to support surface-only triangle mesh: we use a polycube to cover the whole input mesh and then magnify the polycube. Fig. 7.16 compares our method with the result in [119]. After setting the user-selected focus region (red circle in Fig. 7.16(a)), the magnification result generated by Wang's method preserves structure/shape in the focus area, but severely affects the context region (e.g., the upper body, red circle in Fig. 7.16(b)) and introduces visual artifacts, like the distorted proportion of body. By comparison, our technique keeps upper/lower body proportion without obvious shape confusion for easy object cognition (red circle in Fig. 7.16(c)).

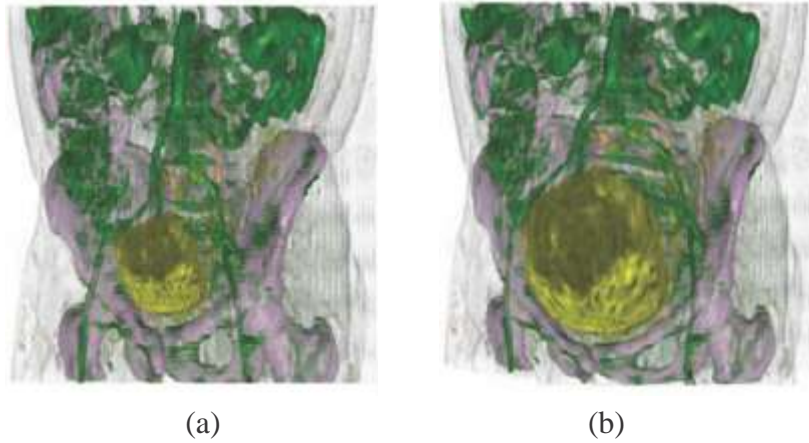

(a)                                     (b)

Figure 7.14: The volumetric fuel dataset with bladder.

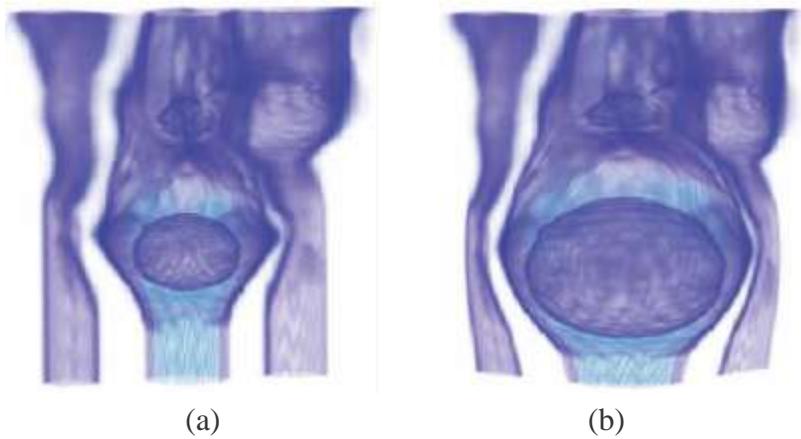

(a)                                     (b)

Figure 7.15: The volumetric fuel dataset with smoke.

Our lens is also similar to the mesh editing based method (which is equivalent to magnifying the surface boundary first and then interpolating the interior texture). However mesh editing techniques are not suitable for the focus+context visualization application because they focus on totally different input and task. First, mesh editing requires users to operate on an exact mesh boundary segment. However, in focus+context visualization applications, the desired regions can not be easily detected, extracted, and described as the triangle mesh model. For example, boundary extraction is extremely difficult for most volumetric/medical datasets. In most practical focus+context visualizations, the user only chooses a general/approximate region via simple user sketch and/or basic geometric primitives (like the region in a drawn circle), enclosing both mesh segment and nearby context space as a reasonable proxy. Second, mesh editing only attempts to preserve the shape of focus region around mesh boundary during magnification. The nearby context region and global shapes will be severely distorted without consideration (In most cases, these regions are just empty space in a typical mesh editing task). Finally, mesh editing only focuses on surface mesh's shape, thus for interior textures/tissues, we still need to design a shape-preserving interpolation technique to preserve the shape after boundary deformation. In Table 7.2, we compare our method with [119, 120] and mesh editing methods. The table clearly shows that our method is a more powerful tool for volume data focus+context visualization.

Table 7.1: Statistics of various test examples: # Tex, # pixels in the texture; # V, # of vertices; Distortion, average distortions by all vertices.

| Model | # Tex | # V | Time | Distortion |
|---|---|---|---|---|
| melon | $64^3$ | 600 | 1.5s | 0.05 |
| kiwi | $64^3$ | 880 | 2.1s | 0.08 |
| aneurism | $256^3$ | $30^3$ | 315s | 0.07 |
| lobb | $41^3$ | $20^3$ | 46s | 0.05 |
| nucleon | $41^3$ | $20^3$ | 44s | 0.04 |
| bonsai | $256^3$ | $40 \times 20^2$ | 127s | 0.03 |
| fuel | $64^3$ | $25^3$ | 110s | 0.08 |
| bladder | $128^3$ | $30^3$ | 275s | 0.03 |
| smoke | $256^3$ | $30^3$ | 340s | 0.04 |

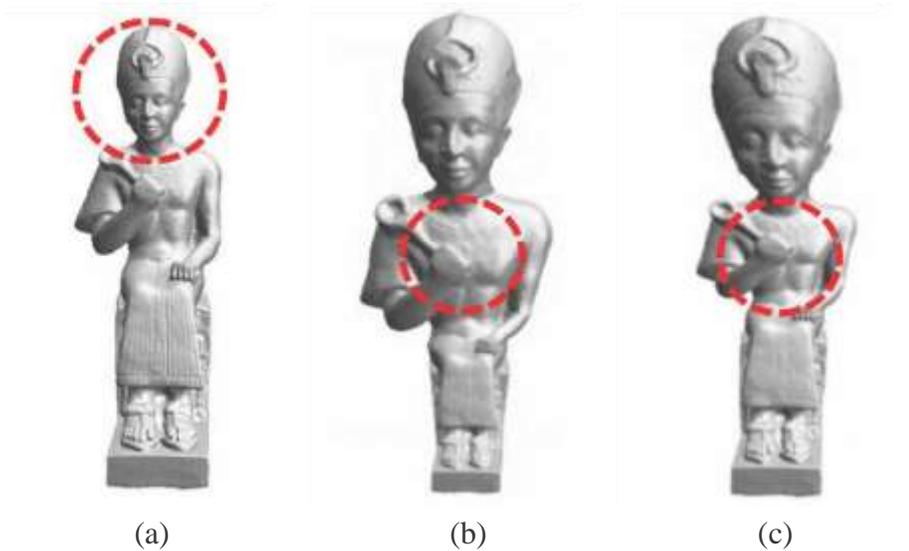

(a)                    (b)                    (c)

Figure 7.16: Comparison between Wang's method. (a) Input and the focus region (red circle). (b) [119]'s method (courtesy to [119]) and its resulting context region (red circle). (c) Our method and the resulting context region (red circle).

Table 7.2: Comparison with various methods. We test their abilities in following aspects: Preserving the shape of focus region (Focus Region); Preserving the shape around context region and/or global shape (Context Region Global Shape); Supporting solid model and/or volume dataset (Solid Texture); Quantifying local distortion (Distortion Quantifying); Allowing simple sketch to choose ROI (Simple Sketch Input).

| Model | [120] | [119] | Mesh Editing | Ours |
|---|---|---|---|---|
| Focus Region | yes | yes | interpolation needed | yes |
| Context Region Global Shape | no | no | empty space | yes |
| Solid Texture | yes | no | no | yes |
| Distortion Quantifying | no | no | yes | yes |
| Simple Sketch Input | yes | yes | no | yes |

## 7.5 Chapter Summary

In this chapter, A novel framework towards designing magnification lens for volumetric datasets is introduced. Specifically, it starts from the input of a 3D tetrahedral mesh and magnify the ROIs in 4D space through the use of dimensional enhancement. The geometry-centric methodology and the prototype system manifest that the 4D geometry greatly empowers the visualization techniques. This approach is expected to transcend the traditional boundary of geometric modeling and is of benefit to data visualization and visual analysis.

From the focus+context visualization application's perspective, this framework outperforms other methods with many unique features. In this system, the geometry-centric techniques offer users the immense power on shape distortion minimization and its quantitative control. Compared with other methods, it can preserve the shape not only around the focus region but also the surrounding context region and global shape. Also, it enables the user to draw either simple sketch (like drawing a sphere) or arbitrary shape as magnifiers to effectively display the entire ROIs. This system affords a wide spectrum of 3D input ranging from volume datasets to solid textured models. All experimental results have demonstrated that 4D lens, as a novel magnification technique, has great potentials in many visualization applications.

In near future, this system can be extended to the exploration of the utility of 4D geometric modeling/processing. At present, this framework still lacks of mechanism to handle sharp features in the input volume, especially when shape features are on the boundary of the context region. For most visualization applications, like medical data visualization, this drawback may not be obvious. However, for visualization involving manufactured objects in game development and traditional CAD, this distortion may cause severe difficulty during object exploration. It is desirable to study how to design better algorithms to support this type of applications to keep meaningful sharp features (e.g., shape crack) unchanged during magnification. Meanwhile, it can also be observed from the examples that the current scheme is capable of handling higher-dimensional datasets, like solid textured models equipped with multiple vector fields. The method could be extended to support multi-scale resolutions, and explore its application on more generalized models like multivariate splines and achieve parallel acceleration on GPU platform.

# Chapter 8

# Conclusion and Future Work

In this dissertation, we present our recent research results, ongoing research and future research direction within our general volumetric spline-based modeling framework. We seek novel modeling techniques based on trivariate tensor-product spline schemes that would allow users to directly construct regular trivariate splines over 3D surface models and preserve all useful properties. Theoretically, it brings fundamental progress in understanding, analyzing and solving volumetric modeling problems. We also demonstrate its great potential in many valuable applications like remeshing, visualization, etc.

## 8.1 Contribution Summary

In this thesis, we have investigated and presented a spline-based volumetric modeling framework to solve 3D objects modeling problems. Particularly, we emphasize our research interest on regular domain ("cuboid") tensor-product splines, because of their favorite advantages. Combining volumetric decomposition, parameterization with trivariate splines, we successfully and effectively solve a variety of problems in the areas of geometric shape design and modeling.

Our specific contributions include:

1. We propose a new concept of *"Generalized poly-cube"* (GPC). A GPC comprises a set of regular cube domains topologically glued together. Compared with conventional poly-cubes (CPCs), GPC is much more powerful and flexible and has improved numerical accuracy and computational efficiency. We propose an automatic method to construct a GPC domain and we develop a novel volumetric parameterization and spline construction framework based on the resulting domain, which is an effective modeling tool for converting surface meshes to volumetric splines.

2. We design a novel component-aware shape modeling methodology based on tensor-product trivariate splines for solids with arbitrary topology. Instead of using conventional top-down method, our framework advocates a divide-and-conquer strategy: The model is first decomposed into a set of components and then each component is naturally modeled as tensor-product trivariate splines. The key novelty lies at our powerful merging strategy that can glue tensor-product spline solids together subject to high-order global continuities, meanwhile preserving boundary restriction and semi-standardness.

3. We propose a systematic framework that transforms discrete volumetric raw data from scanning devices directly into continuous spline representation with regular tensor-product structure. To achieve this goal, we propose a novel volumetric parameterization technique that constructs an as-smooth-as-possible frame field, satisfying a sparse set of directional constraints and compute a globally smooth parameterization with iso-parameter curves following the frame field directions. The proposed method can efficiently reconstruct model with multi-layers and heterogenous materials, which are usually extremely difficult to be handled by the traditional techniques.

4. Aiming to promote new applications of our powerful modeling techniques in visual computing, we present a novel methodology based on geometric deformation metrics to simulate magnification lens that can be utilized for Focus+Context (F+C) visualization. Compared with conventional optical lens design (such as fish-eyes, bi-focal lens), our geometric modeling based method are much more capable of preserving shape features (such as angles, rigidities) and minimizing distortion. We present a novel methodology that integrates 4-Dimensional space deformation to simulate magnification lens on versatile textured solid models.

Practically, we demonstrate their power in many valuable applications, and show their great potential as enabling tools serving for research in broad areas of computer graphics, geometric modeling and processing. Our spline-based framework is endowed with many advantageous properties for modeling continuous quantities defined over multiple domains. Through our extensive experiments, we demonstrate that our framework is more efficient and effective in solving a variety of problems in computer graphics, image processing and other engineering applications.

## 8.2   Future Improvement of Our Work

There are many more immediate and valuable research topics based on our current framework. Here are some research topics that directly extend from work we have done in this dissertation.

We would like to further improve the current stage of our automatic generalized poly-cube construction framework. Our proposed method has certain limitations and demands further improvement in the future. First, the constructed poly-cube mainly depends on the segmentation of the 3D model. Different segmentation may result in very different generalized poly-cubes. In our implementation, we require to generate component-aware segmentation before the poly-cube construction. However, in practice it is always extremely difficult to implement component-aware segmentation. Furthermore, a general component-aware technique may lead to the segmentation result in which many resulting components glue together around the same point or edge. Such a point or edge is extraordinary point it is impossible to approximate continuous planes around such a point. Currently in our existing framework, we use topology-based method (like pants decomposition) or skeleton based method to get component-aware segmentation. These methods has certain limitations and demands further improvement. First, pants decomposition is designed to handle surface modeling and processing like surface mapping. Pants decomposition is directed by topology knowledge only so it is not natural to generate component-aware knowledge. Meanwhile, the skeleton of 3D model could be very complicated with arbitrary branch connection types in real applications. However pants decomposition is suitable for "3 branches merging" only (degree equals to 3 in the skeleton). One potential solution is to first compute the skeleton representation of the given 3D surface. Then we regularize for the generalized skeleton so that we can converting any merging types (with arbitrary branches merging) into the regular cube domain without extraordinary points. By doing so, out generalized poly-cube can handle any shape with very complex skeleton in a divide-and-conquer fashion. An optimized "skeleton-to-cube domain" conversion needs to consider two parameters: the number of branches and the angle between two branches, which will allows us to acquire improved poly-cube mapping and thereby to better spline fitting, texture mapping and synthesis and other further applications.

We also would like to further strengthen our current poly-cube framework. Within the existing framework, users are not allowed to directly specify the extraordinary (corner) points of the poly-cubes on the input 3D surfaces. The cube generation mainly depends on the model's topology. Consequently, no important geometric features exist in the domain representation. We attempt to provide meaningful help to integrate the sharp feature information into the parametric domain. This can also improve the quality of the poly-cube maps. One possible way is to automatically extract the sharp edges and corners first. Then we seek to map the sharp edge to the cube domain edge, corner to the cube domain corner.

During the research of volumetric modeling, we also realize that current existing papers only take surface feature into consideration. We also attempt to integrate the interior feature processing into our generalized poly-cube framework. For 3D surface models, one important interior feature is mid-structure plane. It is analogous to the medial axis of the 2D models. Wee seek to generate the poly-cube domain integrating sharp features on the mid-structure plane. One potential solution is to first compute the skeleton representation of the given 3D surface model. Then we cut the skeleton's orthogonal plane along each point on the skeleton. The cutting plane may include feature edges cutting from the mid-structure plane and we use them to decompose each plane into several quadrilateral meshes. Then we merge the neighboring cutting planes' quadrilateral meshes into poly-cubes.

Hierarchical structure and continuous representation ability are two advantage of our volumetric spline framework. Naturally we want to see their potential application on relative physical-based applications like mechanical analysis, shape deformation, fluid dynamics, collision detection, etc. First, hierarchical structure can allows us to implement a fast simulation on the low resolution model and then generate an accurate result on a high resolution model. This ability enables out framework to provide the flexible performance on the limited computation unite device like smartphone; Continuous representation allows us to implement more direct and accurate physics computation. For example, by doing computing like FEM/FD on this framework, the number of degree of freedom will be much fewer, which will thereby leads to faster and better fluid simulation and collision/detection.

In addition, the regular structure of cubes will for sure facilitate the parallel based applications like volume rendering, optimization, fluid simulation, FEM, etc. The highly data-parallel nature of tensor-product spline computation also enables GPUs to use local memories and multi-cores more directly for computation, achieving higher arithmetic intensity. To utilize it, general volume modeling computations must be recast into hardware-specific terms in order to utilize the underlying hardware. In current popular mobile device architecture, the main hardware system is CPU+GPU. Therefore, it requires specific design to assign different operations on two processing units and minimize the communication between them. However, not every scientific computation in volumetric modeling can take full advantage of the CPU+GPU structure, especially the modeling of complex geometric shapes of arbitrary topology, due to the lack of inherent regularity structure (or parametric domain). Our regular domain can bridge the gap by introducing poly-cube mapping of complex shapes onto regular parametric domain, such that the complex geometric models can be represented as 3D geometric texture in order of the GPUs to perform the general data registration, modeling, and visualization tasks in a high parallel fashion. The GPU-centric data formats and models will enable the efficient implementation of shape registration, solid modeling, multi-scale data modeling via reverse engineering, simulation/analysis, and model visualization. Meanwhile, the efficient GPU-based algorithm will enhance existing algorithm functionalities with improved parallel performance in order to handle large-scale, complicated models.

We also expect to extend our current trivariate generalized poly-cube splines to higher dimensional splines through the volumetric parameterization on volumetric domains, and seek potential applications on heterogeneous volume modeling, simulation, finite element analysis and scientific visualization. The high dimensional model (e.g., 4-Dimensional domain) provides extra flexibility to deform and magnify the volumetric model while still preserving the properties (like shape, geometry, physical laws, etc). The high dimensional framework will provide improved visualization method for solid model and facilitate representations of the design, testing of complicated mechanical objects and will also facilitate the specification of material distributions.

## 8.3   Concluding Remarks

These direction for future work, and the many other open problems that exist, are sure to encourage interesting and exciting research for years to come. As technical difficulties are overcome, and existing computational algorithms are improved, the applications will increase in variety and number. We are pleased to have taken the first step in uncovering the heretofore untapped potential by presenting our framework to the graphics and visual computing. It is our hope that this integrated approach and demonstrated applications will foster continued interest and research in this area. We look forward to the continued exploration of modeling and predict a successful future on it.